\documentclass[11pt]{article}
\pdfoutput=1
\usepackage{jcapmod}
\usepackage{booktabs}
\usepackage{mathtools}
\usepackage{microtype}
\usepackage{bigints}
\usepackage[utf8]{inputenc}
\usepackage[english]{babel}
\usepackage{colortbl}
\usepackage[dvipsnames]{xcolor}
\usepackage{amsmath,amssymb,amsbsy,amstext,amsthm, xcolor}
\usepackage[colorlinks=true]{hyperref}
\usepackage{microtype}
\usepackage{graphicx}
\usepackage{amsfonts}
\usepackage{float}
\usepackage{bbold}
\usepackage{siunitx}
\usepackage{fontawesome5}
\usepackage[most]{tcolorbox}
\RequirePackage{color}

\allowdisplaybreaks[1]

\makeatletter
\newlength{\apb@width}
\newcommand{\autoparbox}[2][c]{\settowidth{\apb@width}{#2}\parbox[#1]{\apb@width}{#2}}

\makeatother

\numberwithin{equation}{section}

\def\beq{\begin{equation}}
\def\eeq{\end{equation}}

\def\bea{\begin{eqnarray}}
\def\eea{\end{eqnarray}}
\def\eg{{\it e.g.~}}

\def\ie{{\it i.e.~}}
\def\d{{\rm d}}

\def\d{{\rm d}}

\def\nn{\nonumber}

\def\sgm{\sigma}

\def\Mpl{M_{\rm pl}}

\def\fr{\frac}

\def\fr{\frac}

\usepackage{setspace} 

\begin{document}

\begin{titlepage}
\setcounter{page}{1} \baselineskip=15.5pt \thispagestyle{empty}
\bigskip\
\vspace{1cm}
\begin{center}
{\fontsize{20}{28}  \selectfont 
\bf Inflation and Primordial Black Holes\\}
\end{center}
\vspace{0.2cm}
\begin{center}
{\fontsize{13}{30}\selectfont Ogan \"Ozsoy$^{1,2}$\note{Correspondence e-mail: \href{mailto:ogan.ozsoy@csic.es}{ogan.ozsoy@csic.es}},\,\, Gianmassimo Tasinato$^{3,4}$
}
\end{center}
\begin{center}
\vskip 8pt
\textsl{$^1$ CEICO, Institute of Physics of the Czech Academy of Sciences, Na Slovance 1999/2, 182 21,
Prague.}
\vskip 3pt
\textsl{$^2$ Instituto de Física Téorica UAM/CSIC, Calle Nicolás Cabrera 13-15, Cantoblanco,
28049, Madrid, Spain.}
\vskip 3pt
\textsl{$^3$ Dipartimento di Fisica e Astronomia, Universit\`a di Bologna, via Irnerio 46, Bologna, Italy.}
\vskip 3pt
\textsl{$^4$ Department of Physics, Swansea University, Swansea, SA2 8PP, United Kingdom.}
\end{center}
\vspace{1.2cm}
\noindent
\begin{abstract}

We review
conceptual aspects  of  
 inflationary scenarios able to produce primordial black holes,  by 
 amplifying
  the size of  curvature fluctuations to the level required for 
   triggering black hole formation.   
   We identify  general mechanisms to do so, 
   both for single and multiple field inflation.
   In
  single field inflation,  
  the spectrum of curvature fluctuations is  enhanced by
pronounced gradients of background
quantities controlling the cosmological dynamics,
which can induce brief phases of non--slow-roll inflationary evolution.
 In multiple field inflation, the amplification occurs through appropriate couplings with additional sectors,  characterized by tachyonic instabilities that enhance the size of their fluctuations. As representative examples, we consider axion inflation, and two-field models of inflation with rapid turns in field space.
We  develop our discussion in a pedagogical manner, by including some of the most relevant calculations, 
 and by  guiding the reader 
 through  the existing theoretical  literature,  emphasizing general themes
common to several models.

\end{abstract}

\vskip 10pt
\vspace{0.6cm}
\end{titlepage}
\tableofcontents
\newpage
\section{Introduction}

\noindent
{\bf Primordial black holes: history of the   concept}

\smallskip
\noindent
Inflation, a short period of accelerated expansion in the very early moments of the universe, has become one of the main pillars of modern cosmology \cite{Guth:1980zm,Linde:1981mu}. Leaving aside its success in addressing the puzzles of the standard hot Big Bang cosmology, inflation provides an explanation for the quantum mechanical origin of structures such as galaxies (including our own!) and the anisotropies in the Cosmic Microwave Background (CMB) radiation \cite{Penzias:1965wn}. In the last two decades, the advances in the observational cosmology and in particular the observations of the CMB and of the large scale structure (LSS) of our universe have so far confirmed the predictions of  inflation, and arguably established its status as the main theoretical framework describing the very early universe \cite{Planck:2018jri,Planck:2019kim}. These successes notwithstanding, CMB and LSS probes only provide us information on the early universe at the largest  cosmological scales ($10^{-4} \lesssim k \,[ \mathrm{Mpc}^{-1}]\, \lesssim 10^{-1}$) corresponding to a small fraction of the early stages of inflationary dynamics. 
Hence,
 while inflation provides us with a consistent, testable framework in understanding the initial conditions in the universe at the largest scales, we do not have direct access to most of 
 the inflationary dynamics, and to the universe evolution in the early post-inflationary era. 
Importantly, these stages could be host to a number of interesting phenomena, including the production of stable relics such as \emph{dark matter} (see \eg \cite{Bertone:2016nfn} for a historical review on dark matter) that is essential in understanding the world we observe today, as well as for establishing new physics. Indeed, the existence of non-luminous, cold dark matter (CDM) that constitutes a quarter of the total energy budget in the universe \cite{Planck:2018vyg} is one of the most glaring evidences for beyond the Standard Model physics \cite{Bergstrom:2000pn}. The absence of signatures from collider experiments, along with unsuccessful  direct and indirect detection searches, have all made the DM puzzle particularly compelling \cite{Bertone:2004pz}.

An intriguing and 
 economical explanation that might account for DM density in our universe is a scenario where DM is made of compact objects, such as primordial black holes (PBHs). Pioneered by the works of Y. Zel'dovic and I. Novikov \cite{Zeldovich:1967lct} and S. Hawking \cite{Hawking:1971ei}, the initial ideas in this direction  began with the realization that PBHs could form by the gravitational collapse of over-dense inhomogeneities in the early universe. In the mid 70's, it was later realized by the works of B. Carr \cite{Carr:1974nx,Carr:1975qj} and G. Chapline \cite{Chapline:1975ojl} that PBHs could contribute to DM density and provide the seeds for the supermassive BHs populating our universe \cite{1984MNRAS.206.801C}. Following these theoretical progresses, the interest of the scientific community on PBHs has risen in the mid 90's by the reported detection of micro-lensing events from MACHO collaboration \cite{Aubourg:1993wb}. An immediate interpretation of these results was suggesting on the possibility that a significant fraction of mass density in our galaxy could be composed of sub-solar mass PBHs. However, these considerations were later rendered invalid by the findings of EROS \cite{EROS-2:2006ryy} and OGLE \cite{Wyrzykowski:2010mh,Wyrzykowski:2011tr} collaborations,  concluding that only a small fraction of mass in the Milky Way could be in the form of PBHs. 

\begin{figure}[t!]
\begin{center}
\includegraphics[scale=0.750]{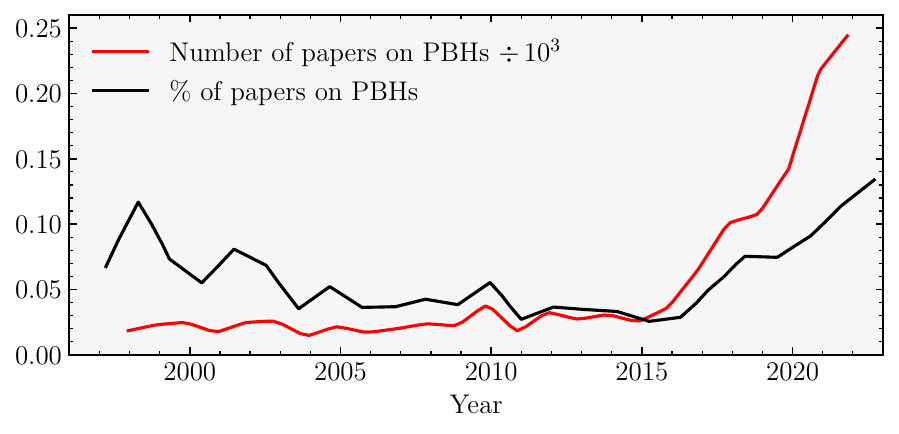}
\end{center}
\caption{Total and relative number of manuscripts appeared on the arXiv from 1996 until today related to various aspects regarding primordial black holes. Spikes of activity in the literature, particularly after the mid 90's due to claimed lensing events by the MACHO collaboration and the GW detection by LIGO in 2015 is clearly visible. Data is extracted from the \href{https://www.benty-fields.com/trending}{website}. \label{fig:pbhdat}}
\end{figure}

Stimulated both by the absence of signals for well-motivated particle DM candidates, and the first detection of gravitational waves (GWs) from merging BHs by the LIGO/VIRGO collaboration \cite{Abbott:2016blz}, a second surge of interest in PBHs was ignited (see Fig. \ref{fig:pbhdat}). In particular, different groups suggested  that merging PBHs could be responsible for the observed GW signals, while constituting a significant fraction of DM density in our universe \cite{Bird:2016dcv,Clesse:2016vqa,Sasaki:2016jop}. Since the first appearance of these articles, a significant amount of effort has been pushed forward by the community,   to search and constrain the abundance of PBHs by utilizing their gravitational and electromagnetic effects on the environment at small scales. Various experiments set stringent constraints on PBH abundance for solar and sub-solar mass range, leaving a viable window for this scenario for tiny PBH masses $10^{-17} \lesssim M_{\mathrm{pbh}} \,\left[M_{\odot}\right] \lesssim 10^{-12}$ ($M_{\odot} \simeq 1.98 \times 10^{33}\, {\rm gr}$) as the totality of DM (see \eg \cite{Niikura:2017zjd,Katz:2018zrn,Montero-Camacho:2019jte}). 
It should be noted that some of the constraints derived in the literature make specific assumptions about the formation process and the subsequent evolution of PBHs (such as monochromatic mass functions, clustering and accretion processes, etc) and on other model dependent specifics (such as non-Gaussianity) which could relax or tighten these constraints \footnote{See \eg the impact of extended mass functions on the PBH abundance constraints \cite{Carr:2017jsz}.}. Since we mostly focus on the subject of inflationary model building,  we will not review these issues and aforementioned constraints, but the interested reader can find more details in  excellent reviews published  recently, see {\it e.g.} \cite{Khlopov:2008qy,Garcia-Bellido:2017fdg,Sasaki:2018dmp,Carr:2020xqk,Carr:2020gox,Green:2020jor,Escriva:2022duf}. 

 PBHs are likely to form well before the end of the radiation dominated era (\ie before the so called matter-radiation equality), and behave like cold and collision-less matter. Therefore they constitute an interesting DM candidate, if they are massive enough $M_{\rm pbh} \gtrsim 10^{15}\, {\rm g} \simeq 10^{-18} M_{\odot}$ to ensure a lifetime comparable with the age of the universe \cite{Page:1976df} \footnote{Ultralight PBHs with $M_{\rm pbh} \ll 10^{15} {\rm g}$ -- although they can not account for the DM density-- may also have interesting observational effects if they come to dominate the energy budget of the universe before BBN, see \eg \cite{Papanikolaou:2020qtd,Papanikolaou:2022chm}.}. In this context, a particularly appealing aspect of PBH dark matter is its economical and minimal structure, in the sense that this scenario does not require any additional beyond Standard Model (BSM) physics (such as new particles and interactions), provided that one alters the not-so-well constrained early universe at small scales by introducing a viable mechanism to account for the production of large density fluctuations required for PBH formation.

Similar to the generation of CMB anisotropies, a compelling and natural source of these perturbations in the early universe could be the quantum fluctuations that are stretched outside the horizon during inflation. However, in order to generate such over-dense regions that can collapse to form PBHs in the post-inflationary universe, one needs to devise a mechanism to enhance by several orders of magnitude the  inflationary scalar perturbations at small scales $k \gg k_{\rm cmb}$ (corresponding to late stages of inflation),  far above the values required to match CMB observations. As the observed temperature anisotropies prefers a red tilted power spectrum at CMB scales, this situation generically requires a blue tilted power spectrum, or some specific features at scales associated with PBH formation. 

In the context of canonical single scalar field inflation, the first ideas in this direction appeared in  works by P. Ivanov, P. Naselsky and I. Novikov \cite{Ivanov:1994pa} (see also \cite{Starobinsky:1992ts}). In particular, these authors  have shown that if the inflaton  potential has a very flat plateau-like region for  field ranges corresponding to the late stages of accelerated expansion, the inflationary dynamics  enters a ``non-attractor" regime called ultra slow-roll (USR) \cite{Dimopoulos:2017ged}. This  leads to super-horizon growth of scalar perturbations \cite{Leach:2000yw,Leach:2001zf,Kinney:2005vj} that can eventually trigger PBH formation in the post-inflationary universe\footnote{Another inflationary background that exhibit similar features is called constant-roll inflation, see \eg \cite{Martin:2012pe,Motohashi:2014ppa}.}. 
Many explicit single-field inflationary models that exhibit similar local features were subsequently studied in the literature: for a partial list of popular works see \eg \cite{Garcia-Bellido:2017mdw,Ezquiaga:2017fvi,Germani:2017bcs,Ballesteros:2017fsr,Hertzberg:2017dkh,Cicoli:2018asa,Ozsoy:2018flq,Mishra:2019pzq, Dimopoulos:2019wew,Ballesteros:2020qam,Inomata:2021tpx} (see also
 \cite{Garcia-Bellido:1996mdl,Kawasaki:1997ju,Yokoyama:1998pt,Kawaguchi:2007fz,Kohri:2007qn,Frampton:2010sw,Drees:2011yz} for earlier constructions). In the context of single scalar field inflation, another possibility to generate an enhancement in the scalar power spectrum is to invoke a variation of the sound speed of scalar fluctuations, for example through a reduction in the speed of sound $c_s^2$ \cite{Ozsoy:2018flq,Ballesteros:2018wlw,Kamenshchik:2021kcw} or through a rapidly oscillating $c_s^2$ which triggers a resonant instability in the scalar sector \cite{Cai:2018tuh,Chen:2020uhe}. 

From a top-down model building perspective, a rich particle content during inflation is not just an interesting possibility, but appears to be a common outcome of many BSM theories \cite{Baumann:2014nda}. 
Since the early days of research on PBHs, multi-field inflationary scenarios has also attracted considerable attention as a natural way to realize enhancement in the scalar perturbations at small scales. 
For instance, large scalar perturbations may arise through  instabilities arising in the scalar sector, \eg during the waterfall phase of hybrid inflation \cite{Randall:1995dj,Garcia-Bellido:1996mdl,Clesse:2015wea} or due to turning trajectories in the multi-scalar inflationary landscape, as reported recently in \cite{Brown:2017osf,Palma:2020ejf,Fumagalli:2020adf,Braglia:2020eai,Zhou:2020kkf,Iacconi:2021ltm,Kallosh:2022vha,Kawai:2022emp}. Another intriguing possibility in this context is by employing axion-gauge field dynamics during inflation \cite{Linde:2012bt,Bugaev:2013fya,Garcia-Bellido:2016dkw,Domcke:2017fix,Cheng:2018yyr,Kawasaki:2019hvt,Ozsoy:2020ccy,Ozsoy:2020kat}. In these models, particle production in the gauge field sector act as a source for the scalar fluctuations, and hence can be responsible for PBH formation.

A common feature of all inflationary scenarios capable of producing PBH populations is the inevitable
production of a stochastic GW background (SGWB) induced through higher order gravitational interactions between enhanced
scalar and tensor fluctuations of the metric \cite{Ananda:2006af,Baumann:2007zm,Kohri:2018awv}. Interestingly, this signal may carry crucial information about the properties of its sources including the amplitude, statistics and spectral shape of scalar perturbations (see \eg \cite{Nakama:2016gzw,Cai:2018dig,Unal:2018yaa,Yuan:2019wwo,Cai:2019cdl,Ozsoy:2019lyy,Pi:2020otn,Yuan:2020iwf,Ragavendra:2021qdu}) and could provide invaluable information on the underlying inflationary production mechanism. Furthermore, since the resulting GW background interacts very weakly with the intervening matter between the time of their formation and today, it leads to a rather clean probe of the underlying PBH formation scenario. This allows us to access inflationary dynamics on scales much smaller than those currently probed with CMB and LSS experiments,  through space and ground based GW interferometers including Laser Interferometer Space Antenna (LISA) \cite{LISA:2017pwj,Barausse:2020rsu}, Pulsar Timing Array (PTA) experiments \cite{Lentati:2015qwp,NANOGrav:2020bcs} and DECIGO \cite{Seto:2001qf,Kawamura:2020pcg}.
 For a detailed review of induced SGWB and the dependence of its properties on the post-inflationary expansion history, see \cite{Domenech:2021ztg}. 

\smallskip
\noindent {\bf The  structure of this review}

\smallskip
\noindent 
If their origin is attributed to the large primordial fluctuations, PBHs may offer us a unique window to probe inflationary dynamics at sub-CMB scales. In this work, focusing mainly on the activity in the literature within the last few years, we aim to revisit and  review different inflationary production mechanisms of PBHs
\footnote{PBHs could also form in the post-inflationary universe through the collapse of cosmic strings \cite{Polnarev:1988dh,Caldwell:1995fu,Helfer:2018qgv} and domain walls \cite{Rubin:2000dq,Garriga:2015fdk,Deng:2016vzb,Kusenko:2020pcg}, phase transitions \cite{Kodama:1982sf,Jedamzik:1996mr}, bubble collisions \cite{Moss:1994iq,Kitajima:2020kig}, scalar field fragmentation via instabilities \cite{Cotner:2016cvr,Cotner:2019ykd}. We note that PBH formation in the post-inflationary can be triggered via the bubble nucleation during inflation or instabilities generated in the final stage of inflation commonly referred as (p)reheating \cite{Kofman:1994rk,Shtanov:1994ce,Kofman:1997yn,Amin:2014eta}. We will not dwell into these possibilities here, for a partial list of recent works in this line of research, see \cite{Ashoorioon:2020hln} and \cite{Martin:2019nuw,Auclair:2020csm}, respectively.} 
and their main predictions, 
 in a heuristic and pedagogical manner. The audience we have in mind are graduate students, or researchers in related fields who wish to learn about inflation and primordial black holes, and to be guided through the large literature on the subject by emphasizing common conceptual themes behind many different realizations. 

The review is organized as follows. In Section \ref{S2}, we present a simplified, intuitive picture of PBH formation in the inflationary universe and give some approximate estimate for the required conditions to produce PBHs from the perspective of inflationary dynamics. In Section \ref{sec_single}, we discuss  ideas  to  enhance the  curvature power spectrum within single-field inflation, as required for PBH formation. These mechanisms exploit large gradients in background
quantities which get converted into an amplification of fluctuations.
Besides reviewing analytic findings, we also develop some numerical
analysis and provide a link to a code for reproducing our results (see Footnote \ref{ftncode}).
In Section \ref{S3p2}, we  focus on multi-field inflationary scenarios that can generate PBH populations including particle production during axion inflation, or
sudden turns in the multi-scalar inflationary landscape.
 Finally, we end with a discussion on future directions in the concluding Section \ref{sec_out}. We supplement this work with several technical Appendices where we provide useful formulas and calculations used in the main body of this work. 

\smallskip

\noindent{\bf Conventions} 

\smallskip
\noindent
Throughout this review, we  work with natural units $\hbar = c = 1$. We will use the reduced Planck mass defined as $\Mpl^2 = (8\pi G)^{-1}$ and retain it in the equations unless otherwise stated. For time dependent quantities, over-dots and primes denote derivatives with respect to cosmological time $t$ and conformal time $\d \tau \equiv  \d t / a(t)$ respectively where $a(t)$ is the scale factor of the background FLRW metric $g_{\mu \nu} = \textrm{diag}(-1, a^2 ,a^2, a^2)$.  

\section{PBH formation in the early universe}\label{S2}

We start providing a  physical description of PBH formation in the early universe, as comprised of an early stage of inflation,  followed by radiation and matter domination (for a mini-review on background cosmology, see Appendix \ref{AppA}). Our aim is to set the stage
and relate basic properties of a PBH population -- as their mass and abundance -- with the features
of primordial curvature fluctuations originating from inflation. 
For this purpose, in Section \ref{S2p1} we describe the  mechanism of PBH formation in the post-inflationary universe,  emphasizing its nature as causal process controlled by the   inflationary quantum fluctuations. In Section \ref{S2p2} we discuss  relevant concepts  such as the threshold for collapse into black holes, and the corresponding  mass and collapse fraction of PBHs,  relevant for a computation of their abundance. Finally, in Section \ref{S2p3}, we relate the PBH abundance to primordial physics during inflation, with the aim to  determine the  amplitude of scalar fluctuations required for producing a population of PBHs with interesting consequences for cosmology. All the concepts we discuss  form the basis and motivations for our analysis of inflationary mechanisms for PBH production, which we develop in Sections \ref{sec_single} and \ref{S3p2}.

\smallskip
\noindent\faIcon{book} {\bf\emph{Main References}}: In compiling the materials of this Section and to set the main framework for our discussion, we have benefited from the ideas presented in the reviews by C. Byrnes and P. Cole \cite{Byrnes:2021jka}, M. Sasaki {\it et al.} \cite{Sasaki:2018dmp} and the Ph.D. thesis by G. Franciolini \cite{Franciolini:2021nvv}. 

\subsection{PBH formation as a causal process}\label{S2p1}

\begin{figure}[t!]
\begin{center}
\includegraphics[scale=0.51]{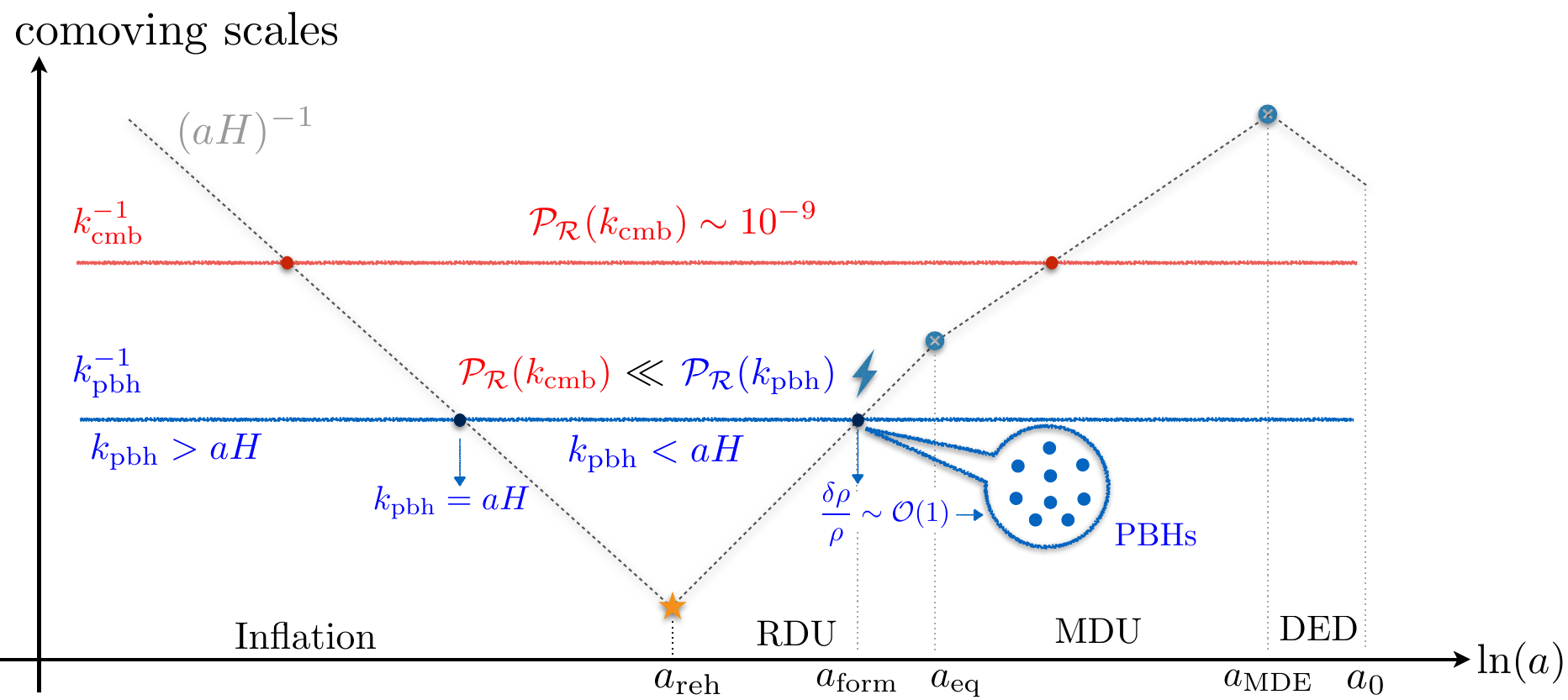}
\end{center}
\caption{A sketch of the time evolution of curvature fluctuations $\cal R$ (labelled by a comoving scale $k^{-1}$) with respect to comoving Hubble horizon (dotted lines) $(aH)^{-1}$ in the early universe. In the post-inflationary universe, $a_{\rm reh}$ denotes the reheating time, $a_{\rm eq}$ refers to the time of matter-radiation equality, $a_{\rm MDE}$ to matter-dark energy equality, and $a_{0}$ denotes to value of scale factor today.     
The blue horizontal line indicates the comoving size of a representative small scale perturbation responsible for PBH formation. If the power spectrum associated with these modes is enhanced during inflation, they can transfer their energy to density perturbations during radiation domination, and ignite PBH formation upon horizon re-entry at $a =a_{\rm form} \equiv a_{\rm f}$. \label{fig:pmc}}
\end{figure}

An important concept in an expanding space-time is the horizon scale,  crucial for understanding the causal properties of the dynamics of perturbations which are  responsible for PBH formation. As an indicator of the rate of our universe expansion, the Hubble rate $H(t) \equiv \dot{a}(t)/a(t)$ has dimensions of inverse length (or time$^{-1}$ in natural units). This makes the quantity $H^{-1}$ (Hubble horizon) as the natural candidate for a physical length scale in an expanding universe.  Commonly referred to as the Hubble distance, the quantity  $1/H$ (or $c/H$, if one wishes to recover physical units) measures the distance that light travels within one Hubble time. Therefore, it can be considered as a good proxy for a (time-dependent) length scale  controlling the size of a causal patch in our universe. Bearing in mind that we relate physical quantities to comoving ones by the scale factor $a(t)$, a useful quantity that guides us in this direction is the comoving Hubble horizon, $(a H)^{-1}$,  and in particular its time evolution. When expressed in terms of the second
derivative  of the scale factor, the time derivative of the comoving horizon can be written as
\beq\label{evhor}
\fr{\d}{\d t}\left(\fr{1}{a H}\right) = - \fr{\ddot{a}}{a^2 H^2}\,.
\eeq
Notice that during inflation $\ddot{a} > 0$: hence, the comoving horizon scale is a decreasing function of time. Whereas, in a decelerating universe with $\ddot{a} < 0$ (\ie in the post-inflationary universe  before dark energy domination), this quantity is an increasing function of time. The property that the comoving horizon decreases during  an accelerated expansion is perhaps the most important element to understand inflation as a solution of the  horizon problem of the Hot Big Bang cosmology, and a framework for the quantum mechanical origin of structures in our universe \footnote{A detailed discussion on these topics can be found in Chapter 4 of Baumann's book \cite{Baumann:2022mni}.}. The time dependence of the comoving Hubble horizon is controlled by the value of the  background equation of state $w$ (EoS) as  (see Appendix \ref{AppA})
\beq
(a H)^{-1} \propto a^{(1+3w)/2}.
\eeq 
Therefore, during inflation $w \simeq -1$ and $(a H)^{-1} \propto a^{-1}$  while, during the subsequent phases of radiation dominated  (RDU) and matter dominated universe (MDU), the comoving horizon  evolves as  $(a H)^{-1} \propto a^{1}$ and $(a H)^{-1} \propto a^{1/2}$ respectively. The evolution of the comoving horizon with respect to logarithm of scale factor $\ln(a)$ is illustrated in Fig \ref{fig:pmc}. 
When we study the statistical properties of fluctuations in Fourier space, we often label a given perturbation mode with a comoving length scale $k^{-1}$,  measured  in units of megaparsecs ($\rm Mpc = 3.26 \times 10^6\,\, \textrm{light years} \simeq 3.1 \times 10^{19}\,\, {\rm km}$). Therefore, a crucial quantity to conceptualize the behavior of perturbations in the inflationary universe is the ratio of the wavelength of a given mode with respect to the size of comoving Hubble horizon: $(aH)/k$. Fluctuations with wavelengths larger than the comoving horizon are referred to as super-horizon modes $k < a H$, while sub-horizon perturbations satisfy $k > a H$. Each mode crosses the horizon at  
 $k = a H$. As shown in Fig. \ref{fig:pmc}, a typical fluctuation with comoving size $k^{-1}$ (horizontal lines) begins its life deep inside the horizon (typically as a quantum fluctuation); then it leaves the horizon to become a super-horizon mode, and finally it re-enters the comoving horizon in the post-inflationary universe. Large scale modes (with smaller $k$) exit the horizon earlier than small scale modes, and re-enter the horizon at a later time in the post-inflationary era. For definiteness, in Fig. \ref{fig:pmc} we represent   the behaviour of the comoving 
 curvature perturbation ${\cal R}$ \cite{Sasaki:1986hm,Mukhanov:1990me}  (see \cite{Baumann:2022mni} for a textbook discussion),  which plays an important role for our discussion.

In fact,
apart from providing seeds for the observed cosmic microwave background (CMB) anisotropies at large scales, the dynamics of  curvature fluctuations ${\cal R}$    may also be at play for PBH formation, provided that cosmological fluctuations exhibit specific `initial conditions' at small scales. For this purpose, denoting $k = k_{\rm pbh} \gg k_{\rm cmb}$  as the comoving momentum  associated with PBH formation, we assume that the curvature  power spectrum at these small scales acquires an amplification well above the  level required to match CMB observations: $\mathcal{P}_{\mathcal{R}}(k_{\rm pbh}) \gg \mathcal{P}_{\mathcal{R}}(k_{\rm cmb}) \sim 10^{-9} $ \cite{Planck:2018jri} (more on this later). Soon after the end of inflation, \ie after reheating\footnote{Throughout this work, we assume an efficient reheating process at the end of inflation such that the universe become radiation dominated shortly after inflation terminates. For a collage of interesting physics that might arise through the reheating stage and alternative post-inflationary histories see the recent review \cite{Allahverdi:2020bys}.}, the modes associated with the enhancement (\eg modes with comoving size of $k_{\rm pbh}^{-1}$) become the seeds of density perturbations in the RDU:

\begin{equation}\label{PRvsdrhovsdelta}
\mathcal{P}_{\mathcal{R}}(k_{\rm pbh})^{1/2} \sim \fr{\delta \rho}{\rho}.
\end{equation}
Since the comoving Hubble scale grows with respect to comoving scales in RDU, the characteristic scale of perturbations eventually becomes  comparable to the comoving horizon at $a = a_{ \rm f}$,  where $k_{\rm pbh} = (a H)_{ \rm f}$ (the subscript $ \rm f$ indicates PBH formation time). At this point, gravitational interactions can trigger the collapse of over-dense regions, if the latter have a sufficient over-density above a certain  collapse threshold (see the discussion below). 

Notice that at sub-horizon scales the radiation pressure can overcome the gravitational collapse: therefore, the  production of PBHs effectively occurs at around horizon re-entry. This implies that the concept of horizon re-entry is crucial for our understanding PBH formation as a causal process:
in fact,  only when the physical wavelength of a  perturbation 
becomes comparable to the causal distance $1/H$,
 gravity is able to communicate the presence of an over-density, and to initiate the gravitational collapse. A schematic diagram that summarizes the discussion above is illustrated in Fig \ref{fig:pmc}.

In what comes next,  we briefly discuss relevant quantities such as the threshold for collapse  and the mass and collapse fraction of PBHs, which are  important for the computation of the current PBH abundance.

\subsection{The relevant quantities for PBH abundance}\label{S2p2}
\smallskip
\noindent{\bf The threshold for collapse} 
\smallskip

\noindent
The first criterion on the collapse threshold --defined as $\delta_{\rm c}$ hereafter-- for PBH formation was formulated by B. Carr in 1975, using a Jeans-type instability  argument within Newtonian gravity \cite{Carr:1975qj}. In Carr's estimate,  an over-density in RDU would collapse upon horizon re-entry if the fractional over-density of the perturbation is larger than the sound speed square of density perturbations $c_s^2$, which is a measure of how fast a pressure wave caused by the over-density can travel from the centre to the edge of a local fluctuation. In RDU, the speed of sound of perturbations satisfies $c_s^2 = w = 1/3$, so that its square is directly related to EoS during RDU. This result implies that a perturbation can collapse to form PBHs if its over-density is larger than the pressure exerted by the radiation pressure\footnote{A simple analytic argument that leads to this result is presented in Appendix \ref{AppB}.}. Implementing general relativistic effects, the second analytical estimate derived in \cite{Harada:2013epa}, obtaining $\delta_{\rm c} \simeq 0.4$ during RDU.  These results notwithstanding, further studies in this direction have found that threshold depends on the initial profile of curvature perturbation \cite{Niemeyer:1999ak,Nakama:2013ica,Musco:2018rwt,Musco:2020jjb} \footnote{Other effects such as anisotropies \cite{Musco:2021sva} in the radiation fluid and a time dependent equation of state parameter $w$ \cite{Papanikolaou:2022cvo}, may also impact the estimate on the threshold for collapse.}.

At this point, we emphasize that a more precise criterion for the threshold of collapse is provided in terms of the so called compaction function (which is a measure of excess mass within a given spatial volume, see \eg \cite{Shibata:1999zs, Musco:2018rwt}) whose value at its maximum gives $\delta_c$. In this prescription, we note that contrary to the simplistic argument presented in Appendix \ref{AppB}, there is no one to one correspondence between the threshold value and the peak over-density \eqref{PRvsdrhovsdelta} $\delta \equiv \delta \rho/\rho$, although they are related see \eg \cite{Germani:2018jgr}. Keeping this in mind, in the rest of this work, we will continue to refer to the threshold using the $\delta_{\rm c}$ notation as commonly practiced in the literature.

An accurate characterization of the threshold requires a dedicated analysis of the evolution of perturbations in the non-linear regime after horizon re-entry, which can be done with the help of numerical simulations (see \eg \cite{Escriva:2021aeh} for a review). Utilizing the criterion of compaction function's peak as the definition of the threshold $\delta_{\rm c}$, simulations performed in the radiation fluid for different primordial perturbation profiles leading to a range of values $\delta_{\rm c} = 0.4 -  2/3$ depending on the shape of the density peak \cite{Musco:2018rwt,Escriva:2019phb}. Although in general threshold depends on the shape of the primordial perturbation, \cite{Escriva:2019phb, Escriva:2020tak} showed the existence of an approximate universal value for the threshold which depends only on the type of the ambient fluid and geometry. For a detailed account on the threshold and its history we refer the reader to the recent review \cite{Escriva:2021aeh}.

\medskip
\noindent{\bf The mass of PBHs} 
\medskip

\noindent
The characteristic mass of PBHs can be related to the mass contained within the Hubble horizon at the time of formation ($a = a_{\rm f}$) through an efficiency factor $\gamma = 0.2$, as suggested by the  analytical model developed in \cite{Carr:1975qj}:
\beq\label{pbhm}
M^{(\rm f)}_{\rm pbh} = \gamma\, M_H \bigg|_{a = a_{\rm f}} = \gamma\, M^{(\rm eq)}_{H}\, \left( \fr{M^{(\rm f)}_{H}}{M^{(\rm eq)}_{H}}\right) = \left(\fr{a_{\rm f}}{a_{\rm eq}}\right)^2 \, \gamma\,  M^{(\rm eq)}_{H}\,. 
\eeq
In this formula,  $M_H (t) \equiv 4\pi \rho(t)/(3H(t)^3)$ is the time-dependent horizon mass, where  the sub/super scripts ``\textrm{f}" and ``\textrm{eq}'' denote quantities evaluated at the time of PBH formation and matter-radiation equality respectively: we   use the standard relations $H^2 \propto \rho \propto a^{-4}$ during RDU. Noting that the horizon mass at the time of equality is given by $M^{(\rm eq)}_H \simeq 2.8 \times 10^{17}\, M_{\odot}$ \cite{Nakama:2016gzw}, Eq. \eqref{pbhm} informs us that PBHs, contrarily to astrophysical black holes, can in principle span a wide range of masses,  depending on their formation time with respect to matter-radiation equality.

Making use of  the time-dependent horizon mass as above, we can relate the PBH mass at formation to the characteristic size of the perturbations that leave the horizon during inflation, and are
 responsible for PBH formation. For this purpose, we first rewrite the PBH mass at formation as
\beq\label{pbhmvsk}
M^{(\rm f)}_{\rm pbh}
= \left(\frac{\rho_{_{\rm f}}}{\rho_{_{\rm{eq}}}}\right)^{1 / 2} \left(\frac{H_{\mathrm{eq}}}{H_{\rm f}}\right)^{2}\,\, \gamma\,  M^{(\rm eq)}_{H}.
\eeq
Using the property of  entropy conservation  $g_{s}(T)\, T^{3}\, a^{3}=\text{constant}$, and the scaling property 
of the energy density with respect to temperature of the plasma during RDU, $\rho \propto g_*(T)\, T^{4}$, Eq. \eqref{pbhmvsk} can then be re-expressed as the following relation 
\begin{align}\label{pbhmvskf}
\nn M^{(\rm f)}_{\rm pbh}(k_{\rm pbh}) &=\left(\frac{g_{*}\left(T_{\rm f}\right)}{g_{*}\left(T_{\mathrm{eq}}\right)}\right)^{1 / 2}\left(\frac{g_{s}\left(T_{\mathrm{eq}}\right)}{g_{s}\left(T_{\rm f}\right)}\right)^{2 / 3}\left(\frac{k_{\mathrm{eq}}}{k_{\rm pbh}}\right)^{2} \, \gamma\,  M^{(\rm eq)}_{H},\\
&\simeq \left(\frac{\gamma}{0.2}\right)\left(\frac{g_{*}\left(T_{\rm f}\right)}{106.75}\right)^{-1 / 6}\left(\frac{k_{\rm pbh}}{3.2 \times 10^{5}\, \mathrm{Mpc}^{-1}}\right)^{-2}\, 30 M_{\odot}\,.
\end{align}
In the second line we assume that the  effective number of relativistic degrees of freedom in energy density and entropy are equal, \ie we set $g_*(T) = g_s (T)$ and take $g_{*} (T_{\rm eq}) \simeq 3.38$ \footnote{Strictly speaking $g_*(T) = g_s (T)$ is only satisfied when species are in thermal equilibrium at the same temperature. For a nice overview on the thermal history of the universe after inflation, see Chapter 3 of \cite{Baumann:2022mni}.} with $k_{\rm eq} \simeq 0.0104\, {\rm Mpc}^{-1}$, accordingly with  the  latest Planck results \cite{Planck:2018vyg}. Equation \eqref{pbhmvskf} indicates  that for masses of PBHs that could be associated with recent LIGO observations, $M_{\rm pbh} \simeq 30 M_{\odot}$, the peak scale of perturbations responsible for PBH formation is much smaller compared to CMB scales $k_{\rm pbh} \gg k_{\rm cmb}$. For sub-solar mass PBHs, the corresponding peak scale for PBH formation gets progressively smaller. 
 For example, considering the currently allowed  sub-lunar range ($M_{\rm moon} \simeq 3.7 \times 10^{-8}\, M_{\odot}$) of PBH masses, $10^{-17} \lesssim M_{\mathrm{pbh}}\left[M_{\odot}\right] \lesssim 10^{-12}$,  which are  objects that can account for the totality of dark matter, the range of  scales associated with PBH formation is quite small: $10^{12} \lesssim k_{\rm pbh} \,[ \mathrm{Mpc}^{-1}]\, \lesssim 10^{15}$. See Table \ref{tab:scales} for an easier-to-visualize summary of these considerations.

Elaborating on Eq. \eqref{pbhmvskf}, we can also derive a rough relation between the PBH mass at formation to the the number of e-folds $N_{{\rm pbh}}$ at which
the PBH-forming
modes leave the horizon during inflation. For this purpose, we first notice that    $k_{\rm pbh}/k_{\rm cmb} = (a H)_{\rm pbh} / (a H)_{\rm cmb}$ where the values of Hubble rate and scale factor should be evaluated at the scales of  horizon exit during inflation (see Fig. \ref{fig:pmc}). Assuming roughly a constant slow-roll parameter $\epsilon \equiv -\dot{H}/H^2 \ll 1$ between the horizon-exit time of modes
associated with CMB and PBH formation respectively, we can relate the Hubble and the scale factor as $H_{\rm pbh} = H_{\rm cmb}\, e^{-\epsilon(N_{{\rm pbh}} - N_{{\rm cmb}})}$ and $a_{\rm pbh} = a_{\rm cmb}\, e^{N_{{\rm pbh}} - N_{{\rm cmb}}}$ where $N_{\rm pbh} > N_{\rm cmb}$ so that we count e-folds forward in time with respect to horizon exit of the CMB mode\footnote{We note that another common convention is to count e-folds with respect to the end of inflation denoting the end point as $N_{\rm end} = 0$.}. Using the last two relations we find $
{k_{\rm pbh}}/{k_{\rm cmb}} \simeq e^{(N_{\rm pbh} - N_{\rm cmb})(1-\epsilon)}$; once plugged  in  Eq. \eqref{pbhmvskf},  assuming $k_{\rm cmb} = 0.002\, {\rm Mpc}^{-1}$),  we find
\beq\label{pbhmvsn}
M^{(\rm f)}_{\rm pbh}(N_{{\rm pbh}}) \approx 7.7 \times 10^{17} M_{\odot}\,\, e^{-2(N_{{\rm pbh}}-N_{{\rm cmb}})(1-\epsilon)} \left(\frac{\gamma}{0.2}\right)\left(\frac{g_{*}\left(T_{\rm f}\right)}{106.75}\right)^{-1 / 6}\,.
\eeq
Modes that leave  the horizon much later compared to CMB scales have $N_{\rm pbh} - N_{\rm cmb}  \gg 1$, therefore the exponential in
Eq. \eqref{pbhmvsn} can considerably reduce the overall large normalization, leading to small PBH masses.

\begin{table}
\begin{center}
\begin{tabular}{| c | c | c | c | c | c |}
\hline
\hline
\cellcolor[gray]{0.9}$ M_{\rm pbh}\, [M_{\odot}] $ & \cellcolor[gray]{0.9}$ \Delta N $&\cellcolor[gray]{0.9}$k_{\rm pbh}\, [{\rm Mpc}^{-1}]$&\cellcolor[gray]{0.9}$M_{\rm pbh}\, [M_{\oplus}] $ &\cellcolor[gray]{0.9}$M_{\rm pbh}\, [M_{\rm Everest}] $  \\
\hline
$10^{6}$ & $14$ & $10^3$ & $10^{11}$ & $10^{21}$\\
\hline
$10^{0} - 10^{2}$ & $18-21$ & $10^{5}-10^{6}$ & $ 10^{5}-10^{7}$& $10^{15}-10^{17}$ \\
\hline
$10^{-17} - 10^{-12}$ &$34-40$ & $10^{12} - 10^{15}$ & $10^{-12}-10^{-7}$& $10^{-2}-10^{3}$\\
\hline
\end{tabular}
\caption{Range of PBH mass(es) vs the corresponding wave-number(s) $k_{\rm pbh}$ (see Eq. \eqref{pbhmvskf}) of the primordial modes together with the approximate horizon crossing time measured with respect to e-folding number the pivot scale $k_{\rm cmb} = 0.002\, {\rm Mpc}^{-1}$ leaves the horizon during inflation, $\Delta N \equiv N_{\rm pbh} - N_{\rm cmb} > 0$ (see Eq. \eqref{pbhmvsn}). First row refers to the corresponding quantities for a typical Super Massive Black Hole (SMBH) like the ${\rm Sagittarius\,A^{*}}$  in the center of our galaxy \cite{EventHorizonTelescope:2022wkp}. The third row refers to astroid-mass PBHs that can still account for a significant fraction (or all) of DM density today \cite{Montero-Camacho:2019jte}. The corresponding mass of the PBHs in terms of the earth's mass $M_{\odot} \simeq 3.33 \times 10^5\, M_{\oplus}$ and the mass of mount Everest $M_{\rm Everest} = 8.1 \times 10^{14}\, {\rm kg} \simeq 4.1 \times 10^{-16}\, M_{\odot}$ are given in the last two columns on the right. \label{tab:scales}}
\end{center}			
\end{table}

\medskip
\noindent{\bf PBH abundance}
\medskip

\noindent
After discussing possible masses for PBH and how they depend on the dynamics of inflation, we analyse the 
notion of  abundance of PBHs relative
to the  energy density of other species.
We can compute this quantity during two epochs: today,
and at PBH formation. 

When considering the present-day fraction of PBH density, 
it is a common practice to relate the  PBH abundance to present-day dark-matter density  introducing the quantity
\beq\label{fpbh}
f_{\mathrm{pbh}} \equiv \frac{\Omega_{\mathrm{pbh}}}{\Omega_{\mathrm{dm}}},
\eeq
where for each species $i$ we define $\Omega_i \equiv \rho_{_{i,0}} / \rho_{_{c,0}}$, with subscript ``$0$" denoting quantities evaluated today and $\rho_{_{c,0}} = 3 H_0^2 \Mpl^2$ is the critical density.  Planck measurements provide the following  value for the dark matter abundance \cite{Planck:2018vyg},
\beq\label{OmDM}
\Omega_{\rm dm} h^2 = 0.120 \pm 0.001,
\eeq 
in terms of $h = 0.6736 \pm 0.0054$, which measures the Hubble rate $H_0$ in units of $100\, \rm km\, s^{-1}\, Mpc^{-1}$. 

We can then relate $f_{\mathrm{pbh}}$ today to the density fraction
of PBH  at the epoch
of their formation, 
denoting this quantity  with $\beta$. In fact,
since we assume that PBH formation takes place during RDU, and since
after  formation the PBH density scales as of like $\rho_{\rm pbh} \propto a^{-3}$,
 we can 
write
\begin{align}\label{beta0}
\nn \beta \equiv \fr{\rho_{_{\rm pbh}}}{\rho} \bigg|_{a = a_{\rm f}} = \fr{\rho_{_{\rm pbh,\rm f}}}{\rho_{_{\rm pbh,0}}} \fr{\rho_{_{c,0}}}{\rho_{_{\rm f}}} \,\,\Omega_{\rm dm} f_{\rm pbh}&= \fr{\rho_{_{\rm pbh,\rm f}}}{\rho_{_{\rm pbh,0}}} \fr{\rho_{_{c,0}}}{\rho_{_{\rm eq}}} \fr{\rho_{_{\rm eq}}}{\rho_{_{\rm f}}} \,\,\Omega_{\rm dm} \,f_{\rm pbh}, \\
&\simeq \fr{1}{2} \fr{a_{\rm f}}{a_{\rm eq}}\,\, \Omega_m^{-1}\,\, \Omega_{\rm dm}\, f_{\rm pbh}
\end{align}
where $\Omega_m$ is the current matter density in the universe. In \eqref{beta0} we normalize the scale factor today as $a_0 = 1$, and
we use the fact that the 
total energy density evolves as $\rho \propto a^{-4}$ for $a_{\rm f} <  a < a_{\rm eq}$, and  $\rho(a_{\rm eq}) = 2\rho_{_{m,0}}\, a_{\rm eq}^{-3}$. 
 Using the conservation of total entropy, $g_{s}(T)\, T^{3}\, a^{3}=\text {constant}$, we can re-express
 the factor ${a_{\rm f}}/a_{\rm eq}$ appearing
 in Eq. \eqref{beta0} as follows:
\begin{align}\label{afovaeq}
\nn \fr{a_{\rm f}}{a_{\rm eq}} &= \fr{T_{\rm eq}}{T_{\rm f}} \left(\fr{g_s (T_{\rm eq})}{g_s(T_{\rm f})}\right)^{1/3},\\
& \simeq 3.17 \times 10^{-9} \left(\fr{\gamma}{0.2}\right)^{-1/2} \left(\frac{g_{*}\left(T_{\rm f}\right)}{106.75}\right)^{-1 / 12} \left(\fr{M^{(\rm f)}_{\rm pbh}}{M_{\odot}}\right)^{1/2},
\end{align}
where   we make use of Eq. \eqref{pbhmvsk} to relate $T_{\rm eq}/T_{\rm f}$ to the mass of PBH at formation, and as before  we assume $g_*(T) = g_s (T)$. Finally, plugging \eqref{afovaeq} in \eqref{beta0},  and implementing  Planck measurements on $\Omega_{\rm dm}$ \eqref{OmDM} and $\Omega_{m}$ (see Eq. \eqref{PP}), we  directly relate the PBH abundance at formation, $\beta$, to their present-day fraction $f_{\rm pbh}$, in terms of the PBH mass at formation: 
\beq\label{betaf}
\beta \simeq 1.33 \times 10^{-9}  \left(\fr{\gamma}{0.2}\right)^{-1/2} \left(\frac{g_{*}\left(T_{\rm f}\right)}{106.75}\right)^{-1 / 12} \left(\fr{M^{(\rm f)}_{\rm pbh}}{M_{\odot}}\right)^{1/2}\, f_{\rm pbh}\,.
\eeq
 Therefore, we learn that in the case when PBH abundance account for the total DM density today, $f_{\rm pbh} \to 1$, the fraction of the total density in the form of PBHs ($\beta$) at the time of their formation takes extremely small values, when considering  an interesting range of masses $M^{(\rm f)}_{\rm pbh}$. This situation reflects the fact that PBH formation in the early universe is a very rare event. There is also another way to parametrize $\beta$ in terms of (relative) number of collapsing regions to form PBHs. This approach is especially useful to relate the PBH abundance to the statistical properties of primordial fluctuations as we discuss below.   

\vspace{0.25cm}
\noindent{\bf Collapse fraction of PBHs at formation}
\medskip

\noindent
 The PBH abundance at formation can be also interpreted as the fraction $\beta$ of local regions in the universe that has a density larger than a certain threshold. The standard treatment of estimating $\beta$ is based on the so-called Press-Schechter model of gravitational collapse,  widely used in the literature on large-scale structure formation  \cite{Press:1973iz} \footnote{
 Contrarily to the original approach by Press-Schechter \cite{Press:1973iz}, we do not take into account a symmetry factor of 2 in the right hand side of \eqref{beta} that accounts for all the mass in the universe, since  it is not clear whether such a factor makes sense when considering non-symmetric PDFs of $\delta$ (\eg non-Gaussian cases). Furthermore, the error introduced by omitting this factor is comparable with the other uncertainties in the computation of $\beta$ such as fraction of horizon mass which collapse to form a PBH (see \eg \cite{Evans:1994pj,Niemeyer:1997mt,Koike:1995jm}).
 }, 
\begin{figure}[t!]
\begin{center}
\includegraphics[scale=1]{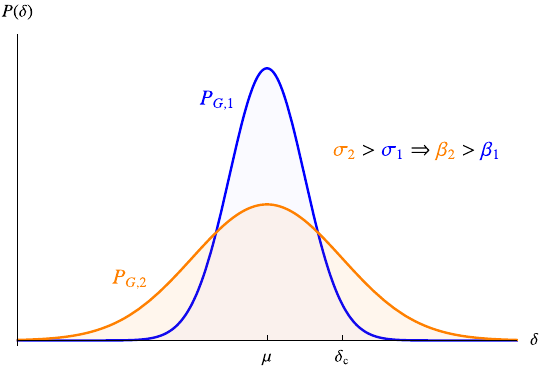}
\end{center}
\caption{Two Gaussian PDFs of over-density field $\delta$ with different variances $\sigma_2^2 > \sigma_1^2$. Since the second distribution have a larger variance, the area under the curve above the critical threshold ($\delta_{\rm c}\leq \delta \leq \infty$) is larger, leading to larger PBH abundance $\beta$ \eqref{beta} at formation. \label{fig:pg}}
\end{figure}
\beq\label{beta}
\beta \equiv \frac{\rho_{_\mathrm{{pbh}}}}{\rho} \bigg|_{a = a_{\rm f}} \equiv \int_{\Delta_{\rm c}}^{\infty}P(\delta)\,\, \d \delta,
\eeq
where $P(\delta)$ is the probability distribution function (PDF), which describes how likely that a given fluctuation have an over-density $\delta$ and we assume that a perturbation will collapse to form BH if its amplitude is larger than a critical value $\Delta_c$. Notice that $\Delta_c$ is the critical density contrast and in general can not be identified with the threshold, $\Delta_{\rm c} \neq \delta_{\rm c}$ where the latter can be defined as the peak value of the compaction function and can be related to an averaged density contrast \cite{Escriva:2021aeh}.  Note also that an alternative, more accurate approach to compute the PBH abundance requires to use the PDF of the compaction function,
see e.g \cite{Germani:2019zez}. A discussion of this approach and the actual relation between $\Delta_{\rm c}$ and $\delta_{\rm c}$ would bring us outside the pedagogical purposes of this review. For a detailed account on the use PDF of compaction function and the influence of resulting non-linearities on the PBH abundance we refer the reader to \cite{Germani:2019zez} and references therein. For more details on the relation between $\Delta_{\rm c}$ and $\delta_{\rm c}$, see also the review \cite{Escriva:2021aeh}. Keeping an eye on the recent progress in the literature on these issues, we will continue to identify $\Delta_{\rm c} = \delta_{\rm c}$ for the estimates on $\beta$ in this and the next section and continue to work with a PDF on over-density $\delta$.

Let's assume that the latter follows a Gaussian distribution, 
\beq\label{pdfG}
P_{G}(\delta)= \fr{1}{\sqrt{2\pi}\sigma} e^{-{(\delta -\mu)^2}/{2 \sigma^2}},
\eeq
where $\mu$ is the mean and $\sigma^2$ is the variance of the distribution. In Fig. \ref{fig:pg}, we represent two Gaussian  PDFs that have the same mean value, and  two different variances satisfying $\sigma_2^2 > \sigma_1^2$. As the second distribution is more ``spread" with a larger variance compared to the first one, the probability of an over-density to be larger than the critical threshold $\delta_{\rm c}$, is larger, and so does PBH abundance $\beta$ at formation, since  the integral in Eq. \eqref{beta}  has more support within the integration limits  $\delta \in [\delta_{\rm c}, \infty]$. Hence, we can expect that the quantity $\sigma$ plays an important role for estimating the PBH abundance. 

Using \eqref{betaf} and \eqref{beta}, we can 
estimate the required variance $\sgm^2$ of \eqref{pdfG} that can give rise to large population of PBH today, as
controlled by the quantity $\beta$. Focusing on a distribution with zero mean $\mu = 0$ in \eqref{pdfG} and integrating \eqref{beta}, we have 
\beq\label{betaG}
\beta = \int_{\delta_{\rm c}}^{\infty} \frac{\mathrm{d} \delta}{\sqrt{2 \pi} \sigma} \exp \left(-\frac{\delta^{2}}{2 \sigma^{2}}\right) = \fr{1}{2} \rm{Erfc}\left(\fr{\delta_{\rm c}}{\sqrt{2}\sigma}\right)\simeq \frac{ \sigma}{\sqrt{2 \pi} \delta_{\rm c} } \exp \left(-\frac{\delta_{\rm c} ^{2}}{2 \sigma^2}\right),
\eeq
where ${\rm Erfc}(x) = 1 - {\rm Erf}(x)$ is the complementary error function, and in the last equality we take $\delta_{\rm c} \gg \sigma$. As we will learn shortly,  this is a good approximation for all practical purposes. As concrete examples, 
substituting Eq. \eqref{betaG} into Eq. \eqref{betaf} we   infer that a solar mass PBH population with $f_{\rm pbh} = 10^{-3}$ requires $\delta_{\rm c}/\sigma \simeq 7$, whereas for a population with $M^{(\rm f)}_{\rm pbh}= 10^{-12} M_{\odot}$ and $f_{\rm pbh} = 1$ we need $\delta_{\rm c}/\sigma \simeq 7.9$. Assuming a threshold of $\delta_{\rm c} = 0.4$, these results translate into $\sigma \simeq 0.06$ and $\sigma \simeq 0.05$ respectively. Notice also from \eqref{betaG} that the PBH abundance at the time of formation is exponentially sensitive to the variance of the distribution. We will dwell more on this dependence  below. But first, we discuss  the implications of these findings in terms of the amplitude of scalar fluctuations generated during the phase of cosmic inflation.

\subsection{Relating PBH properties with  primordial scalar fluctuations
}\label{S2p3}

We now examine how to relate the notion of PBH abundance
with the properties of the comoving curvature fluctuation $\mathcal{R}$, \cite{Sasaki:1986hm,Mukhanov:1990me}, as 
produced in the early universe by cosmic inflation.  $\mathcal{R}$ is conserved on super Hubble scales as the modes evolve from the inflationary phase to RDU. 
We start by connecting the amplitude of  $\mathcal{R}$ with the fractional over-density $\delta$, which 
triggers PBH formation as we learned  in our previous discussion.
Working in Fourier space, Taylor expanding 
at leading order in   a gradient expansion (controlled by   the small  parameter $k / a H$) and
at linear order in  $\mathcal{R}$, one finds  \cite{Musco:2018rwt}: 
\beq\label{dvsR0}
\delta(\vec{x},t)\simeq \fr{2(1+w)}{(5+3w)} \fr{\nabla^{2} \mathcal{R}(\vec{x})}{(a H)^{2}}\, + \,\dots \quad \Longrightarrow \quad \delta_k \simeq - \fr{4}{9} \left(\fr{k}{a H}\right)^2 \mathcal{R}_k,
\eeq
where used $w = 1/3$ during RDU and $\dots$ denote terms of higher order $\mathcal{O}(\mathcal{R}^2)$ in the curvature perturbation\footnote{Non-linearities that we neglect in the expression \eqref{dvsR0} can be important to understand intrinsic non-Gaussianity present in the PBH formation process, see \eg \cite{DeLuca:2019qsy,Young:2019yug} and references therein.}. Defining the power spectrum of a Fourier variable $X_k$ as 
\beq
\langle X_k X_{k'} \rangle =  \fr{2\pi^2}{k^3} \mathcal{P}_X (k) \,\delta^{(3)}(\vec{k}+\vec{k}'),
\eeq
the relation between the power spectrum of over-density and curvature perturbation is then given by
\beq\label{dvsR}
\mathcal{P}_\delta (k) \simeq \fr{16}{81} \left(\fr{k}{a H}\right)^4 \mathcal{P}_{\mathcal{R}} (k).
\eeq
In the computation of the density contrast, one should typically use a window function $\mathcal W$ to smooth $\delta(\vec{x},t)$ on a scale $R \approx k^{-1} \approx (aH)^{-1}$ (\eg on scales of size $k_{\rm pbh}^{-1}$ at horizon re-entry, as shown in Fig. \ref{fig:pmc}) relevant for PBH formation. Therefore, the variance of density contrast can be related to the primordial power spectrum as \cite{Young:2014ana,Sasaki:2018dmp} \footnote{The variance \eqref{vardef} can be equivalently written as $\sigma^2 = \sigma^2(k)$ or $\sigma^2 = \sigma^2(M)$ using the relation between the peak scale of PBH formation and PBH mass \eqref{pbhmvsk}.}
\beq\label{vardef}
\sigma^2 (R) \equiv \langle \delta^2 \rangle_{R} = \int_{0}^{\infty} \d \ln q\,\, \mathcal{W}^2(q, R)\, \mathcal{P}_{\delta} (q) \simeq \fr{16}{81} \int_{0}^{\infty} \d \ln q\,\, \mathcal{W}^2(q, R)\, (qR)^4\,\, \mathcal{P}_{\mathcal{R}}(q)
\eeq
where $\mathcal{W}$ is the Fourier transform of a real space window function. Popular choices of $\mathcal{W}$ include a volume-normalized Gaussian, or a top hat window function, whose    Fourier transforms are respectively given by
\beq
\mathcal{W}(k, R) = \exp\left({-\fr{k^2 R^2}{2}}\right), \quad\quad \mathcal{W}(k,R) =  \frac{3\sin (k R)-3k R \cos (k R)}{(k R)^{3}}\,.
\eeq
When
selecting a curvature power spectrum $\mathcal{P}_{\delta}$ characterized by a narrow peak around the wave-number $k_{\rm pbh}$, the integral in \eqref{vardef} can be approximated as $\sigma^2 \sim \mathcal{P}_{\delta}(k_{\rm pbh})$. Then, utilizing \eqref{dvsR} at  horizon entry $k \simeq aH$ (\ie at the time of PBH formation), since $81/16\sim 5$, we can roughly relate the variance $\sigma$ to the primordial curvature power spectrum as 
\begin{equation}
\mathcal{P}_{\mathcal{R}}(k_{\rm pbh})\sim 
5\,\sigma^2\,.
\end{equation}
 Finally, recall  our considerations after Eq. \eqref{betaG}: a Gaussian PDF of $\delta$ requires $\sigma \simeq 0.06$ ($\sigma \simeq 0.05$) for $M_{\mathrm{pbh}}^{(\mathrm{f})}=M_{\odot}$  ($M_{\mathrm{pbh}}^{(\mathrm{f})}=10^{-12} M_{\odot}$) to generate a population of  $f_{\rm pbh} = 10^{-3}$ ($f_{\rm pbh} = 1$) today.
 Hence, we can estimate the amplitude of the scalar power spectrum needed at scales relevant for PBH formation:
\beq\label{PvsS}
\mathcal{P}_{\mathcal{R}} (k_{\rm pbh}) \sim 5\, \sigma^2 \sim 10^{-2} \quad\quad \textrm{for Gaussian fluctuations}.
\eeq
This estimate holds
 for a wide range of sub-solar PBH masses. This implies that we need a very large
amplification of the curvature spectrum between large CMB and small PBH-formation scales:
\beq
\label{req_ampl1}
\Delta \mathcal{P}_{\mathcal{R}} \equiv \fr{\mathcal{P}_{\mathcal{R}}(k_{\rm pbh})}{{\mathcal{P}_{\mathcal{R}}(k_{\rm cmb})}} \sim 10^{7}\,,
\eeq
and the task is to produce such amplification in a controllable way
by an appropriate inflationary mechanism.

\medskip

It is worth pointing out that this estimate  does not change much for even smaller mass PBHs with $M_{\mathrm{pbh}}^{(\mathrm{f})} <10^{-12} M_{\odot}$, because the power spectrum has a logarithmic sensitivity to the PBH fraction $\beta$. In order to see this, we can invert the expression \eqref{betaG}, and use \eqref{PvsS} to relate the primordial power spectrum of curvature perturbations to $\beta$ as 
\beq\label{PvsS2}
\mathcal{P}_{\mathcal{R}}(k_{\rm pbh}) \sim 5\sigma^2 \sim \fr{5\,\delta_{\rm c}^2}{2\ln (1/\beta)}\, .
\eeq
Now, as an extreme case, we can consider the smallest PBHs $M_{\mathrm{pbh}}^{(\mathrm{f})} \simeq 10^{-18} M_{\odot}$ that can survive until today (not yet eliminated by Hawking radiation) which have the tightest available observational constraints, restricting their current abundance to $f_{\rm pbh} \lesssim 10^{-9}$ \cite{Carr:2020gox}. Plugging these values in \eqref{betaf}, PBH fraction at formation gives $\beta \lesssim 10^{-28}$ which in turn leads to the constraint $ \mathcal{P}_{\mathcal{R}}(k_{\rm pbh}) \lesssim 6 \times 10^{-3}$ in \eqref{PvsS2} for a threshold of $\delta_{\rm c} = 0.4$. Therefore, we conclude that for Gaussian perturbations and for any PBH mass of interest, the amplitude of scalar power spectrum relevant for PBH formation requires $\mathcal{P}_{\mathcal{R}} \sim 10^{-2}$ for any potentially observable PBH fraction $f_{\rm PBH}$ today. The discussion above informs us that a small change in the amplitude of power spectrum leads to many order of magnitude difference in the fraction of regions collapsing into PBHs, as clearly implied by the exponential dependence of $\beta$ to $\mathcal{P}_{\mathcal{R}}$ in \eqref{PvsS2}. Similarly, a small change in the choice of threshold $\delta_{\rm c}$ could lead to very different estimates in terms of $\beta$. For example, focusing on fixed value of variance $\sigma^2 \simeq 0.05$ as relevant for PBH formation, $\beta$ \eqref{betaG} can chance by various orders of magnitude, if we reduce the threshold $\delta_{\rm c}$ by just about $20 \% $. In fact, 
\beq
\frac{\beta(\delta_{\rm c} = 0.4)}{\beta(\delta_{\rm c} = 1/3)} \simeq 10^{-5}\,,
\eeq 
 demonstrating how tuned the conditions
are for producing a cosmologically interesting population
of PBHs. 

\medskip
\noindent{\bf Collapse fraction vs curvature perturbation}
\medskip

\noindent
While it is customary to use the smoothed density contrast at horizon crossing to estimate the number of collapsing regions, it is also possible to work directly with the comoving curvature perturbation to approximately compute the PBH 
fraction $\beta$ at time of formation. In this case,
 there is no need of   relying on the smoothing procedure of sub-horizon fluctuations provided by the window functions \cite{Byrnes:2012yx,Young:2014ana}. Interestingly, as we will see later this approach also provides a  way to assess the effects of large primordial non-Gaussianity that might be present in some of the PBH-forming inflationary scenarios. 

For understanding the role of the primordial curvature fluctuation $\mathcal{R}$, we approximate its variance  with the power spectrum $\sigma_{\mathcal{R}}^2 \approx \mathcal{P}_{\mathcal{R}}$. Using the Press-Schechter approach with a Gaussian PDF 
for the curvature fluctuation spectrum, the fraction of collapsing regions at formation can be estimated as
\beq\label{betaRG}
\beta_{G} = \int_{\mathcal{R}_{\rm c}}^{\infty} \d \mathcal{R}\,  \frac{e^{-\mathcal{R}^{2} /\left(2 \sigma_{\mathcal{R}}^2\right)}}{\sqrt{2 \pi} \sigma_{\mathcal{R}}} \simeq  \fr{1}{2} \rm{Erfc}\left(\fr{\mathcal{R}_{\rm c}}{\sqrt{2 \mathcal{P}_{\mathcal{R}}}}\right),
\eeq 
where $\mathcal{R}_{\rm c}$ is the threshold. To roughly estimate $\mathcal{R}_{\rm c}$, we can assume an almost scale invariant power spectrum of $\mathcal{R}$, for a logarithmic range of wave-numbers relevant for PBH formation. Making use of a Gaussian window function in \eqref{vardef} gives in this case  $\sigma_{\mathcal{R}}^2 = \langle \mathcal{R}^2 \rangle \simeq 8 \mathcal{P}_{\mathcal{R}}/81$. Finally, plugging the latter in \eqref{betaG}, and comparing the resulting expression with \eqref{betaRG}, we obtain  \cite{Young:2014ana}:
\beq
\mathcal{R}_{\rm c}\approx\frac{9}{2 \sqrt{2}} \delta_{\rm c}. 
\eeq
For a density threshold of $\delta_{\rm c} = 0.4$, the relation above gives $\mathcal{R}_{\rm c} \simeq 1.3$, which we set as fiducial value for the estimates below. Following these considerations and using the formulas derived so far -- in particular Eqs. \eqref{betaRG} and \eqref{betaf} -- we can then repeat the previous estimates, and  determine the approximate amplitude of power spectrum required for PBH formation. The result is that for Gaussian primordial fluctuations a sensible PBH population today requires $\mathcal{P}_{\mathcal{R}}\sim 10^{-2}$. These findings   confirm our earlier results of Eq.  \eqref{PvsS}. 

\medskip
\noindent{\bf The case of non-Gaussian curvature fluctuations}
\medskip

\noindent
So far, we assumed that primordial fluctuations obey Gaussian statistics in order to estimate the amplitude of the power spectrum required for PBH formation. Since PBHs are expected to form through extremely rare large fluctuations (see Fig. \ref{fig:pg}), any small deviation in the shape of the tail of the fluctuation distribution -- which essentially depend on the amount of non-Gaussianity (\ie skewness of the PDF) -- can have a significant impact on the PBH abundance \cite{Byrnes:2012yx,Young:2013oia,Passaglia:2018ixg,Biagetti:2021eep,Atal:2018neu,Taoso:2021uvl,Young:2022phe,Ferrante:2022mui,Gow:2022jfb}. For the sake of obtaining a lower limit on the amplitude of the PBH-forming curvature power spectrum, we now consider scenarios with large primordial non-Gaussianity. A particularly interesting case of this type occurs if the main source of the curvature perturbation results from a higher order interaction, where the distribution of $\mathcal{R}$ can be modeled as a $\chi^2$ distribution (see \eg \cite{Byrnes:2012yx, Lyth:2012yp,Linde:2012bt}):
\beq\label{chisq}
\mathcal{R} = g^2 - \langle\, g^2 \,\rangle,
\eeq
where $g$ is a Gaussian random variable ($\langle g \rangle =0$) with variance $\sigma_g^2 \equiv \langle \,g^2\,\rangle$. The PDF of $\mathcal{R}$ in this case can be determined by making a change of variable  $P_{\rm NG}(\mathcal{R}) = P_G(g)|\d g / \d \mathcal{R}|$,  which takes the following form
\beq
P_{\rm NG}(\mathcal{R}) = \fr{e^{-(\mathcal{R}+\sigma_g^2)/2\sigma_g^2}}{2 \sqrt{2\pi (\mathcal{R}+\sigma_g^2)}\, \sigma_g}.
\eeq
Making a change of variable to a quantity $t$ through the definition $\sigma_g^2 \,t  = \mathcal{R} + \sigma_g^2 \longrightarrow  \d \mathcal{R} = \sigma_g^2 \,  \d t$, the fraction of regions in the universe that can collapse to form PBHs can be estimated as
\beq\label{betaRNG}
\beta_{\rm NG} = \int_{\mathcal{R}_{\rm c}}^{\infty} \d \mathcal{R}\,\, P_{\rm NG}(\mathcal{R}) \simeq \fr{1}{2}{\rm Erfc}\left(\sqrt{\fr{1}{2} + \fr{\mathcal{R}_{\rm c}}{\sqrt{2\mathcal{P}_{\mathcal{R}}}}}\right),
\eeq
where in the last step we approximate the variance as $\langle \mathcal{R}^2 \rangle = 2 \langle g^2 \rangle^2 = 2 \sigma_g^4 \approx  \mathcal{P}_{\mathcal{R}}$, in order   to express $\beta_{\rm NG}$ in terms of the curvature power spectrum\footnote{Note that power spectrum is the variance of curvature perturbation per logarithmic interval in $k$, \ie $\langle \mathcal{R}^2\rangle \equiv \int \d \ln k\, \mathcal{P}_\mathcal{R}(k)$. Therefore the approximate signs $\simeq$ in the expressions in \eqref{betaRG} and \eqref{betaRNG} can be turned into an equality  if we consider the $\beta$'s defined in those expressions as the collapse fraction per $\d \ln k$ in the spectrum, namely $\beta = \beta(k)$.}. We can now compute the amplitude of the curvature power spectrum required for PBH formation,  when the statistics of fluctuations  is strongly non-Gaussian. Using \eqref{betaf} together with \eqref{betaRNG}, a population of solar mass PBHs with $f_{\rm pbh} = 10^{-3}$ requires $\mathcal{P}_\mathcal{R} \simeq 1.5 \times 10^{-3}$, whereas for a population of PBHs with $M^{(\rm f)}_{\rm pbh} = 10^{-12} M_{\odot}$ and $f_{\rm pbh} = 1$, we find $\mathcal{P}_{\mathcal{R}} \simeq 9 \times 10^{-4} \simeq 10^{-3}$. Hence we conclude that the required amplitude  of power spectrum is roughly given by
\beq\label{PNG}
\mathcal{P}_{\mathcal{R}} (k_{\rm pbh}) \sim 10^{-3},\quad\quad \textrm{for non-Gaussian fluctuations.}
\eeq
Compared to the case of Gaussian distributed curvature perturbation (see Eq. \eqref{PvsS}) we learn that the required amplitude of the power spectrum is reduced by about one order of magnitude. Therefore, a curvature perturbation with a smaller amplitude  can produce the same amount of PBH abundance, if  non-Gaussianity is present (see Eq. \eqref{chisq}). In particular, for $\mathcal{R}_{\rm c}^2 \gg \mathcal{P}_{\mathcal{R}}$, which is typically satisfied to a very good approximation, we can disregard the $1/2$ factor in \eqref{betaRNG}.  Comparing  with Eq. \eqref{betaRG}, the power spectrum required to generate the same collapse fraction of PBHs in both cases can be related as 

\beq
\mathcal{P}_{\mathcal{R}_{\rm NG}} \simeq \fr{2}{\mathcal{R}_{\rm c}^2} \mathcal{P}_{\mathcal{R}_{\rm G}}^2.
\eeq
\begin{figure}[t!]
\begin{center}
\includegraphics[scale=0.7]{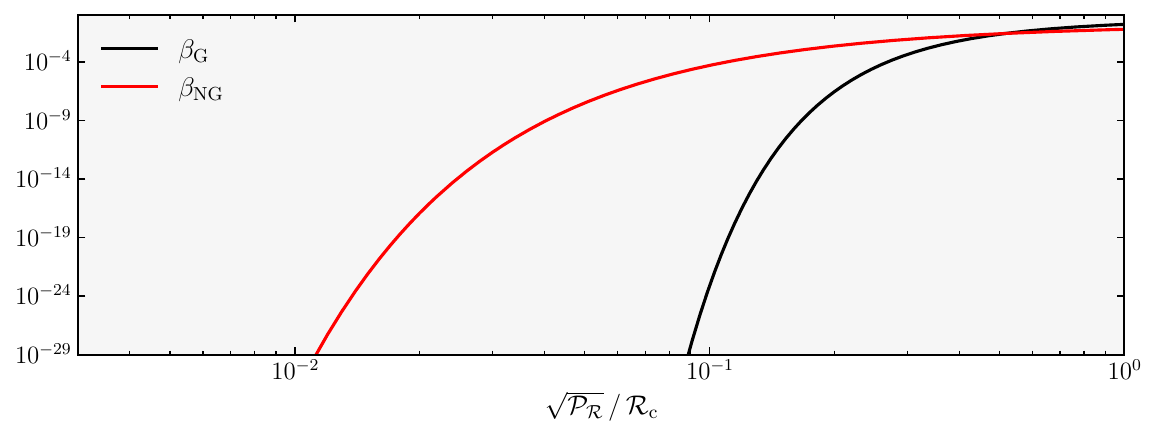}
\end{center}
\caption{Fraction of the universe that collapses into PBHs as a function of the power spectrum. For phenomenologically interesting interval of $\beta$ (see \eg \eqref{betaf}) values, in the non-Gaussian case we need a smaller amplitude of power spectrum in order to generate the same amount of PBHs at horizon re-entry.  \label{fig:beta}}
\end{figure}

To further illustrate these points, in Fig. \ref{fig:beta} we show the quantity $\beta$ both for Gaussian and non-Gaussian cases, represented as a function of the curvature power spectrum. We learn from the figure that, in
 the non-Gaussian case,  within a phenomenologically relevant interval of $\beta$ (see \eg \eqref{betaf})  a given value of the power spectrum leads to  a much larger value of $\beta$. We  emphasize that  we focused on a specific type of non-Gaussian distribution (namely $\chi^2$) to estimate the amplitude of the power spectrum required for PBH formation, and so the results we derived could change for milder cases depending on the amplitude and sign of the non-Gaussianity (\ie depending on whether the PDF in \eqref{betaRG} is positively or negatively skewed) \cite{Byrnes:2012yx}.

\subsection{Brief summary, and the path ahead}

Let us summarize the arguments we reviewed  so far. We computed  the required amplitude of small-scale primordial power spectrum  $\mathcal{P}_{\mathcal{R}}(k_{\rm pbh})$ to generate a sizeable population of PBHs that can account for all or a fraction of DM density. The typical small scale of PBH  formation $k_{\rm pbh}$ is related with the BH mass through equation \eqref{pbhmvskf}: see
Table \ref{tab:scales} for examples. 
Comparing with power spectrum at large
CMB scales, we need a

\beq
\label{reqencps}
\Delta \mathcal{P}_{\mathcal{R}} \equiv \fr{\mathcal{P}_{\mathcal{R}}(k_{\rm pbh})}{{\mathcal{P}_{\mathcal{R}}(k_{\rm cmb})}} \sim 10^{6} - 10^{7}
\eeq
enhancement in the spectrum amplitude between small and large scales,  depending on the statistics obeyed by the primordial curvature perturbation (see Eqs. \eqref{PvsS} and \eqref{PNG}).

\smallskip
We  also learned that the PBH abundance is extremely sensitive to
the amplitude of the primordial curvature spectrum.  Notice that the 
results we reviewed  are derived for the case of PBHs produced during RDU: if early phase transitions
or  early phases of non-standard cosmology occur, the corresponding  modified
 equations of state  can also considerably influence 
 the properties of the PBH population \cite{Musco:2012au,Escriva:2020tak}. An interesting example
 is the  QCD phase transition, which can lead to a high peak
 in the distribution of solar mass PBHs, several orders of magnitude larger than 
 the corresponding values in RDU \cite{Byrnes:2018clq,Carr:2019kxo}. 

\smallskip
There are various opportunities for improving and elaborating on these results. 
In our considerations, we assumed that PBHs form at a particular mass (Eqs. \eqref{pbhm} and \eqref{pbhmvskf}), by means  for example  of a sharply peaked primordial power spectrum; moreover we  ignored the effects of clustering \cite{Clesse:2016vqa,Ali-Haimoud:2018dau,Young:2019gfc,DeLuca:2020jug,DeLuca:2022uvz} and mergers \cite{Ali-Haimoud:2017rtz,Ballesteros:2018swv,Vaskonen:2019jpv} on the PBH abundance and evolution. As shown in \cite{Germani:2018jgr,DeLuca:2020ioi} and \cite{MoradinezhadDizgah:2019wjf}, assumptions on the shape of the primordial spectrum may alter the PBH abundance and the corresponding clustering properties. 
Another topic of debate concerns the use of over-density $\delta$ versus the curvature perturbation $\mathcal{R}$ when computing the  PBH abundance: see \eg \cite{DeLuca:2022rfz} for a discussion on these issues. In light of the discussion above,  we emphasize that the calculations we carried on in this section should be regarded as  rough  order-of-magnitude estimates, in need of more precise numerical analysis. Furthermore, in discussing the effects of non-Gaussianities on PBH formation, we stress that we computed the  corresponding  power spectrum only for an extreme example of $\chi^2$  non-Gaussian  statistics. For a detailed analysis of the impact of primordial non-Gaussianity on PBH formation and abundance, we refer the reader to the general discussion in 
\cite{Young:2016mxm}. An additional phenomenological consequence
of PBH-forming inflationary scenarios is the inevitable  production of a stochastic gravitational wave (GW)
background \cite{Ananda:2006af,Baumann:2007zm}. In fact,
 an enhanced spectrum 
 of curvature fluctuations, as needed to produce PBH, acts as a source
 for GW \cite{Kohri:2018awv}. The characterization of  
this GW background (see \eg \cite{Cai:2019cdl,Yuan:2019wwo,Pi:2020otn}) along with the properties of its scalar sources (see \eg \cite{Garcia-Bellido:2017aan,Unal:2018yaa,Yuan:2020iwf}), and the corresponding forecasts for its detection, is an important avenue for the experimental probe of PBH forming inflationary models. 
We refer the reader to \cite{Domenech:2021ztg} for a detailed recent review on the scalar-induced GW backgrounds.

 \smallskip
 All the topics  mentioned above are being  actively developed by the PBH community. 
 The
arguments and results we reviewed in this Section are sufficient for introducing   our specific purpose,
which is
 reviewing the theoretical foundation of inflationary scenarios leading to PBH.
 From now on, we discuss different   conceptual ideas and concrete inflationary
mechanisms for   obtaining the   enhancement \eqref{reqencps} of the curvature power spectrum, as
  needed to generate PBHs. We focus on the 
inflationary theory aspects only, 
without computing 
 the resulting PBH abundance, as well as other phenomenological properties
 which are already covered in various recent complementary reviews \cite{Khlopov:2008qy,Garcia-Bellido:2017fdg,Sasaki:2018dmp,Carr:2020xqk,Carr:2020gox,Green:2020jor,Escriva:2022duf}.

\section{Enhancement of scalar perturbations during single-field inflation}
\label{sec_single}

We now focus our attention to inflationary scenarios able to lead to PBH formation. As we learned in the previous section,
 they are characterized by a significant enhancement in the curvature power spectrum at a  scale  $k_{\rm pbh}$ (which depends on the PBH mass) much smaller with respect to CMB
 scales $k_{\rm cmb}$. The condition to satisfy is Eq. \eqref{req_ampl1}, which we rewrite here:
\beq
\frac{\mathcal{P}_{\mathcal{R}}(k_{\rm pbh})}{{\mathcal{P}_{\mathcal{R}}(k_{\rm cmb})}} \sim 10^{7}\,.
\eeq 
We classify inflationary models into single-field (this section) and multi-field type (next section), depending on whether  the mechanism responsible for the enhancement in the scalar fluctuations  respectively  relies on a single   or multi-field scenario.
In general, existing inflationary mechanisms
amplify the spectrum of curvature fluctuations by means of significant gradients in the background evolution of fields responsible
for inflation.
 In this section  
 we   phrase our discussion as model-independent  as possible, mostly focusing on conceptual aspects of the problem. We aim to   discuss   the dynamics and the general properties of  curvature fluctuations 
 in inflationary models leading to PBHs,  
 and refer to representative specific scenarios when necessary. 
 
\smallskip
\noindent\faIcon{book} {\bf\emph{Main References}}: 
Our discussion in this section 
is based on the papers \cite{Leach:2000yw,Leach:2001zf,Ozsoy:2018flq,Ballesteros:2018wlw}.
\subsection{The dynamics of curvature perturbation}\label{S3p1}
In order to analyse  the behavior of the scalar power spectrum
in single-field  scenarios,  we consider  the second-order action of scalar perturbations around an inflationary phase
of evolution. The background  metric corresponds to a
(quasi) de Sitter background, with nearly constant
Hubble parameter $H$. 
Cosmological inflation  is controlled by 
a slow-roll parameter $\epsilon \equiv \dot{H}/H^2$   satisfying $\epsilon \ll 1$, with $\epsilon = 1$ corresponding to the condition to conclude  the inflationary process. We work with conformal time, $\tau \le 0$ during inflation. (See e.g. \cite{Lyth:1998xn} for a classic survey of inflationary models.) 

The dynamics of scalar fluctuations can be formulated in terms of the comoving curvature perturbation $\mathcal{R}$ \cite{Sasaki:1986hm,Mukhanov:1990me}, whose  second order action (at lowest order in derivatives)\footnote{One can also introduce a time dependent mass in the action \eqref{cps2} which may arise through broken spatial translations as in solid  \cite{Endlich:2012pz} and supersolid \cite{Cannone:2014uqa,Bartolo:2015qvr} inflation. Another possibility is to include higher derivative terms in the quadratic action to modify the dispersion relation of curvature perturbation (see \eg \cite{Ashoorioon:2019xqc}) as in Ghost inflation \cite{Arkani-Hamed:2003juy}. We will not consider these possibilities here. For a discussion on PBH formation in solid and ghost inflation see Section 4 and 6 of \cite{Ballesteros:2018wlw} and \cite{Ballesteros:2021fsp}.} takes the following form (see e.g. \cite{Ballesteros:2018wlw})
\beq\label{cps2}
S^{(2)}_{\mathcal{R}} = \fr{1}{2}\int \d \tau\, \d^3 x\, \fr{2\,a^2(\tau) M^2(\tau) \,\epsilon(\tau)}{c_s^2(\tau)}\, \bigg[\mathcal{R}'^2 - c_s^2(\tau) (\vec{\nabla} \mathcal{R})^2\bigg]\,.
\eeq
In this formula,
  $c_s$ is the sound speed  of the curvature perturbation, $M$ is an effective time-dependent Planck mass, and $\epsilon$ the
  aforementioned slow-roll parameter.  

\smallskip
Few initial words for contextualising 
 single-field models aimed to produce
 PBHs, leading to a dynamics 
 of curvature perturbation controlled by 
  action \eqref{cps2}. 
  The simplest option
   to consider are  PBH-forming models
  with unit sound speed and constant Planck mass, characterized only by the shape  of the potential $V(\phi)$. As mentioned in the Introduction,  models in this class
  require a potential characterized by 
  flat  plateau-like region, see e.g.  \cite{Garcia-Bellido:2017mdw,Germani:2017bcs,Ballesteros:2017fsr,Hertzberg:2017dkh,Ezquiaga:2017fvi,Cicoli:2018asa,Ozsoy:2018flq,Mishra:2019pzq,Ballesteros:2020qam}
  for a choice of  works studying this possibility (We will  discuss  its implications for the dynamics of curvature perturbations in the next subsection.). PBH-forming potentials with the required characteristics can find explicit  realisations for example
   in models of Higgs inflation
\cite{Ezquiaga:2017fvi,Drees:2019xpp,Cheong:2019vzl,Rasanen:2018fom}, alpha-attractors \cite{Dalianis:2018frf,Dalianis:2019asr}, and string inflation \cite{Cicoli:2018asa,Ozsoy:2018flq,Cicoli:2022sih}.
  Considering more complex possibilities, PBH-generating 
 models which exploit a time-dependence for the sound speed  are based on non-canonical kinetic terms for the inflaton scalar, 
as K-inflation \cite{Armendariz-Picon:1999hyi,GM}: see e.g. \cite{Solbi:2021wbo,Solbi:2021rse,Teimoori:2021pte,Ahmed:2021ucx,Ballesteros:2018wlw,Kamenshchik:2018sig,Kamenshchik:2021kcw,Cai:2018tuh,Chen:2020uhe}
for concrete examples, and Section \ref{s3p3} for  some of their implications. Finally, scenarios with a 
  time-dependent effective Planck mass can be generated by  
  non-minimal couplings of the inflaton scalar
  with gravity, as in the Horndeski action \cite{Horndeski:1974wa} and its cosmological
  applications to G-inflation scenarios \cite{Kobayashi:2011nu}.
   Realisations of PBH-forming models in set-up with non-minimal
   couplings belonging to the Horndeski sector
include \cite{Frolovsky:2022ewg,Fu:2019ttf,Heydari:2021gea,Kawai:2021edk}.
  To the best of our knowledge, early universe models
  based on the more recent covariant DHOST 
  actions \cite{Langlois:2015cwa,Crisostomi:2016czh,BenAchour:2016fzp},
   have not been  explored  so far in the context of  PBH model building.

\bigskip
Interestingly, despite the many distinct 
concrete realisations, all single-field scenarios
rely in few common mechanisms for enhancing the spectrum
of curvature fluctuations, which exploit the
behaviour of background quantities. We are now going to discuss
these mechanisms in a model-independent way. 
We treat $M$, $c_s$ and $\epsilon$ as appearing in action \ref{cps2}  as time-dependent quantities, controlled by the single scalar background profile that drives inflation. To start with,
    it is convenient to redefine the time variable
    in action \eqref{cps2}, so to adsorb the   time-dependent $c_s$ into a re-scaled conformal time and impose
 an equal-scaling condition of  time and space coordinates: 
\beq\label{cps2m}
 \d \bar{\tau} = c_s \d \tau \quad \Longrightarrow \quad S^{(2)}_{\mathcal{R}} = \fr{1}{2}\int \d \bar{\tau}\, \d^3 x\, z^2(\bar{\tau})\, \bigg[\mathcal{R}'^2 - (\vec{\nabla} \mathcal{R})^2\bigg],
\eeq
with a prime  indicating a derivative with respect to $\bar{\tau}$, the re-scaled conformal time. Importantly, we introduce a so-called  time-dependent `pump field' $z(\bar \tau)$ as
\beq\label{zsq}
z^2(\bar \tau) \, = \,\frac{2\,a^2(\bar \tau)\, M^2(\bar \tau)\,\epsilon (\bar \tau)}{c_s(\bar \tau)}.
\eeq
 The dynamics of $\mathcal{R}$ is strongly tied to the time dependence of the pump field $z(\bar \tau)$, and more generally to the behavior of the background quantities $M,\epsilon, c_s$ that  constitute it. 
 
 To analyze  the  evolution
 mode by mode, we work in 
 Fourier space,  and write the Euler-Lagrange mode equation for  curvature perturbation, derived
 from the action \eqref{cps2m}:
\beq\label{cpeom}
\frac{1}{z^{2}(\bar{\tau})}\left[z^{2}(\bar{\tau}) \mathcal{R}_{k}^{\prime}(\bar{\tau})\right]^{\prime}=-k^{2} \mathcal{R}_{k}(\bar{\tau}),
\eeq
where $k \equiv |\vec{k}|$ is the magnitude of the wave-number that labels a given mode.
 This is a differential equation involving derivatives along the time direction, acting on the function $\mathcal{R}_{k}(\bar{\tau})$ depending both on time and momentum $k$. 
 
 To express its solution, we implement  a gradient expansion approach (see e.g. \cite{Leach:2000yw,Leach:2001zf,Ozsoy:2018flq}), starting from the solution in the limit of small $k/(a H)$, and including its momentum-dependent corrections which solve \eqref{cpeom} order-by-order
 in a $k/(a H)$ expansion. This approach is particularly
 suitable for our purpose of describing scenarios
 where the size of small-scale curvature fluctuations ($k/(a H)$ large)
 differs considerably from large-scale ones ($k/(a H)$ small): see condition \eqref{req_ampl1}.
  Indeed, a gradient
  expansion  allows us to better  understand the physical origin of possible mechanisms which raise the curvature spectrum at small scales. 

\smallskip  
 The most general solution of
 Eq. \eqref{cpeom}, up to second order in powers of  $k/(a H)$, is formally given by the following  integral equation \footnote{In fact, if the time evolution of the pump field  is known, up to second order in the gradient expansion we can generate a solution for the curvature perturbation 
 by replacing $\mathcal{R}_k(\bar{\tau}'')$ in the last  integral of Eq. \eqref{cpgs} with the leading growing mode solution of the homogeneous part of Eq. \eqref{cpeom}, which we can identify as $\mathcal{R}^{(0)}_k$. }
\beq\label{cpgs}
\mathcal{R}_k(\bar{\tau})\,=\,\mathcal{R}^{(0)}_{k}\left[1+\frac{\mathcal{R}^{(0)\,\prime}_{k}}{\mathcal{R}^{(0)}_{k}} \bigintsss_{\bar{\tau}_{0}}^{\bar{\tau}} \frac{\d \bar{\tau}'}{\tilde{z}^{2}(\bar{\tau}')}- k^{2} \bigintsss_{\bar{\tau}_{0}}^{\bar{\tau}} \frac{\d \bar{\tau}'}{\tilde{z}(\bar{\tau}')^{2}} \bigintsss_{\bar{\tau}_{0}}^{\bar{\tau}'} \mathrm{d} \bar{\tau}'' \,\tilde{z}^{2}(\bar{\tau}'')\, \frac{\mathcal{R}_k (\bar{\tau}'')}{\mathcal{R}^{(0)}_{k}} \right],
\eeq
where the  sub and super-scripts $0$ denote a reference time, and tilde over a time-dependent quantity indicates that it is normalized with respect to its value  at $\bar{\tau} = \bar{\tau}_0$.

Typically, we are interested in relating the late time curvature perturbation at $\bar \tau$ to the same quantity computed at some earlier time $\tau_0$.   For this purpose, it is convenient to identify $\bar{\tau}_0$ as the time coordinate evaluated soon after horizon crossing, and  $\mathcal{R}^{(0)}_{k}$ as the mode function computed at  $\bar{\tau}_0$. In order for enhancing
the spectrum of curvature fluctuations at small scales (recall the PBH-forming condition of Eq. \eqref{req_ampl1}) we can envisage two possibilities. 
One option is to exploit the structure of Eq. \eqref{cpgs}, making sure that its contributions within the square parenthesis become more and more important
as time proceeds after modes leave the horizon. In this way, we generate a
sizeable scale-dependence for $\mathcal{R}_k(\bar{\tau})$ after horizon crossing, with the possibility of amplifying the small-scale curvature spectrum. Alternatively, we can design methods that lead to significant scale dependence
already at horizon crossing, {\it i.e.} for the quantity $\mathcal{R}^{(0)}_{k}$, which then maintains  frozen its value  at super-horizon scales. In what follows, we  explore both
these two options, 
in Sections \ref{s3p2} and \ref{s3p3} respectively.
 
 To develop a quantitative discussion, it is convenient to introduce the so-called slow-roll parameters as
 \beq
\eta \equiv \frac{\d \ln \epsilon}{\d N}, \,\, s \equiv  \frac{\d \ln c_s}{\d N}, \,\, \mu \equiv \frac{\d \ln M^2}{\d N}
\label{src}
 \eeq
where in our
definition
we make use of
 the relation between e-foldings and the time coordinate $\bar{\tau}$: $\d N = H \d t = (a H /c_s) \d \bar{\tau}$. 

 In standard models of inflation based on an inflationary attractor dynamics, one imposes the so-called
 slow-roll conditions throughout the entire inflationary period, corresponding
 to the requirements
 \beq
 \eta ,\, s,\, \mu,\, \frac{\d \eta}{\d N},\, \frac{\d s}{\d N},\, \frac{\d \mu}{\d N}\,\ll \,1,
 \label{src2}
 \eeq
 which 
 imply  that the pump field  {\it always grows in time} $z^2 \propto (-\bar{\tau})^{-2}$ as $\bar{\tau} \to 0$ (see Eq. \eqref{zsq}). 
 As a consequence, the second and  third terms in the general solution \eqref{cpgs} decay respectively as $(-\tau)^3$ and $(-\tau)^2$ in the late time limit $(-\tau) \to 0$. Hence they can be identified as  decaying modes \footnote{In particular, the standard decaying mode is given by the last term in \eqref{cpgs} as it decays slowly,  \ie $\propto (-k\bar{\tau})^2$, compared to the second.} that rapidly cease to play any role in 
 the dynamics of curvature perturbations. This is a regime of {\it slow-roll attractor}, where   soon after horizon crossing the curvature perturbation  settles into a nearly-constant   configuration $\mathcal{R}^{(0)}_k$, whose spectrum is almost scale-invariant.  In this case, the momentum-dependent  terms in Eq. \eqref{cpgs} do not have the opportunity to raise the curvature spectrum at small scales.
 
\smallskip
Hence, for producing PBH we need to go beyond the slow-roll conditions
of Eq. \eqref{src2}, as first emphasized in \cite{Motohashi:2017kbs}.
Before discussing concrete ideas to do so, in view of numerical implementations, as well as for improving our  physical
understanding, 
 it is  convenient
 to express
the curvature perturbation 
equation  \eqref{cpeom} in a way that makes more manifest the role 
of slow-roll parameters in controlling the mode evolution. We introduce a canonical variable $v_k$ defined as 
\beq \label{msre}
v_k (\bar{\tau}) = z(\bar{\tau}) \mathcal{R}_k(\bar{\tau})
\,.
\eeq
Plugging this definition in Eq. \eqref{cpeom},
we obtain  the so-called Mukhanov-Sasaki equation, which reads
\beq\label{MS}
v_k''(\bar{\tau}) + \left(k^2 - \fr{z''}{z}\right)v_k(\bar{\tau}) = 0,
\eeq
where 
\beq\label{zppoz}
\fr{z''}{z} = 2\left(\fr{aH}{c_s}\right)^2\, \big[ 1 + \Theta\big].\\
\eeq
Expanding the derivatives of $z$ \eqref{zsq} in terms of the slow-roll parameters of Eq. \eqref{src}, we define 
\begin{align}\label{theta}
\Theta &\equiv -\frac{1}{2}(\epsilon + s) + \frac{3}{4}(\eta - s + \mu) + \frac{1}{8}(\eta -s + \mu)^2 -\frac{1}{4}(\epsilon + s) (\eta - s + \mu) + \frac{1}{4}\left(\fr{\d \eta}{\d N}- \fr{\d s}{\d N} + \fr{\d \mu}{\d N}\right).
\end{align}
Standard slow-roll attractor scenarios correspond to  situations
where $\Theta$ is negligibly small:  the quantity in Eq. \eqref{zppoz}
then reads $2\,\bar \tau^{-2}$, leading to a scale-invariant curvature power
spectrum. To break scale-invariance of curvature perturbation, we need 
to consider a sizeable time dependent $\Theta$. 
We note that the expression \eqref{theta} is exact, and does not  assume any slow-roll hierarchy as Eq. \eqref{src2}. Hence it can be used to study the system beyond slow-roll, as we are going to do in what comes next.

\subsection{Enhancement through the resurrection of the decaying mode}\label{s3p2}

\noindent{\bf The idea}
\medskip

\noindent
An interesting mechanism to enhance the curvature perturbation at super-horizon scales is suggested by
the structure of the integrals within the  square parenthesis of Eq. 
 \eqref{cpgs}. Suppose that, for a brief time interval,  a given mode $k$ experiences a background evolution during which the pump field  $z$ \emph{rapidly decreases} after the  horizon exit epoch $\bar{\tau}_0$. Then,  
 the would be  `decaying' mode can grow large, and 
 the integrals in the parenthesis of Eq  
 \eqref{cpgs} can  contaminate the nearly constant  solution $\mathcal{R}^{(0)}_k$,   eventually leading to a late-time value $\mathcal{R}_k (\bar{\tau}) \gg \mathcal{R}^{(0)}_k$ on super-horizon scales. This situation signals a significant departure from the attractor, slow-roll regime discussed after Eq. \eqref{src2}. 
In fact, in this case the criterion for the enhancement of the curvature perturbation  can be  explicitly phrased in terms of the derivative of the pump field, transiently changing sign during some short time interval  during inflation: 
\beq\label{zpoz}
\frac{z'}{z} = \fr{a H}{c_s} \left[1 + \frac{\eta - s + \mu}{2}\right] < 0 \,.
\eeq
This condition implies that the combination of the slow-roll parameters, $\eta - s + \mu$ should be of order ${\cal O}(1)$ and negative during some e-folds during inflation, violating the slow-roll conditions \eqref{src2}. In particular, we require 
\beq\label{dmc}
\eta - s + \mu < -2\,.
\eeq
 If Eq. \eqref{zpoz} is satisfied,
the slow-roll conditions \eqref{src2} are not satisfied, and the contributions
within parenthesis of Eq. 
 \eqref{cpgs} can grow large.  Strong time gradients of homogeneous
 background quantities, which lead to condition \eqref{dmc}, can
 then be converted into a small-scale amplification of the curvature
 power spectrum. 
As    discussed in \cite{Ballesteros:2018wlw}, the expression  
\eqref{zpoz}, along with the considerations  above, generalizes to a time-dependent sound speed and Planck mass  the arguments first  developed in \cite{Leach:2000yw,Leach:2001zf}.

\medskip
\noindent{\bf Model building, and a parametrization of the non-attractor  phase}
\medskip

\noindent
To illustrate a viable model that can generate a seven-order of magnitude enhancement required for PBH formation -- see Eq. \eqref{req_ampl1} -- we  focus on canonical single-field models, $c_s \to 1,\, M \to \Mpl$ (and $s \to 0,\, \mu \to 0$), in order to simplify our analysis. The  background evolution for the single scalar field driving 
inflation is
\begin{equation}
\label{kgeq1}
\ddot{\phi} + 3H\dot{\phi} + V'(\phi) = 0\,,
\end{equation}
with $V(\phi)$ the scalar potential, and the time-derivatives are carried on in coordinate time $t$. The non--slow-roll dynamics is controlled
 by the properties of the potential $V$, as
 we are going to discuss, and by its consequences for the behaviour
 of the inflaton velocity $\dot \phi$. 

Since in these scenarios the pump field can be parametrized purely in terms of the slow-roll parameter $\epsilon$ as $z = a \sqrt{2\epsilon}\, \Mpl$ (see \eg Eq. \eqref{zsq}), the linear dynamics of $\mathcal{R}_k$ (Eq. \eqref{cpgs}) is  dictated by the first slow-roll parameter, whose evolution is in turn determined by the sign and amplitude of  the slow-roll parameter $\eta$. Hence, the criterion required to realize the desired growth in the spectrum  can be simply parametrized as a condition on the second slow-roll parameter, as $\eta < -2$ in \eqref{dmc}. 

From a concrete model building perspective, scalar potentials $V(\phi)$ that can induce this type of dynamics include a characteristic  `plateau' within a  non-vanishing field range $\Delta \phi \neq 0$ \cite{Starobinsky:1992ts,Ivanov:1994pa,Pi:2022zxs}. This property  gives rise to  phases of transient non-attractor dynamics, of ultra slow-roll (abbreviated USR) \cite{Kinney:2005vj,Martin:2012pe,Dimopoulos:2017ged} or constant roll (CR) \cite{Inoue:2001zt,Tzirakis:2007bf,Motohashi:2014ppa} evolution,  depending on the  shape profile of the potential around the aforementioned feature. In particular, for USR the potential typically has a very flat plateau with $V' \simeq 0$, whereas for constant roll $V' < 0$, so that the field climbs a hill by overshooting a local minimum\footnote{Here we assume field rolls down on its potential from large to small values ($\dot{\phi} < 0$) with $V'(\phi)$ before it encounters with feature required for the enhancement. Since $V' < 0$ during the feature, there must be a point in the potential where the second derivative of the field vanishes $V'' = 0$.}. As the scalar field, during its evolution, traverses such a flat region
with negligible potential gradient, the acceleration term $\ddot \phi$ is balanced by the Hubble damping term in the Klein-Gordon equation \eqref{kgeq1}, and the inflaton speed is no longer controlled by the  scalar potential. This phenomenon 
 changes significantly the values of the inflaton velocity $\dot \phi$
 during the transient non-attractor phase, and 
inevitably leads to  the violation of one of the slow-roll conditions:
\beq\label{KG}
\ddot{\phi} + 3H\dot{\phi} + V'(\phi) = 0 \,\,\, \Longrightarrow \,\,\, \eta = -6 - \fr{2 V'}{\dot{\phi} H} + 2\epsilon,
\eeq
\noindent hence $\eta \simeq -6$ ($\eta < -6$) for transient USR $V' = 0$ (CR $V' < 0$) phases respectively\footnote{Note that for single-field inflation we have $\epsilon \equiv \dot{\phi}^2/(2H^2\Mpl^2)$ and using $\eta$ \eqref{src} we get $\eta = 2\ddot{\phi}/(\dot{\phi}H) + 2\epsilon$. Using the Klein-Gordon equation in the last expression gives the relation on the right hand side of \eqref{KG}.}.
 We emphasize  that since the non-slow-roll inflationary era is characterized by a large negative $\eta$ for a brief interval of e-folds, the pump field, as well as the first slow-roll parameter $\epsilon$, quickly decay during this stage as required for the activation of the decaying modes. In fact, 
\beq
\fr{\d \ln \epsilon}{\d N} \equiv \eta \quad \Longrightarrow \quad z^2 \propto \epsilon \propto e^{-|\eta|\Delta N},
\eeq
where for simplicity we assume a constant $\eta$ during the non-attractor phase.
For explicit inflationary scenarios that can realize such transient phases in the context of PBH formation, see \eg \cite{Ballesteros:2017fsr,Cicoli:2018asa,Ozsoy:2018flq,Motohashi:2019rhu}. 
 Nevertheless, it is worth pointing out that, although possible, explicit
 constructions of suitable inflationary potentials involve a high degree of
 tuning to render the potentials extremely flat for a small region in field
 range, and ensure an appropriate transition for the scalar velocity among different epochs. See e.g. the discussion in \cite{Hertzberg:2017dkh}, as well as the comments at the end of this section.

\smallskip

After this general discussion on model building,
in the analysis that follows we do not need to work with an explicit form of potential $V(\phi)$ to analyze the enhancement through the non-attractor dynamics. Instead, we 
exploit the general idea we are discussing in a model-independent way, and we 
model PBH forming inflationary scenarios as a succession  of distinct phases which connect smoothly one with the other, each parametrized by a constant $\eta$ (Related
approaches are developed in \cite{Byrnes:2018txb,Ragavendra:2020sop,Ng:2021hll,Karam:2022nym,Franciolini:2022pav}). Our perspective catches  the important features of  scenarios based on the idea of transiently resurrecting the decaying mode at super-horizon scales, satisfying Eqs. \eqref{zpoz} and \eqref{dmc}.
In order to capture the transitions among phases, we multiply each phase by the smoothing function \cite{Cole:2022xqc}:  
\beq\label{sf}
\sigma(N, \Delta)=\frac{1}{2}\left[\tanh \left(\frac{N-N_{i}}{\Delta}\right)-\tanh \left(\frac{N-N_{f}}{\Delta}\right)\right],
\eeq
\begin{table}
\begin{center}
\begin{tabular}{| c | c | c | c |}
\hline
\hline
\cellcolor[gray]{0.9}&\cellcolor[gray]{0.9} Phase I &\cellcolor[gray]{0.9} Phase II &\cellcolor[gray]{0.9} Phase  III \\
\hline
\cellcolor[gray]{0.9}$\eta$& $ 0.02 $ & $-6.30$ & $ 0.30\,\, \big|\,\, 3.00 $ \\\hline
\cellcolor[gray]{0.9}$N_i$ & $0.00$ & $33.2$& $ 35.7\,\, \big|\,\,55.0$\\\hline
\cellcolor[gray]{0.9}$N_j$ & $33.2$ & $35.7$ & $ 55.0\,\, \big|\,\,65.0$  \\\hline
\cellcolor[gray]{0.9}$\nu$ & $1.00 $ & $0.50$ & $1.00\,\, \big|\,\,2.00$ \\\hline
\hline
\end{tabular}
\caption{\label{tab:par} Parameter choices that characterize the background evolution of $\eta$ smoothed by the function \eqref{sf} at each phase. Note that the final phase of evolution divided into two in order to accommodate the end of inflation with $\epsilon = 1$ at $N_{\rm end} = 60$.}
\end{center}			
\end{table}
\noindent where $N$ denote e-folds, $N_i$ and $N_f$ are the e-folding numbers at the beginning and end of the constant $\eta$ phase, and $\Delta$ signifies the duration of the smoothing procedure. Keeping this smoothing prescription in mind, the inflationary evolution can be divided into three phases:
\begin{itemize}
    \item {\bf Phase I}. The initial phase of  inflationary evolution is characterized by a standard slow-roll (SR) regime, where $\epsilon,\eta \ll 1$ and $\epsilon < \eta$ at the pivot scale $k_{\rm cmb} = 0.05\, {\rm Mpc}^{-1}$ (assuming that modes at the pivot scale exits the horizon at the beginning of evolution $N_{\rm b} = 0$), in order to match  Planck observations \cite{Planck:2018jri}.
    \item {\bf Phase II}. As the scalar field starts to traverse the flat plateau-like region in its potential, its dynamics eventually enter the non-attractor era lasting  some  e-folds of evolution. This phase is characterized by a large negative $\eta < -6$, during which the first slow-roll parameter $\epsilon$ decays exponentially: 
    \beq\label{deps}
    \epsilon(N) \equiv \exp{\left[\int_{60}^{N} \eta(N')\, \d N'\right]}.
    \eeq
    
    \item {\bf Phase III}. The final phase of evolution ensures a graceful exit from the non-attractor phase into a final slow-roll epoch, leading to the end of inflation. Since $\epsilon$ decays quickly in the non-attractor era, this final phase is characterized by a hierarchy between the slow-roll parameters: \begin{equation}\eta \gg \epsilon\,
    \end{equation}
    where $\eta > 0$. We typically require a large positive $\eta$ to bring back $\epsilon$ from its tiny values at the end of the non-attractor era,
    towards the value $\epsilon = 1$  needed to conclude  inflation. To capture this behavior accurately, we split the final phase of evolution into two parts, parametrizing $\eta$ as   
    \beq
    \label{deps3}
    \eta(N) = \eta^{(1)}_{\mathrm{III}}\, \sigma_1(N,\Delta)\, + \,\eta^{(2)}_{\mathrm{III}} \,\sigma_2(N,\Delta)\,.
    \eeq
The relevant parameter choices to model the dynamics can be found in the third column in Table \ref{tab:par}. 

We note that our choice of $\eta$ in the initial stage of the Phase III and in Phase II is not a coincidence: most of the single-field modes there exist a correspondence that relates $\eta$'s in Phase II and Phase III: $\eta_{\mathrm{III}} = -6 - \eta_{\mathrm{II}}$, which
is a consequence of Wands' duality \cite{Wands:1998yp}. We will elaborate below on the consequence of this correspondence in the context of the power spectrum, in particular for modes that exit the horizon as the background evolves from Phase II to Phase III.
\end{itemize}

\begin{figure}[t!]
\begin{center}
\includegraphics[scale=0.6]{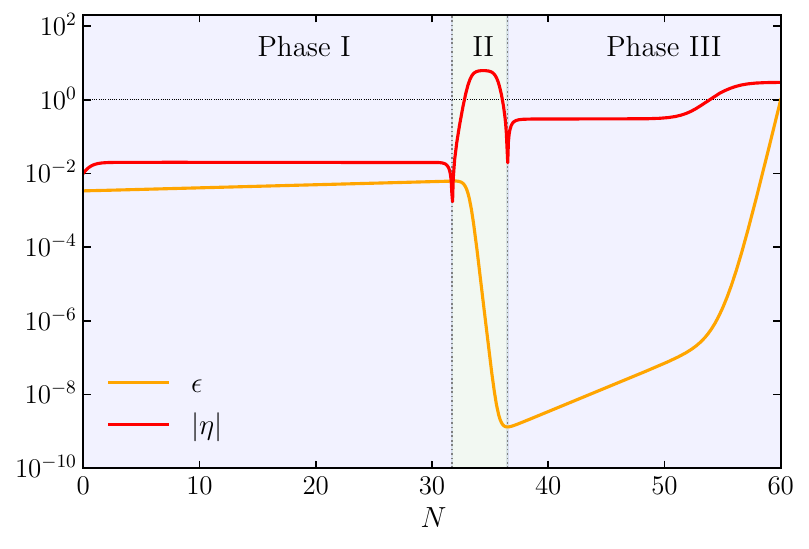}\,\,\includegraphics[scale=0.6]{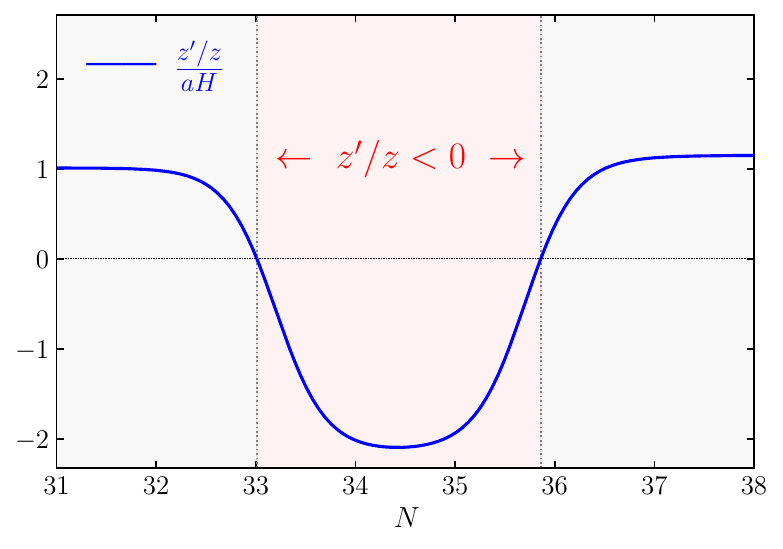}
\end{center}
\caption{Left panel: Evolution of $\epsilon$ and $\eta$ in e-folds through the successive phases outlined in the main text. The green colored region indicates the range of e-fold numbers where $\eta <0$, corresponding roughly to the beginning and end of the non-attractor phase. Right panel: the time evolution of $z'/z = a H (1 + \eta/2)$, with $z'/z < 0$ in the region highlighted with red color. \label{fig:cb}}
\end{figure}

Following the discussion above, we can characterize the full background evolution using the Hubble hierarchy in \eqref{deps} and $H(N) = H_{\rm end} \exp{[-\int_{60}^{N} \epsilon(N') \d N']}$, where $H_{\rm end}$ denotes the Hubble rate at the end of inflation, where $N_{\rm end} = 60$. 

For a representative set of parameter choices (see  Table \ref{tab:par}), 
 we show in Fig. \ref{fig:cb} 
  an example of background evolution,  in which  we plot $\epsilon,\eta$ and $z'/z$.  The right panel of the figure makes manifest  that the background evolution leads to $z'/z < 0$ for a short interval of e-folds  ($N = 33 - 35.7$), as highlighted by the red region in the plot. In accord with our discussion so far, this behavior is appropriate for triggering a significant  enhancement in the power spectrum of curvature perturbation through the resurrection of the decaying mode.

\medskip
\noindent{\bf Numerical analysis}
\medskip

\noindent
Having obtained the background evolution, we are ready to describe mode evolution to obtain power spectrum of curvature perturbation towards the end of inflation \footnote{Note that evaluating the power spectrum at the end of inflation is necessary when modes evolve outside the horizon, as in the example background we are focusing in this section.}:
\beq\label{psf}
\mathcal{P}_{\mathcal{R}}(k, N_{\rm end}) = \fr{k^3}{2\pi^2} \bigg|\frac{v_k (N_{\rm end})}{z(N_{\rm end})}\bigg|^2 ,
\eeq

\begin{figure}[t!]
\begin{center}
\includegraphics[scale=0.6]{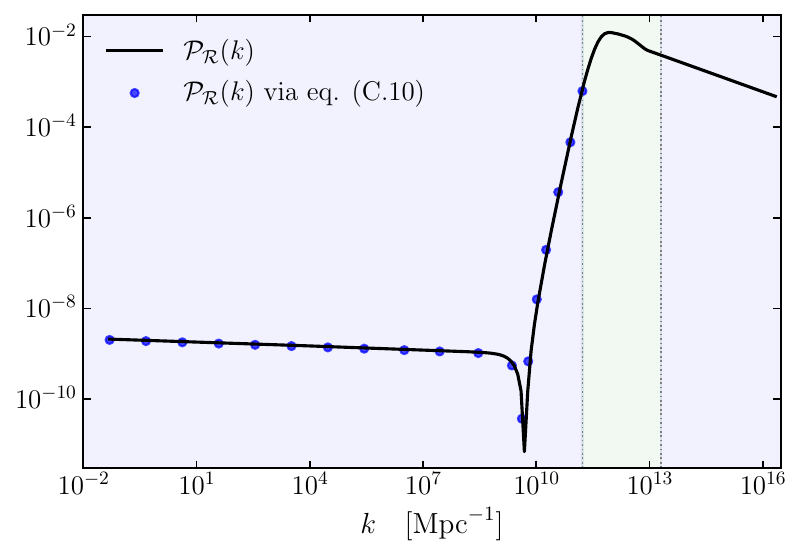}\,\,\includegraphics[scale=0.6]{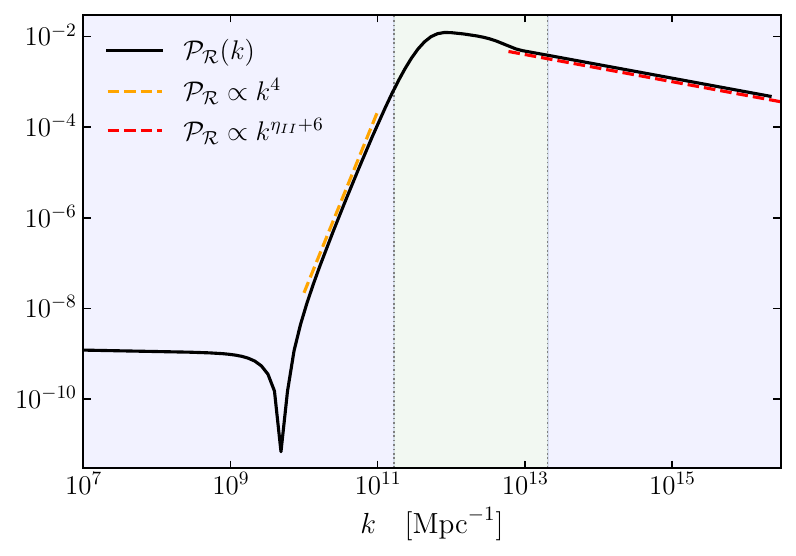}
\end{center}
\caption{Power spectrum of curvature perturbation in the three-phase model described in the main text. Pale green region separated by the vertical lines denote the range of modes that exit the horizon during the non-attractor era whereas the light blue regions denote range of modes that cross the horizon during the initial and final slow-roll era respectively. \label{fig:psc}}
\end{figure}

\noindent 
where 
to study the  evolution of curvature perturbations, we
make use of the canonical variable $v_k$ and consider the Mukhanov-Sasaki  system of equations \eqref{MS}-\eqref{theta}
after setting $s=\mu=0$. In general, it is not possible to find full analytic solutions for this system of equations, and a  numerical analysis is needed~\footnote{Although, as we will explain soon, interesting  properties of the resulting curvature spectrum can be derived and understood analytically.}.  We implement the numerical procedure explained in detail in the technical Appendix \ref{AppC}, which solves  the Mukhanov-Sasaki equation with Bunch-Davies initial conditions, and we provide a Python code that reproduces our numerical findings~\footnote{\label{ftncode} In fact, the general procedure outlined in Appendix \ref{AppC} can be generalized to accurately solve Mukhanov-Sasaki equation a broad class of single-field models of inflation. In the context of phenomenological models we discuss in this and the next section, {\sf jupyter} notebook files that compute the power spectrum is
available at the link \href{https://github.com/oozsoy/SingleFieldINF_Powerspec_PBH}{github}. We acknowledge the use of the python libraries: {\sf matplotlib} \cite{4160265}, {\sf numpy} \cite{Harris_2020}, {\sf scipy} \cite{Virtanen_2020}, {\sf pandas} \cite{inproceedings} along with {\sf jupyter} notebooks \cite{soton403913}. }.
 The resulting power spectrum  is represented in Fig. \ref{fig:psc}: it  manifestly  grows in amplitude towards small scales, exhibiting a peak at around $k_{\rm peak} \simeq 10^{12}\,\,{\rm Mpc}^{-1} \gg k_{\rm cmb}$. Notice that the spectrum 
 grows as $k^4$ towards its peak, and is characterized by a dip preceding
 the phase of steady growth \cite{Byrnes:2018txb}. We will have more to say soon about these
 features.

 Interestingly, for the system under consideration  the bulk of the enhancement can be attributed to the active dynamics of the would-be  `decaying modes', the second and third term of Eq. \eqref{cpgs}. To show this explicitly, we study super-horizon solution of the curvature perturbation in Appendix \ref{AppC} by applying the  formula \eqref{cpgec}, a special case  of Eq. \eqref{cpgs}, to the canonical single-field scenario we discuss here. For a grid of wave-numbers that exit the horizon during the initial slow-roll era, the amplitude of power spectrum obtained in this way is shown by blue dots in Fig. \ref{fig:psc}. The accuracy of these locations with respect to the full numerical result (black solid line) confirms our expectation that decaying modes in  \eqref{cpgs}  play a crucial role for the enhancement of the curvature perturbation for this scenario.
In the right panel of Fig. \ref{fig:psc}, we zoom in to the growth and the subsequent decay of power spectrum following the peak. 

 \medskip
 \noindent{\bf The features of the spectrum: analytic considerations}
  \medskip

\noindent
Besides
 the numerical findings  presented above, we can
  derive general analytic results for the spectrum of curvature fluctuations
 in scenarios  activating  the would-be
 decaying modes through a brief non-attractor era.

 \smallskip
 
  We start noticing that for modes that leave the horizon during the initial slow-roll stage (leftmost region colored by light blue in Fig. \ref{fig:psc}), the spectrum shows characteristic features such as the presence of a dip, followed by an enhancement parametrized by a spectral index of $n_s - 1 = 4$ during the bulk of the growth \cite{Byrnes:2018txb, Carrilho:2019oqg, Ozsoy:2019lyy, Liu:2020oqe, Tasinato:2020vdk}.
  The dip is physically due to a disruptive  interference between the `constant' mode of curvature fluctuation at super-horizon scales, and the `decaying' mode that is becoming active and ready to contribute to  the enhancement of the spectrum. 
  The position and depth of the dip is analytically calculable in terms of other features of the spectrum, at least in a limit of short duration
of the non--slow-roll epoch. It is found that the position of the dip  in momentum space is proportional
to the inverse fourth root of the enhancement of the spectrum, and the depth
of the dip is proportional to  the inverse square root of the enhancement of the spectrum \cite{Tasinato:2020vdk}. These relations are valid for any single-field models that enhance the spectrum  through  a brief deviation from the standard attractor era, including
 cases with  a time-varying sound speed and Planck mass.  
 They are accompanied by consistency conditions on the squeezed limit of non-Gaussian higher-order point functions around the dip \cite{Ozsoy:2021qrg,Ozsoy:2021pws,Zegeye:2021yml}, as expected in single-field scenarios \footnote{The distinct behavior of the cosmological correlators around the dip feature may also be probed by the 21-cm signal of the Hydrogen atom \cite{Balaji:2022zur}.}.  
 
\medskip

While in the considerations of the previous paragraph we considered modes leaving the horizon
during the first  stage of slow-roll evolution, we can also derive
analytic results for what happens during the non-attractor epoch. In fact,  for modes that exit the horizon deep in the non-attractor era (light green region in the middle of  Fig. \ref{fig:psc}) and the following final slow-roll era, the spectrum behaves as  expected in a standard slow-roll phase, with  spectral index 
\begin{equation}
\label{prewd}
n_s - 1 = -2\epsilon - \eta \simeq -\eta_{\rm{III}} = 6 + \eta_{\rm II}\,.
\end{equation}
(Recall that the latin numbers ${\rm II}$ and ${\rm III}$ relate with the phases of evolution, see Eqs. \eqref{deps} and  \eqref{deps3}.) This behavior is  a manifestation of the duality invariance of perturbation spectra within distinct inflationary backgrounds, called Wands duality (see \eg \cite{Wands:1998yp,Morse:2018kda}). Wands duality can be understood by noticing
that the structure of Mukhanov-Sasaki
equation, Eq. \eqref{MS}, is unchanged by a redefinition
of the pump field that leaves the combination $z''/z$ invariant:
\begin{eqnarray}
\label{def_tiz}
z(\bar \tau)&\to&\tilde z(\bar \tau)\,\equiv\,z(\bar \tau)\left(c_1+c_2\,\int^{\bar \tau}\,\frac{d \tilde \tau}{z (\tilde \tau)} \right)\hskip0.5cm\Rightarrow\hskip0.5cm \frac{\tilde z'' }{\tilde z}\,=\,\frac{ z'' }{ z}
\end{eqnarray}
where $c_{1,2}$ are
arbitrary constants. If $z(\bar \tau)$ controls a phase of slow-roll attractor, $\tilde z \propto 1/\bar \tau$, a dual phase whose pump field $z \propto \bar \tau^2$ as given by Eq. \eqref{def_tiz}  describes a non-attractor era. Although the statistics of the canonical variable $v_k$ is identical in the two regimes, the amplitude of the curvature perturbation spectrum ${\cal R}_k$
 increases in the non-attractor epoch. In scenarios where the parameter $\eta$ is well larger than the other slow-roll parameters, Wands duality \eqref{def_tiz} analytically prescribes the relation \eqref{prewd}, in agreement with the numerical findings plotted in Fig. \ref{fig:psc}. Subtleties can arise in joining attractor and non-attractor phases, since consistency conditions can be violated \cite{Suyama:2021adn} due to the effects  of boundary conditions at the transitions between different epochs. All these considerations are  relevant for our topic, given the sensitivity of PBH formation and properties on the shape of the spectrum near the peak.

For further detailed accounts on the characterization of the interesting  features in the power spectrum of PBH forming single-field scenarios, we refer the reader to  \cite{Byrnes:2018txb, Carrilho:2019oqg, Ozsoy:2019lyy, Liu:2020oqe, Tasinato:2020vdk, Ng:2021hll, Davies:2021loj, Karam:2022nym, Cole:2022xqc}. 

 \medskip
 \noindent{\bf Stochastic inflation and quantum diffusion} 
  \medskip
 
 \noindent
 While, so far, we focused on the predictions of the second order action \eqref{cps2},
non-linearities and non-Gaussian effects can play an important role in the
production of PBHs, as we learned in Section \ref{S2p3}. For the case of ultra--slow-roll (USR)
models based on non-attractor phases of inflation, there are   sources of 
non-Gaussianity  associated with  stochastic effects during inflation.

The stochastic approach to inflation, pioneered by Starobinsky 
\cite{Starobinsky:1986fx}, constitutes 
a powerful formalism for describing the evolution of coarse-grained, super-horizon
fluctuations during inflation. It is based on a classical (but stochastic) Langevin equation,
which reads in canonical single-field inflation  ($N$ is the number of e-folds, and we assume constant sound speed and Planck mass):
\begin{equation}
\label{eq_langv}
\frac{d \phi}{d N}\,=\,-\frac{V'}{3\,H^2}+\frac{H}{2 \pi}\,\xi(N)\,.
\end{equation}
Here, $\phi$ represents a coarse-grained version of  super-horizon scalar fluctuations;
 $V'$ is the derivative
 of the inflationary potential, which leads  to a deterministic drift for the coarse-grained super-horizon
 mode; $\xi$ is a source of stochastic noise acting
 on long wavelength fluctuations, caused by the continuous
 kicks of modes that cross the cosmological horizon, and pass from sub to super-horizon
 scales during inflation. 
 
 Besides the physical insights that it offers,  the
  inflationary stochastic 
 formalism  \cite{Starobinsky:1986fx,Nambu:1987ef,Kandrup:1988sc,Nambu:1989uf,Starobinsky:1994bd,Finelli:2008zg,Burgess:2014eoa,Vennin:2015hra,Burgess:2015ajz} offers  the opportunity to obtain accurate  results  for 
 the probability distribution function controlling coarse-grained  super-horizon
 modes, beyond any Gaussian approximation. As a classic example, by solving
 the Fokker-Planck equation associated with  \eqref{eq_langv}, the
 seminal work \cite{Starobinsky:1994bd}  analytically obtained the full  non-Gaussian 
  distribution functions for certain  representative  inflationary potentials, 
   going beyond the reach of a perturbative treatment of the problem. 

Returning to the discussion of  an USR inflationary evolution for PBH scenarios, we can expect that stochastic effects can be very relevant in this context, see e.g. \cite{Ando:2020fjm,Biagetti:2018pjj,Ezquiaga:2018gbw,Ballesteros:2020sre,Vennin:2015hra,Cruces:2018cvq,Firouzjahi:2018vet,Pattison:2019hef,Vennin:2020kng,Rigopoulos:2021nhv}. 
In fact, since the amplitude of scalar
fluctuations gets amplified,  the stochastic noise can become 
much larger than what occurs in slow-roll inflation. Moreover, during USR, the derivative of the
potential $V'=0$, the classical drift is absent, and the stochastic
evolution is  driven by  stochastic effects only. Various works  studied
the topic by solving the stochastic evolution equations,
 and \cite{Pattison:2017mbe,Ezquiaga:2019ftu,Figueroa:2020jkf,Figueroa:2021zah} find  that non-Gaussian effects can 
change the predictions of PBH formation, depending on the duration
of the USR phase. In fact, the stochastic noise modifies
the tails for the curvature probability distribution function, which decays
with an exponential (instead of a Gaussian) profile \footnote{Notice that an  exponential tail in the PDF  of 
curvature fluctuations,  similar to the one  arising in the context of a stochastic approach
to inflation, has been recently shown \cite{Pi:2022ysn} to be a property 
of all single field models  whose potential is up to quadratic. Such a feature
is physically due to the logarithmic relation between curvature fluctuation
and the (Gaussian) inflaton field fluctuations. See \cite{Pi:2022ysn} for details.}, and consequently tends to overproduce PBHs. 
 \cite{Pattison:2017mbe,Ezquiaga:2019ftu,Figueroa:2020jkf,Figueroa:2021zah} set constraints on the duration of the USR phase, which (depending on the
 scenarios) can last at most few e-folds before overproducing PBH. There is a growing
  activity 
 on these subjects, and 
 we refer the readers
 to the aforementioned literature  for details on the state of the art on  this important topic.

\subsection{Growth in the power spectrum when the decaying modes are slacking}\label{s3p3}

\noindent{\bf Slow-roll violation without triggering decaying modes} 

\medskip

\noindent
 We learned in the previous subsections
 that a possible way
for enhancing the  spectrum of fluctuations at
small scales, with respect to its large-scale
counterpart, is to amplify the $k$-dependent corrections
to the  constant-mode  solution ${\cal R}^{(0)}_k$ 
within the parenthesis of  Eq. \eqref{cpgs}.

But, as we anticipated in the paragraph following Eq. \eqref{cpgs}, we
can also design scenarios 
where an enhanced  time-dependence of the
  slow-roll parameters leads to a
 scale-dependent   curvature power spectrum at horizon crossing,
 even without exciting the decaying mode at super-horizon scales. 
 The idea is 
 to still make sure  that the pump field $z(\tau)$ increases with time -- hence conditions \eqref{zpoz} and \eqref{dmc} are {\it not} satisfied,   the decaying mode keeps inactive, and the terms within parenthesis of  Eq. \eqref{cpgs} can be neglected. However,   at the same time,  each individual slow-roll
 parameter changes considerably during a short time interval during inflation. The derivatives of slow-roll parameters
  can be large: they can contribute significantly to the quantity $\Theta$ controlling the Mukhanov-Sasaki equation, and they can influence the scale-dependence of the 
  the curvature spectrum at horizon crossing (see Eqs 
  \eqref{MS} and \eqref{theta}).
 
 We start this section by setting the stage, and derive formulas for 
 describing this possibility. We will  then we present an explicit realization of this scenario.  
 It is convenient to work with the canonical variable $v_k$ defined through Eq. \eqref{msre}, 
 and 
 solve the Mukhanov-Sasaki system in the form of the set of equations  \eqref{MS}-\eqref{bds}. 
We assume that the pump field $z$  is monotonic and always increasing with time,
 and we identify the sound horizon of fluctuations as $aH/c_s \simeq \sqrt{z''/(2 z)}$ \footnote{\label{footmultiple}Notice that if the background dynamics have  localized features, phases of slow-roll violation might occur for some time interval, and this exact identification  ceases to be valid. In particular, in this case $z''/z$ might loose its monotonic  nature for some time interval, leading to multiple horizon crossing for a range of modes, resulting in  oscillations of the spectrum \cite{Ballesteros:2018wlw}. For the time being, we set a discussion on this possibility aside,  and continue identifying the quantity $z'' / z$ as an effective horizon for slow-roll violating scenarios we are interested in.}.  
   We can identify two asymptotic regimes for each mode $k$: i) an early-time regime, when each mode is deep inside the horizon and ii) a late-time one, when the modes get stretched to become super-horizon. On the one hand, in the former regime the  modes  satisfy $k^2 \gg z''/z$, and behave as the standard vacuum fluctuations in Minskowski space-time 
\beq\label{bds}
v_k(\bar{\tau}) = \frac{e^{-i k \bar{\tau}}}{\sqrt{2k}}, \quad\quad\quad \textrm{for} \quad k^2 \gg \frac{z''}{z}\,.
\eeq
On the other hand, later during inflation the fluctuations get  stretched outside the horizon, entering the second regime, and eventually satisfying $k^2 \ll z''/z$, with a solution given by 
\beq\label{vsl}
v_k(\bar{\tau}) \simeq \mathcal{C}_{1,k}\, z + \mathcal{D}_{2,k}\, z \int \fr{\d \bar{\tau}'}{z^2(\bar{\tau}')} + \mathcal{O}(k^2),
\eeq
where the finite $k^2$ corrections to this solution can be derived in a similar fashion as in \eqref{cpgs}. Recall that  we are now   interested in attractor  background configurations, hence
we can neglect the last two terms in Eq. \eqref{vsl}
that rapidly decay. 
Shortly after horizon crossing, the canonical variable will settle into the solution $v_k = z\, \mathcal{C}_{1,k}$. Using the field redefinition \eqref{msre}  we   can identify the constant mode as the curvature perturbation at late times $\mathcal{C}_{1,k} = \mathcal{R}_k = \mathcal{R}^{(0)}_k$. In order to determine its expression, we  match the solutions  some time around horizon crossing $\bar{\tau} = \bar{\tau}_0$,
and we obtain 
\beq\label{R0sq}
|\mathcal{C}_{1,k}|^2 = |\mathcal{R}^{(0)}_k|^2 = \fr{1}{2k}\fr{1}{z(\bar{\tau}_0)^2} = \fr{1}{2k} \fr{c_s}{a^2 M^2 2 \epsilon} \bigg|_{\bar{\tau} = \bar{\tau}_0}. 
\eeq
The horizon-crossing time can be conveniently expressed  as leading contribution in a 
WKB  approximation \cite{Ballesteros:2018wlw}:
\beq\label{mt}
k^2 = \fr{z''}{2 z} \bigg|_{\bar{\tau}_0} = \left(\fr{aH}{c_s}\right)^2 (1 + \Theta) \bigg|_{\bar{\tau}_0}\,,
\eeq
with $\Theta$ given in Eq. \eqref{theta}. Collecting
there results, we can write the late-time  power spectrum for curvature fluctuations as 
\beq\label{psg}
\mathcal{P}_{\mathcal{R}}(k) \equiv \fr{k^3}{2\pi^2} \left|\fr{v_k(\bar{\tau})}{z}\right|^2 = \fr{k^3}{2\pi^2} \left|\mathcal{R}^{(0)}_k\right|^2 = \fr{H^2}{8\pi^2\, \epsilon\, c_s \,M^2}\left(1 + \Theta\right) \bigg|_{\bar{\tau}_0}.
\eeq
From \eqref{psg} we observe that rapid changes in the background quantities $\epsilon, c_s, M$ and slow-roll parameters constituting
the quantity $\Theta$ of Eq. \eqref{theta} as a function of $\bar \tau_0$ can then translate
into a scale-dependent amplification of the power spectrum. As we will see, this situation leads to a scale dependent enhancement in the power spectrum realized through the `constructive interference' of the time-dependent background parameters $\epsilon, c_s, M$. 

\medskip
\noindent{\bf An explicit realization}
\medskip

\noindent
We now review a possible realization of this scenario,
 following closely the discussion of   \cite{Ballesteros:2018wlw}.  
We  focus on the generalized single-field framework  discussed in Section \eqref{S3p1}, and we consider a background dynamics that includes  simultaneous pronounced  dips in the time dependent profiles for the parameters  $\epsilon, c_s, M^2$. We then study the power spectrum by solving numerically the  Mukhanov-Sasaki equation for curvature perturbations (see Appendix \ref{AppC}), and we compare the result with the analytical expressions  discussed in Section \ref{S3p1} (see \eg \eqref{psg}).

\begin{table}
\begin{center}
\begin{tabular}{| c | c | c | c | c |}
\hline
\hline
\cellcolor[gray]{0.9}$X$ & \cellcolor[gray]{0.9}$n_{\rm a}$ &\cellcolor[gray]{0.9} $n_*$ & \cellcolor[gray]{0.9}$N_*$ &\cellcolor[gray]{0.9} $\sigma$ \\
\hline
\cellcolor[gray]{0.9}$\epsilon$& $-2. $ & $-4.$ & $ 36.0 $ & $2.0$ \\\hline
\cellcolor[gray]{0.9}$c_s$ & $0.$ & $-2.$& $ 35.5$ & $2.5$\\\hline
\cellcolor[gray]{0.9}${\tilde{M}^2}$ & $0.0$ & $-3.0$ & $34.8$ & $ 4.0$\\\hline
\hline
\end{tabular}
\caption{\label{tab:par2} Parameter choices that characterize the background evolution of the time dependent parameters $\epsilon$, $c_s$, $\tilde{M} = M/\Mpl$.}
\end{center}			
\end{table}
\begin{figure}[t!]
\begin{center}
\includegraphics[scale=0.6]{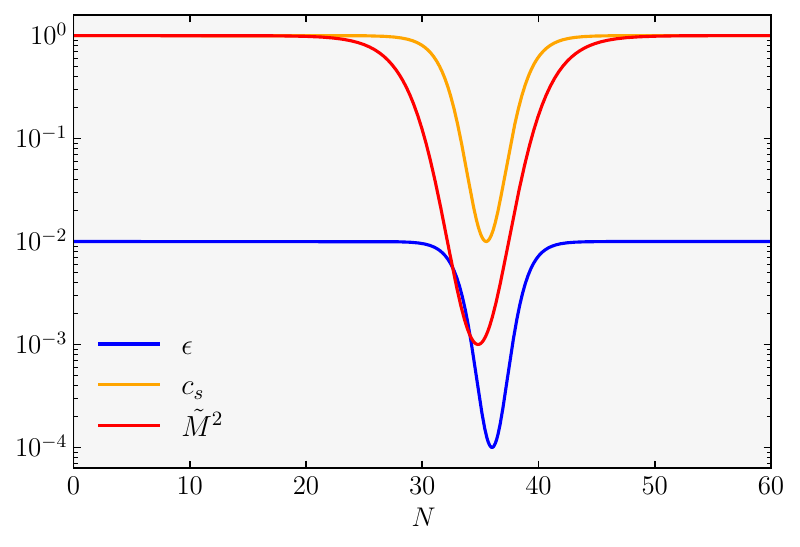}\,\,\includegraphics[scale=0.6]{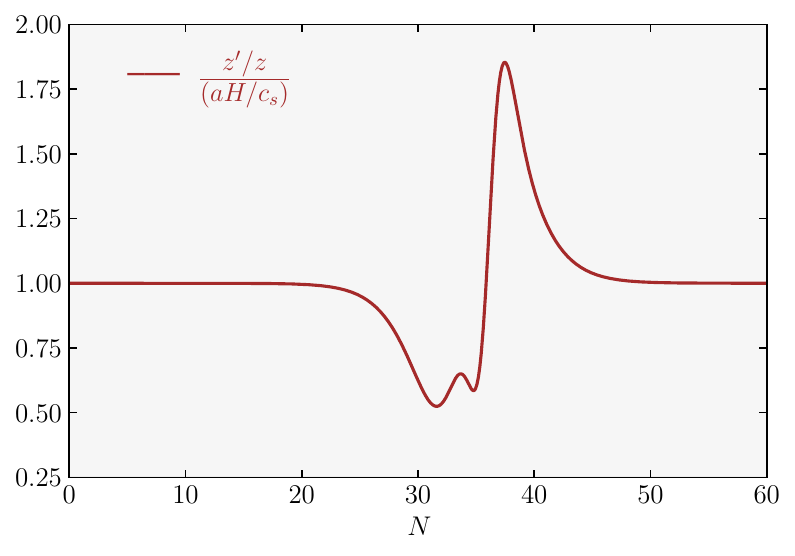}
\end{center}
\caption{Left panel: Evolution of $\epsilon,c_s, \tilde{M}^2 = M^2/\Mpl^2$ in e-folds characterized by the expression \eqref{X} with the parameter choices presented in Table \ref{tab:par2}. Right panel:  time evolution of \eqref{zpoz} in units of $aH/c_s$ to illustrate the fact that the decaying modes do not grow for the background presented in the left panel.\label{fig:gback}}
\end{figure}
To analyze a representative scenario in this category, we parameterize the three time dependent quantities $ X = \{\epsilon,c_s, M^2\}$ as \cite{Ballesteros:2018wlw}
\beq\label{X}
\log_{10} X(N) 
= \left(n_{\rm a}-n_{*}\right) \tanh ^{2}\left(\frac{N-N_{*}}{\sigma}\right)+n_{*}.
\eeq
\begin{figure}
\begin{center}
\includegraphics[scale=0.6]{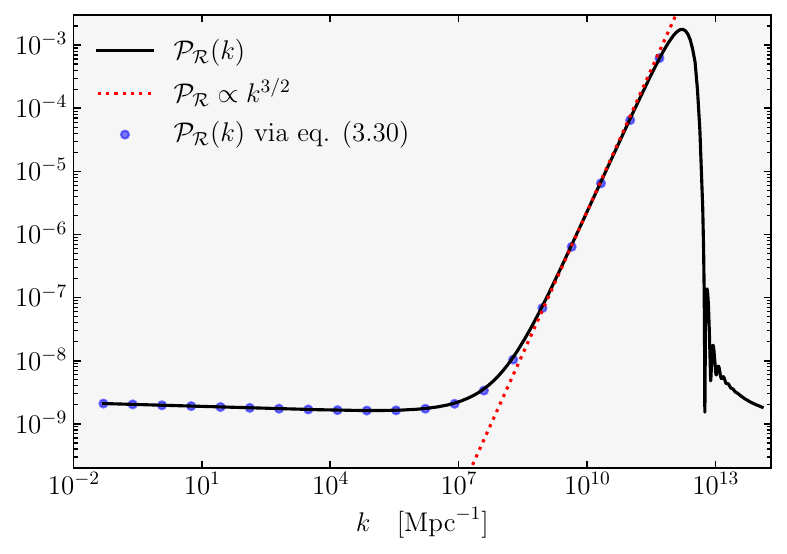}\,\,\includegraphics[scale=0.6]{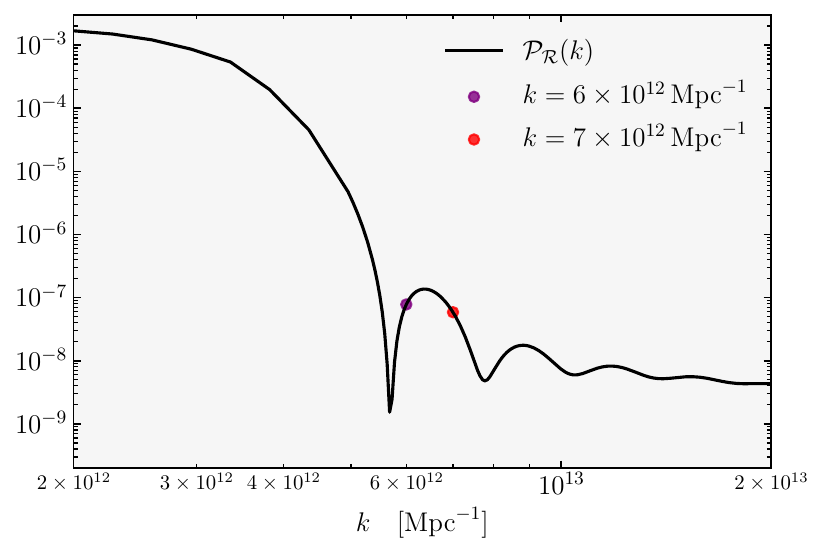}
\end{center}
\caption{The full power spectrum of curvature perturbation for the background model presented in Fig. \ref{fig:gback} (Left). The blue dots represents the accuracy of the formula \eqref{psg} in describing the rise of the $\mathcal{P}_\mathcal{R}$ towards its peak. As shown by the red-dotted line, the spectral index on the rise satisfies $n_s - 1\lesssim 3/2$. Power spectrum for scales following the peak to illustrate the oscillations in the amplitude at those scales (Right).\label{fig:gps}}
\end{figure}
Each of these quantities tend to $10^{n_{\rm a}}$ asymptotically away from $N_*$ in both directions, $|N-N_*| \gg \sigma$ and become equal to $10^{n_*}$ at $N_*$ while staying around the neighborhood of this value for a number of e-folds determined by $\sigma$. Notice that we are interested in features of the inflationary dynamics that affect modes at scales much smaller than the CMB pivot scale. Hence we can assume 
$N_* \gg 1$, where  modes associated with CMB scales leave the horizon at $N\simeq 0$, while inflation ends at $N_{\rm end} = 60$. A representative set of parameter choices that leads to a localized decrease in the slow-roll parameters is presented in Table \ref{tab:par2}. The resulting background evolution together with the behavior of $z'/z$ is shown in the left and right panels of Fig. \ref{fig:gback}. We observe from the right panel of the figure that $z'/z >0$ is always satisfied during inflation, suggesting (as expected) that the decaying modes do not grow in time.  Using the  parametrization \eqref{X} we then  numerically solve the  Mukhanov-Sasaki equation, following Appendix \ref{AppC}. 

\begin{figure}[t!]
\begin{center}
\includegraphics[scale=0.6]{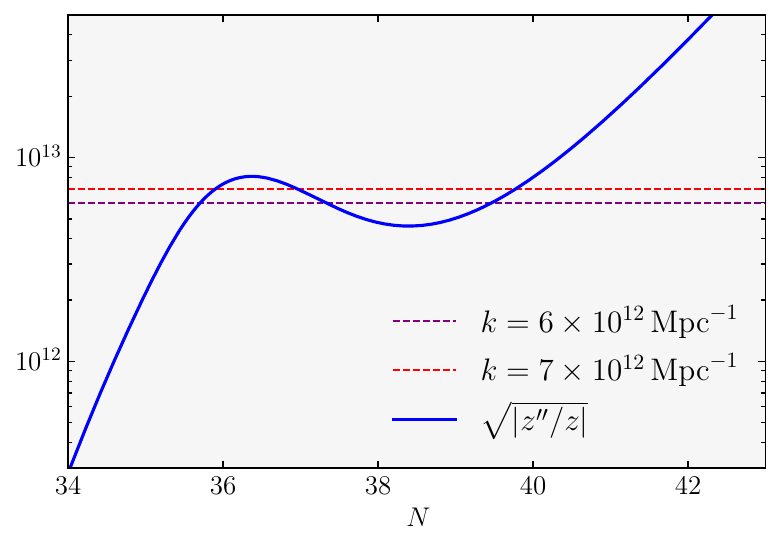}\includegraphics[scale=0.6]{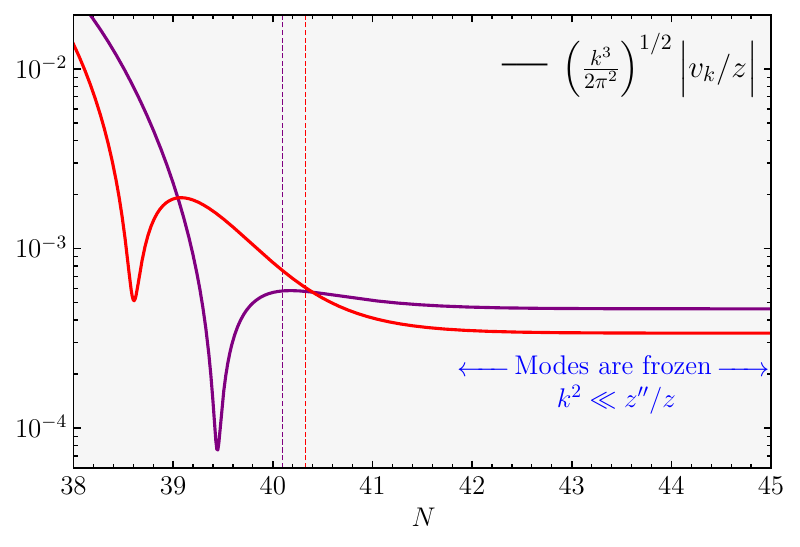}
\end{center}
\caption{Left: The occurrence of multiple horizon crossing for the neighboring modes labeled by red and purple dot in Fig. \ref{fig:gps}. Right:  Evolution of the modes that exhibit multiple horizon crossing (note the same color coding with the left panel). The vertical lines illustrates the final horizon crossing time for each mode.\label{fig:zpozvsk}}
\end{figure}

Our results for  $\mathcal{P}_{\mathcal{R}}(k,N_{\rm end})$ are presented in Fig. \ref{fig:gps}. In the left panel we notice  that the expression \eqref{psg} accurately describes the behavior of the power spectrum towards its peak, confirming our expectation that the constant growing mode is responsible for the enhancement. Notice the absence
of the dip proceeding the growth (which characterized instead
the scenario of Fig \ref{fig:psc}). This in
agreement with our  interpretation in the previous section:  the dip is due
to disruptive interference between `growing' and `decaying' modes -- while in this context the decaying mode is not active.
On the other hand, as shown in the right panel of the figure, the power spectrum exhibits oscillations for scales following the peak. As we  discussed in footnote \ref{footmultiple}, this is due to multiple horizon crossing of  modes within certain momentum scales, leading to excited states 
 and  oscillations in the spectrum. 
 We illustrate this phenomenon in Fig. \ref{fig:zpozvsk} ---for two neighboring modes labeled by a red and purple dots in Fig. \ref{fig:gps}--- where  we plot $k$ versus $\sqrt{|z''/z|}$ as a function of e-folds. As shown in the right panel of Fig. \ref{fig:zpozvsk}, although these modes are neighboring, they exhibit non-trivial behaviour before their final horizon exit $N \simeq 40$ so that their asymptotic values ($N \to N_{\rm end}$) differ considerably, giving rise to the sizeable modulations in the late time power spectrum spectrum (see Fig. \ref{fig:gps}), as discussed in \cite{Ballesteros:2018wlw}. 
    
   Besides  \cite{Ballesteros:2018wlw}, the ideas discussed in this subsection for 
   enhancing the spectrum at small scales found further realizations in \cite{Kamenshchik:2021kcw}. Conceptually similar frameworks include
   sound speed resonance scenario proposed in \cite{Cai:2018tuh} --which can be explicitly realized 
   within a DBI inflation model  
    \cite{Chen:2020uhe}-- and parametric resonance model discussed within the single field canonical inflation framework \cite{Cai:2019bmk}.

\subsection{Brief summary}
\label{sec_sinsum}

We find it remarkable that many distinct single-field models of inflation, built with the
aim of  producing PBH, 
share  common features in the properties
of the resulting power spectrum. 
 The reason being that  the enhancement of the curvature  spectrum at small scales  is due to few mechanisms common to several scenarios. We identified the idea of 
resurrecting the 
decaying mode of curvature fluctuations at super Hubble scales,
by increasing the absolute value
 of the slow-roll parameter $\eta$ -- see Section \ref{s3p2} --
 and the idea of having very rapid  
  changes in the values of slow-roll parameters (keeping them relatively
  small) -- see Section \ref{s3p3}.

 The resulting  properties of the curvature spectrum might result essential for testing  single-field models of inflation at small scales,  thanks to their  predictions for the  PBH population properties, as well as   for the spectrum of gravitational waves  induced at second order in perturbations \cite{Ananda:2006af,Baumann:2007zm}. In fact, the latter is a very interesting topic, relevant and well-studied for  single-field models: we refer the reader to \cite{Domenech:2021ztg} for a detailed review. 
 
 Nevertheless, the existing explicit scenarios of single-field inflation, which are able to generate an 
appreciable PBH abundance,
 typically suffer of severe fine-tunings on the choices of the parameters characterizing their Lagrangians. For example, in producing sufficiently 
  flat, plateau regions of their potential, and in ensuring regular transitions between attractor and non-attractor eras during the inflation  process. See e.g. the discussion in \cite{Hertzberg:2017dkh}
 (but also \cite{Ballesteros:2019hus} for a scenario that can partially ameliorate the tuning involved). Moreover, as we learned in 
 Section \ref{S2p2}, the resulting PBH population can be very sensitive
 to the details of the spectrum profile, and to the presence of primordial non-Gaussianities. The latter  should at certain extent be generated in many concrete single-field realizations (see Section \ref{s3p2}), and
 render particularly delicate the task of making precise theoretical predictions. In fact, large non-Gaussianities might not only change the predictions for PBH production, but can also impose  restrictions on single-field model building, due to large one-loop corrections \cite{Kristiano:2021urj,Meng:2022ixx,Kristiano:2022maq,Inomata:2022yte} (see however \cite{Riotto:2023hoz}).
 
 It is then interesting to try to consider PBH inflationary scenarios in  contexts where more than one field acquire dynamics during inflation. The hope being to find qualitatively new ideas for producing PBH, or alternatively  more natural realizations of known mechanisms, in order  to amplify the primordial curvature spectrum at small scales.  Possibly, 
  multiple fields affect the predictions on the statistics of curvature
  fluctuations with respect to single-field models, 
  with important implications for PBH formation.  We turn to  review this topic in what comes next. 

\section{Enhanced primordial power spectrum in multi-field models}\label{S3p2}

Despite the remarkable agreement of single-field, slow-roll inflation with the CMB and the LSS data \cite{Planck:2018jri}, the fundamental nature of inflation continues to elude us.  In particular, it would be very important
to know whether additional fields took part  in the  dynamics of inflation, besides the scalar driving cosmic
expansion.  In fact, while
current cosmological observations   do not provide
hints  of iso-curvature fluctuations associated with extra inflationary degrees of freedom, it might also be that additional fields play a role during epochs of inflation that are not well probed by current large-scale surveys.
 The formation of PBHs, occurring at small scales, can be sensitive to iso-curvature fluctuations, and represent a valuable probe of 
 inflationary multi-field dynamics.
  Multi-field inflation can offer 
  new possibilities for producing PBH, exploiting  novel, distinctive 
  ways for converting large gradients
  of background quantities into the spectrum
  of curvature fluctuations. Moreover, when used in tandem with the techniques
   discussed in the previous section, 
  multi-field models might alleviate  some of the fine-tuning issues occurring in  single-field  scenarios (see the discussion in Section \ref{sec_sinsum}). For these reasons,
 in this section we review a selected
 choice of theoretical frameworks aimed at enhancing the curvature spectrum at small scales \footnote{Some other notable models we won't cover in this context utilize i) dissipative dynamics as in warm inflation type scenarios discussed in \cite{Arya:2019wck,Bastero-Gil:2021fac,Ballesteros:2022hjk} ii) the conversion of resonantly produced spectator fluctuations to curvature perturbation during \cite{Zhou:2020kkf} or after inflation, \eg in curvaton type scenarios discussed in \cite{Pi:2021dft,Meng:2022low}.}, by using
 tools that specifically involve more than one field during inflation, in particular the tachyonic behaviour of field dynamics in some extra 
 sectors during inflation.

 First, in Section \ref{s4p1}, we review PBH scenarios based on axions
 interacting with gauge fields during inflation, exploiting a mechanism based on particle
 production during inflation. This possibility
 is  well motivated by particle physics 
 constructions, and it has been much explored in the literature: we make efforts for carefully reviewing
 the status of the art,  discussing  opportunities and challenges for PBH formation in this context. Then, in Section \ref{s4p2} we review multi-field inflationary  scenarios where the rise in the spectrum is  a result of large iso-curvature perturbations,  induced by
 a curved
 inflationary  trajectory  traversing    a rich  
 multi-field moduli space. 
 
\smallskip

\noindent\faIcon{book} {\bf\emph{Main References}}: 
In the discussion of Section \ref{s4p1} we  benefit from the works presented in  
 \cite{Linde:2012bt,Bugaev:2013fya,Garcia-Bellido:2016dkw,Domcke:2017fix,Ozsoy:2020ccy,Ozsoy:2020kat} while the material presented in Section \ref{s4p2} is based on \cite{Palma:2020ejf,Fumagalli:2020adf}. 
 
 \subsection{Enhanced scalar perturbations from axion-gauge field dynamics}\label{s4p1}

Axions \cite{Wilczek:1977pj, Weinberg:1977ma} are pseudo-scalar
particles,
theoretically  introduced  in  studies of  particle physics theories 
beyond the Standard Model. First motivated
in scenarios addressing the strong CP problem
in terms of a Peccei-Quinn mechanism \cite{Peccei:1977hh,Peccei:1977ur}, they 
find natural realizations in string theory (see e.g. \cite{Svrcek:2006yi}), as well as many useful applications
to cosmology -- see e.g. \cite{Marsh:2015xka} for a comprehensive
review. The possibility of using the physics of axions for producing
PBH is then very interesting, given their strong theoretical and experimental motivations
which allow  to make  connections between particle physics, quantum gravity, and cosmology. We do so in this section, reviewing several representative ideas and their realizations in this context.

Axion fields, denoted here as $\chi$, 
correspond to Goldstone bosons of the  Peccei-Quinn (PQ) global symmetry, spontaneously broken at a scale dubbed $f$. Their Goldstone nature equips
axions with   
a shift symmetry $\chi\,\to\,\chi+$constant,    valid
at all orders in perturbations. The shift symmetry  makes them interesting
inflaton 
candidates, being their potential  naturally very flat \cite{Freese:1990rb}.  
The PQ symmetry suffers from a chiral anomaly, which breaks the shift  symmetry and assign axions a potential $V(\chi)$ through non-perturbative contributions in field theory, or monodromy effects in string theory 
(see {\it e.g.} \cite{Baumann:2014nda} for a review).
The chiral anomaly implies that axions interact with gauge vectors through higher-dimensional operators. This property is essential for our arguments. The typical structure
of an axion Lagrangian considered for cosmological purposes reads
\begin{equation}
\label{LAG}
 \mathcal{L} = -\frac{1}{2}\partial_\mu \chi\, \partial^{\mu} \chi - V(\chi) -\frac{1}{4}F_{\mu\nu}\,F^{\mu\nu} - \frac{g_{\rm cs}}{4 f}\,\chi\, F_{\mu\nu}\,\tilde{F}^{\mu\nu}\,.
 \eeq
 The axion $\chi$ is equipped by a kinetic 
 term and a potential term $V(\chi)$.
 The potential is often   a polynomial function of the axion field, at least for small values of $\chi$ \footnote{See \eg \cite{McAllister:2008hb,Silverstein:2008sg,McAllister:2014mpa,Flauger:2014ana} for string theory motivated constructions in the context of  inflation.}. As anticipated, the Lagrangian \eqref{LAG} also includes
 a $U(1)$ gauge field $A_\mu$ with field
 strength $F_{\mu\nu}$,  coupling
 to the axion through a dimension-5, gauge-preserving
 operator 
 \begin{equation}
 \label{linta}
 {\cal L}_{\rm int}\,=\,
 - \frac{g_{\rm cs}}{4 f}\,
 \chi\, F_{\mu\nu} \, \tilde{F}^{\mu\nu}\,,
 \end{equation}
 where $f$ is the PQ symmetry-breaking scale, and
 $g_{\rm cs}$ is a dimensionless coupling constant.
 The dual field strength appearing in Eq. \eqref{linta} is  
$\tilde{F}^{\mu\nu} \equiv \eta^{\mu\nu\rho\sigma} F_{\rho\sigma} / (2\sqrt{-g})$, and the alternating symbol $\eta$ is $1$ for even permutations of its indices, and $-1$ otherwise (starting with $\eta^{0123}\,=\,1$). 
 The  dimension-5 operator  \eqref{linta} is especially
 important for our  arguments. As we  review below, this axion-vector coupling leads 
 to a tachyonic instability in the gauge sector, which exponentially enhances one of the helicities 
 of the gauge vector modes, and produces a large amount of gauge quanta \cite{Anber:2006xt}. 
 The energy adsorbed in the production of vector modes is carried away from the 
 axion kinetic energy, that slows down its evolution: this is good news
 for axion inflationary models, since the axion can slowly roll even along steep
 potentials
 \cite{Anber:2009ua}. Moreover, the inverse decay of enhanced gauge quanta
 into axion fluctuations (schematically, $ A +  A \to  \chi$),  can amplify curvature  perturbations \cite{Barnaby:2010vf,Barnaby:2011vw}, and lead
 to large curvature spectrum at small scales, at a level
 able to produce PBH \cite{Linde:2012bt}. We refer the reader
 to 
\cite{Pajer:2013fsa} for a  review on axion inflation  written one  decade ago, which  includes  many of the topics mentioned above. We now focus on reviewing
the consequences of these ideas for scenarios amplifying curvature fluctuations, also covering   opportunities and challenges pushed
forward in the most
recent literature on this subject.   
 
 \medskip
 
 \noindent{\bf Amplification of gauge field fluctuations}

  \medskip

\noindent
 When considering a dynamical axion field with a time dependent background profile $\bar{\chi}(t)$, the dimension-5 axion-vector interaction of Eq. \eqref{linta} 
  leads  to a copious production of vector fluctuations. 
   This property can be understood by considering  the equation of motion (EoM) for the gauge field mode functions (See Appendix \ref{AppD}) \cite{Anber:2009ua} 
\beq\label{meqgf}
\partial_x^2 A_{\pm}+\left(1 \pm \frac{2 \xi}{x}\right) A_{\pm}=0, \quad\quad\quad\quad \xi \equiv-\frac{g_{\rm cs}\, \dot{\bar{\chi}}}{2 H f}
\eeq
where $A_{\pm}$
correspond to the gauge vector polarizations, and we defined a dimensionless time variable $x \equiv-k \tau = k/(aH)$, as well as  the effective dimensionless coupling $\xi$ between the spectator axion and the gauge field. Without  loss of generality, we work within the conditions
\begin{equation}\label{negvel}
\xi\,>\,0 \hskip0.5cm {\rm and} \hskip0.5cm \dot{\bar{\chi}}<0
\,.
\end{equation}
Hence, the  axion rolls along  its potential from  large positive to small values of $\bar{\chi} \geq 0$. 

Notice from \eqref{meqgf} that the dimension five operator of Eq. \eqref{linta} introduces a time dependent mass term in the dispersion relation of ${\rm U (1)}$ field, which  changes sign depending on the helicity $\lambda = \pm$. This property reflects the parity violating nature of the dimension-5 interaction.  When the gauge modes are deep inside the horizon, $(x=k /(a H) \gg 1)$, the time-dependent correction term is negligible,  and both vector polarizations obey a standard dispersion relation as in Minkowski space. However, as the modes stretch outside the horizon, the correction becomes dominant for $x=k /(a H) \lesssim 2 \xi$, leading to an instability for one of the circular polarizations of the gauge fields. In our conventions \eqref{negvel}, only $A_{-}$ state experiences a tachyonic instability, while $A_{+}$ is unaffected. The dynamics of the axion like field, controlled by the axion  velocity $\dot{\bar{\chi}} \neq 0$, therefore induces a significant production of helical vector fields. 

\smallskip

The nature of gauge field production, and its consequences  as a source for  scalar perturbations,  is sensitive to the scalar potential $V(\chi)$,  since this quantity determines the  profile of the axion background velocity $\dot{\bar{\chi}}$.  The dynamics of the curvature perturbations, as generated through the axion-gauge field dynamics,  depends on whether we identify the axion  field as the inflaton, or whether it  belongs to some extra  spectator sector during inflation. In what comes next, we  arrange our discussion so to clearly distinguish among these possibilities. We consider 
the following scenarios:
\begin{itemize}
    \item[{\bf 1.}] {\bf Section \ref{s4p1p1}} As first possibility, we identify the axion $\chi$ with the inflaton  $\phi$
   that drives inflation: $\chi\,=\,\phi$. Then, the order parameter $\dot{\bar{\phi}}$  controlling the gauge-field production increases with time, generating 
   scalar perturbations through an inverse vector decay: 
    $\delta A + \delta A \to \delta \phi$. 
     We  refer to this scenario as ``Smooth Axion Inflation'', and models that can be considered in this category are studied in \cite{Linde:2012bt,Bugaev:2013fya,Garcia-Bellido:2016dkw,Domcke:2017fix}. Such scenarios 
      suffer of dynamical instabilities associated with large back-reaction effects from the gauge
     fields on the axion evolution.
    \item[{\bf 2.}] {\bf Section \ref{s4p1p2}} In certain axion-inflation models, the axion potential  has special features,  located far away from the field range 
     corresponding to scales affecting CMB physics. 
    A sudden increase in the axion inflaton velocity occurs at their location, with  enhanced scalar perturbations amplified at small PBH scales through inverse vector decay. 
     This possibility is first discussed  in \cite{Garcia-Bellido:2016dkw}, while further  developments
      are studied in \cite{Cheng:2018yyr,Ozsoy:2020kat}. We  refer to  scenarios producing PBH by exploiting  localized particle production as ``Bumpy Axion Inflation'', and we analyze
      how they address the instabilities mentioned above. 
     \item[{\bf 3.}] {\bf Section \ref{s4p1p3}} A final possibility  corresponds 
     to
      gauge field production  in a hidden sector, through the
     dynamics of an  axion spectator field.
      Also this case allows one to address the aforementioned
      dynamical instabilites, given that the  back-reaction effects from the vector sector can be placed under control. 
      Depending on the shape of the spectator axion  potential, such a scenario 
      can lead to localized peaks in the scalar curvature power spectrum \cite{Garcia-Bellido:2016dkw,Ozsoy:2020ccy}. We  refer to this scenario as ``Spectator axion-gauge field dynamics''.
\end{itemize}
We build our discussion mainly in terms of representative, concrete examples, presenting the main results and relegating technical details to Appendixes. 

\subsubsection{Smooth Axion Inflation}\label{s4p1p1}

In the first scenario we consider, we identify the axion $\chi$
with the scalar inflaton that drives inflation, dubbed $\phi$: $\chi\to\phi$. We  
study the behaviour of scalar perturbations in a set-up
described by Lagrangian \eqref{LAG}. We assume that the profile for the scalar potential $V(\phi)$ 
 is  sufficiently flat, so to support inflation. In this case,  the effective coupling $\xi$, as defined in Eq. \eqref{meqgf},  adiabatically increases during the inflationary process: 
 \begin{equation}
 \xi \equiv g_{\rm cs} \sqrt{\frac{\epsilon}{2}} \frac{\Mpl}{f}\,,
 \end{equation}
 where $\epsilon$ is the standard slow-roll parameter. The amplified gauge-field mode function can be analytically expressed as \cite{Anber:2009ua}
 (recall that the axion speed has negative sign, see Eq. \eqref{negvel})
\beq\label{meqsol}
A_{-}(\tau, k) \simeq \frac{1}{\sqrt{2 k}}\left(\frac{-k \tau}{2 \xi}\right)^{1 / 4} \exp (\pi \xi-2 \sqrt{-2 \xi k \tau}), \quad\quad\quad \xi \equiv -\frac{g_{\rm cs}\, \dot{\bar{\phi}}}{2Hf}\,.
\eeq
The slowly changing time-dependent parameters $\xi$ and $H$
in Eq. \eqref{meqsol} are evaluated at the epoch of 
 horizon crossing. The amplification of gauge field modes is maximized when the size of the mode is comparable to the horizon, $-k\tau \sim \mathcal{O}(1)$.
 
  In fact, the analytic expression in \eqref{meqsol} is valid within the interval $(8\xi)^{-1} \ll -k\tau \ll 2\xi$. For the values $\xi \sim \mathcal{O}(1)$ we will be interested on, this range  corresponds to a phase during which  the gauge  modes grow, and then remain frozen to their maximal value before being diluted away by the universe expansion. We proceed
  discussing the time evolution of the relevant quantities in this set-up. For concreteness, we focus on a representative example. 
   For a more detailed account on the gauge field amplification by the slowly rolling scalar  we refer the reader to Section 2.1 of \cite{Peloso:2016gqs}, or Appendix B of \cite{Ozsoy:2017blg}. 

\medskip
\noindent{\bf Background evolution in a concrete example} 
\medskip

\noindent
As the effective coupling $\xi$ of Eq. \eqref{meqsol} increases during inflation, the enhanced vector modes  eventually lead to a sizeable back-reaction  on the background evolution of the axionic inflaton field. This fact can
 significantly affect the homogeneous dynamics of the system.

The back-reaction is mainly controlled by  the vector-dependent friction term in the equation of motion for the homogeneous inflaton field \cite{Anber:2009ua}. 
 This phenomenon implies  that  the gauge field amplification occurs at the expense of  the inflaton velocity $\dot{\bar{\phi}}$. 
We consider a situation characterized by a  
transition between a standard slow-roll dynamics in the early stages
of inflation, and  a new attractor regime 
at late times, when the gauge field enhancement dominates over the Hubble friction \cite{Barnaby:2011qe} (but see the cautionary remark towards the end of this section).

The aforementioned  friction effect --  induced by the gauge mode production --  can be analyzed considering  modified Klein-Gordon and Friedmann equations:
\begin{align}\label{mkgafe}
\nn \ddot{\bar{\phi}} + 3 H \dot{\bar{\phi}}&+V'(\bar{\phi})=\frac{g_{\rm cs}}{f}\,\langle\vec{E} \cdot \vec{B}\rangle \,\,,\\
3 H^2\Mpl^2&= \frac{1}{2} \dot{\bar{\phi}}^2+V(\bar{\phi})+\frac{1}{2}\left\langle\vec{E}^2+\vec{B}^2\right\rangle,
\end{align}
where we introduced  `electric' and `magnetic' field contributions\footnote{Be aware that, despite the terminology we adopt, the gauge fields $A_\mu$ do not need to correspond to Standard Model photons.},  as discussed 
in detail in Appendix \ref{AppD} (see in particular Eq. 
\eqref{EandB}). 
Using the solutions  \eqref{meqsol} for gauge-field modes, the expectation values for the 
 electric and magnetic fields can be computed analytically  \cite{Anber:2009ua,Ozsoy:2017blg}, finding (see Appendix \ref{AppD})
\beq\label{ebarhoA}
\langle \vec{E}\cdot\vec{B}\rangle \simeq 2.1 \times 10^{-4}\, \frac{H^4}{\xi^4}\, e^{2\pi\xi}, \quad\quad \rho_A \equiv \frac{1}{2}\langle \vec{E}^2 + \vec{B}^2\rangle = 1.4 \times 10^{-4}\, \frac{H^4}{\xi^3}\, e^{2\pi\xi}\,.
\eeq
The quantity $\rho_A$ corresponds the total energy density contained in the gauge field fluctuations. Using the evolution equations \eqref{mkgafe}, we
can identify the conditions corresponding to a small  back-reaction of the vector modes into the background evolution. They are $3H|\dot{\bar{\phi}}|\gg g_{\rm cs} \langle \vec{E}\cdot \vec{B}\rangle /f$ and $\rho_A \ll 3H^2\Mpl^2$, which   can be expressed as 
\begin{align}\label{br1}
\nn \xi^{-3/2} e^{\pi \xi} &\ll 79 \frac{\dot{\bar{\phi}}}{H^2}, \quad\,\,\; \longrightarrow \quad \textrm{negligible back-reaction on $\phi$ equation\,,}\\
\xi^{-3/2} e^{\pi \xi} &\ll 146 \frac{\Mpl}{H}, \quad \longrightarrow \quad \textrm{negligible back-reaction on the Friedmann equation\,.}
\end{align}
Assuming that  inflation starts  in a standard slow-roll regime, the relation $\dot{\phi} = \sqrt{2\epsilon} H \Mpl \ll H \Mpl$ implies that the first condition in \eqref{br1} is more demanding than the second. When these conditions are not met, we enter into an inflationary phase  characterized by a strong back-reaction of gauge modes, which goes beyond the standard slow-roll inflationary attractor.

 To concretely illustrate the back-reaction of the gauge field production on the homogeneous background evolution of the inflaton, we focus on a modified version of the linear monodromy type potential \cite{McAllister:2008hb} that interpolates between $V \propto \phi$ and $V \propto \phi^2$ from large to small field values around the global minimum:
\beq\label{potam}
V(\phi) = \lambda\Mpl^4 \left[\sqrt{1+\frac{\phi^2}{\Mpl^2}}-1\right],
\eeq
where $\lambda$ is a dimensionless parameter that fixes the overall amplitude of the potential. Assuming negligible back-reaction in the beginning of inflation, $\lambda$ can determined by enforcing the standard normalization of power spectrum in the slow-roll regime for a given choice of $\phi_{\rm in}$ using
\beq
\mathcal{P}_{\mathcal{R}}(k_{\rm cmb}) = \frac{H^2}{8\pi^2 \epsilon \Mpl^2} \simeq \frac{V(\phi_{\rm in})}{24\pi^2 \epsilon_V(\phi_{\rm in})\Mpl^4} \equiv 2.1 \times 10^{-9}.
\eeq
We then solve numerically the equations \eqref{mkgafe} utilizing \eqref{ebarhoA} by setting the coupling between the inflaton and the gauge fields to its maximally allowed value~\footnote{In fact, 
at large CMB scales, described
by modes leaving the horizon in the early stages of inflation,
the value of the effective coupling $\xi$ is  restricted by existing information and  constraints on the scalar power spectrum and non-Gaussianity \cite{Meerburg:2012id,Barnaby:2010vf,Barnaby:2011vw}. Depending on  priors and on the shape of the inflaton potential, these constraints give $\xi_{\rm cmb} < 2.2 - 2.5$ \cite{Ade:2015lrj}.} by Planck: $g_{\rm cs} \Mpl / f = 48$, see \eg \cite{Ade:2015lrj}. Note that for a given initial value of the scalar field, this number can be translated to an initial value of the effective coupling which we label as $\xi_{\rm cmb}$:
\begin{figure}
\begin{center}
\includegraphics[scale=0.6]{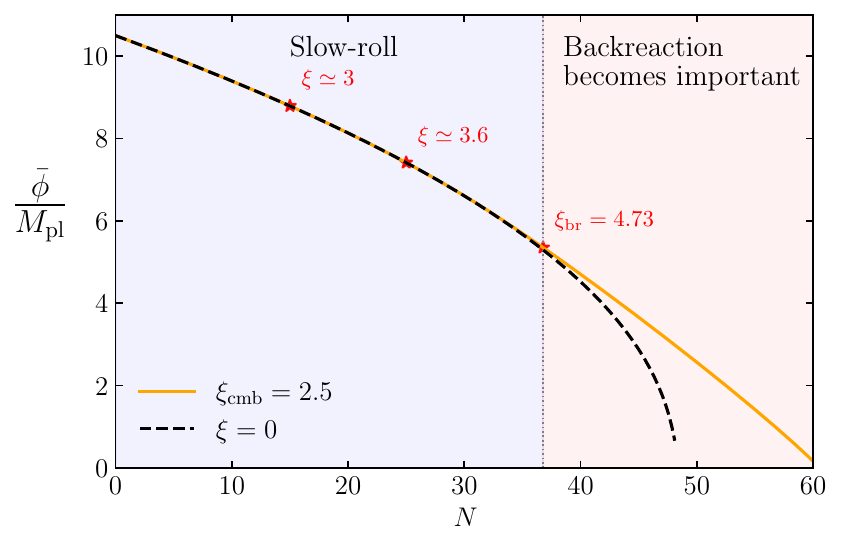}\includegraphics[scale=0.6]{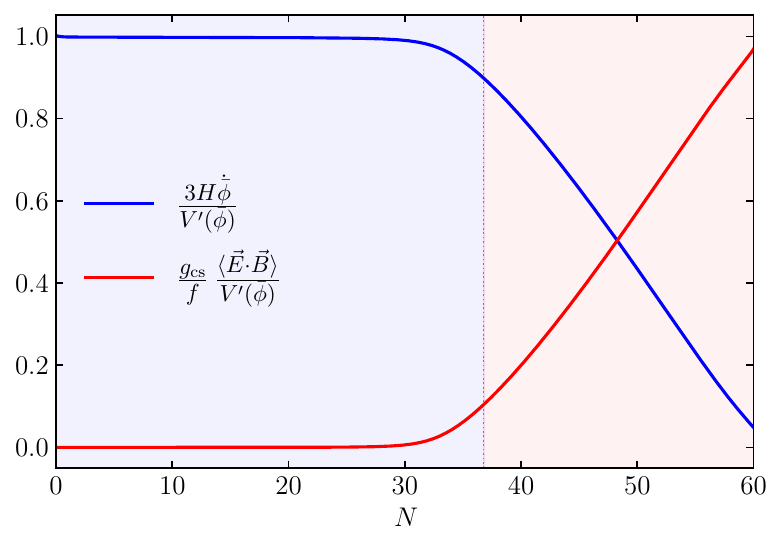}
\end{center}
\caption{Field profile $\bar{\phi}(N)$ with respect to e-folding number $N$ in smooth axion inflation for the initial choice $\xi_{\rm cmb} = \xi(N=0) = 2.5$ (Left). At early times, the dynamics is in the standard slow-roll regime (light blue region) but as $\xi$ increases, it enters into a stage dominated by the friction generated due to gauge field production (light red region) which has the effect of prolonging inflation with respect to the background evolution with $\xi = 0$. For the choice of $\xi_{\rm cmb} = 2.5$, the behavior of the Hubble damping term $3 H \dot{\bar{\phi}}$ and the friction induced by the source term $\propto \langle \vec{E}\cdot \vec{B}\rangle$ are shown in the right panel. \label{fig:sai}}
\end{figure}
\beq
\xi_{\rm cmb} \simeq  \frac{g_{\rm cs}\,\Mpl}{f} \frac{1}{2\phi_{\rm in}}.
\eeq
For $\phi_{\rm in} = 10.5\, \Mpl$ ($\xi_{\rm cmb} = 2.5$), the results of the numerical computation are shown in Fig. \ref{fig:sai}. In the left panel we show the inflaton profile as a function of e-folds until the end of inflation where $\dot{H} = -H^2$. We can clearly observe that in the absence of coupling to the gauge fields, $g_{\rm cs} = 0$ (black dashed line with $\xi = 0$), inflation lasts for about $\approx 48$ e-folds. Turning on the interactions with gauge fields at the beginning of inflation (solid orange line with $\xi_{\rm cmb} = 2.5$), the back-reaction induced by the vector fluctuations become noticeable around $\approx 36$ e-folds, which in turn has the overall effect of extending the duration of inflation for about $\approx 12$ e-folds compared to the standard slow-roll case ($\xi = 0$). 

These findings are also supported by the right panel of Fig. \ref{fig:sai}, where we compare the standard Hubble damping term $3H\dot{\phi}$ with the friction term $g_{\rm cs} \langle \vec{E}\cdot\vec{B} \rangle / f$, induced by the gauge field production as a function of e-folds until the end of inflation. We observe that the standard Hubble friction dominates the dynamics at early times, but as the effective coupling $\xi$ increases the dynamics gradually evolves into a regime dominated by the back-reaction of the produced gauge quanta. For the parameter choices we adopt, the onset of this regime starts around $N_{\rm br} \simeq 36$ e-folds, and at around $N \simeq \mathcal{O}(50)$ the system enters into a strong back-reaction regime where the dynamics becomes completely controlled  by the  gauge modes \cite{Anber:2009ua}. 

Therefore, in any viable model of axion-inflation, back-reaction effects induced by  gauge field production  become important only towards the latest stages of inflation. For example, for a rough estimate of the orders of magnitude involved we can adopt a linear potential $V \propto \phi$. Assuming the dynamics is initially in the slow-roll regime, we have $\xi \simeq g_{\rm cs} \sqrt{2 \epsilon_V} M_{\rm pl}/f \propto 1/\phi$ and $\phi_{\rm cmb}/\phi_{\rm br} = \xi_{\rm br}/\xi_{\rm cmb} \simeq 2.2 - 1.9$. For typical initial field values that can sustain enough inflationary evolution, we have $\phi_{\rm cmb}\sim \mathcal{O}(10) \Mpl$, which implies that $\phi_{\rm br} \sim \mathcal{O}(5) \Mpl$. Hence back-reaction   becomes dominant in the latest stages, but before the end of inflation corresponding to  $\phi_{\rm end}\sim \mathcal{O}(1)$.
  \begin{figure}
\begin{center}
\includegraphics[scale=0.75]{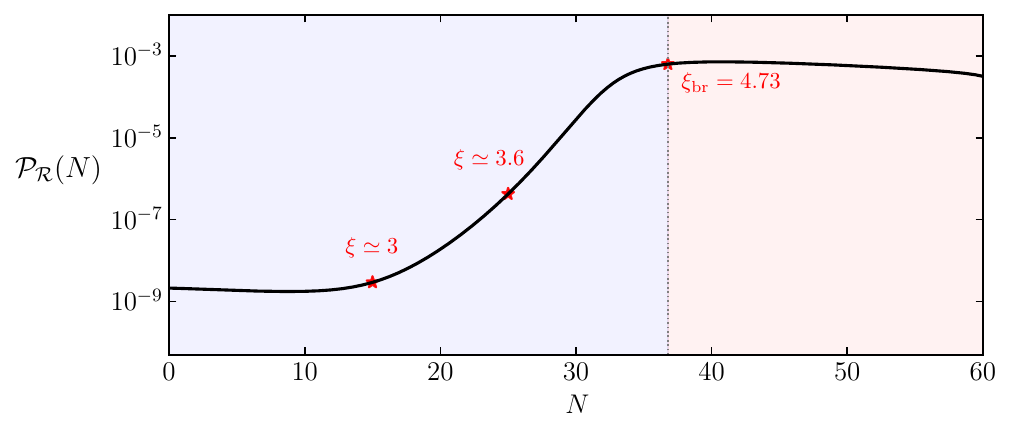}
\end{center}
\caption{Total power spectrum \eqref{SAIPS} as a function of e-folds in smooth axion inflation for the potential in \eqref{potam} and $\xi(N=0) \equiv \xi_{\rm cmb} = 2.5$ as we discuss in the main text.\label{fig:pssai}}
\end{figure}

\medskip
\noindent{\bf Scalar fluctuations sourced by gauge fields} 
\medskip

\noindent
As inflation progresses, the particle production becomes more and more efficient. Hence,  the vector modes start to act as an additional source of inflaton fluctuations $\delta \phi$,  due to the presence of tri-linear coupling $\delta \mathcal{L}_{\rm int}\propto \delta\phi F\tilde{F}$ (See Appendix \ref{AppD2}, and references therein). In particular, a tachyonic amplification of the vector fields leads --  at  second order  in perturbations -- to an enhancement in the scalar sector. Since these contributions to the scalar fluctuations are statistically independent from their vacuum counterpart, the resulting  curvature power spectrum   acquires two  separate contributions (see Appendix \ref{AppE}) that can be parameterized as \cite{Anber:2009ua,Linde:2012bt} 
\begin{align}\label{SAIPS}
\nn \mathcal{P}_{\mathcal{R}} &= \mathcal{P}^{(\rm v)}_{\mathcal{R}} + \mathcal{P}^{(\rm s)}_{\mathcal{R}},\\ &\simeq \frac{H^2}{8\pi^2 \epsilon \Mpl^2} + \left(\frac{g_{\rm cs}}{f}\frac{\langle\vec{E} \cdot \vec{B}\rangle}{3 H \beta \dot{\bar{\phi}}}\right)^2 \mathcal{F}^2\,.
\end{align}
 The first term in the second line denotes the standard vacuum contribution to the curvature power spectrum. The second  accounts for the vector part, with the time dependent quantities $\beta$ and $\mathcal{F}$  defined  in Eqs. \eqref{betaff} and \eqref{RAGF}.
 In the weak 
  back-reaction regime, the power spectrum of the curvature perturbation is dominated by the  vacuum contribution in the first term of Eq. \eqref{SAIPS}. As inflation progresses, the effective coupling between the inflaton and the vector fields increases:  the source contribution to the power spectrum starts to kick in, and the power spectrum grows as $\mathcal{P}_{\mathcal{R}}\propto \langle \vec{E}\cdot \vec{B}\rangle^2 \sim {\mathrm e}^{4\pi \xi}$ (the system is still in the weak back-reaction regime where $\beta \simeq 1$). Then the scalar fluctuations grow, they start influencing the gauge fields sources, and lead to an increase in the damping factor $\beta \sim \langle \vec{E}\cdot \vec{B}\rangle / 3 H \dot{\bar{\phi}}$. Therefore, eventually the sourced contribution to the power spectrum saturates towards late times during inflation.  
   In fact,  the time-dependent factor $\mathcal{F}$ is introduced to capture the modified definition of the curvature perturbation in the strong back-reaction regime. As discussed in detail in \cite{Garcia-Bellido:2016dkw}, this factor is responsible for  an order one correction to the power spectrum amplitude towards the end of inflation. 

  We summarize these arguments with a plot, Fig. \ref{fig:pssai}, of the total power spectrum with respect to the e-folding number. We notice that 
  as the effective coupling $\xi$ between the scalar and the gauge fields increases along the inflationary trajectory, the power spectrum acquires  a large  contribution from the vector source. As desired, it grows towards smaller scales,  eventually saturating towards a constant value. 
   Since the source of the peak in the power spectrum originates from a tri-linear coupling between the inflaton and the gauge fields, the statistics of the curvature perturbation at those scales is non-Gaussian (see also \cite{Caravano:2022epk}). The values of the curvature
  spectrum amplitude at the peak appearing in Fig. \ref{fig:pssai} is sufficiently large  to generate a sizeable population of the PBHs  (recall the discussion in Section \ref{S2p3}).

\medskip
\noindent{\bf Cautionary remarks: an instability in the inflaton dynamics}
\medskip

\noindent
We  warn the reader that the findings  reviewed in this section have been  recently questioned. In particular, \cite{Cheng:2015oqa,DallAgata:2019yrr,Domcke:2020zez,Gorbar:2021rlt,Caravano:2022epk} study the axion-gauge field dynamics for different choices of potentials and parameters, focusing  in the strong back-reaction regime. By implementing sophisticated numerical techniques, these works go  beyond the constant-velocity approximation assumed in our discussion above. Their  findings  indicate that once entering  the strong back-reaction regime,  the inflaton velocity is characterized by oscillations of increasing amplitude, in sharp contrast with  analytical studies approximating $\dot{\bar{\phi}}$ (and $\xi$) as constant. Finally, the recent analytic study of  \cite{Peloso:2022ovc} fully support the numerical results. 
The physical source of this
instability seems to be due to a delayed response of the vector
source  to changes in the axion velocity $\dot{\bar \phi}$, whenever
the system enters in a 
 strong back-reaction  regime. This phenomenon causes
the aforementioned  oscillations with increasing amplitude
in the inflaton velocity.

\smallskip
It is not clear at the moment whether these  secular effects can be tamed by adding ingredients to the set-up discussed in this section.
 A potential way out is to modify 
  the inflationary dynamics  in such a 
  way to push the strong vector back-reaction regime
   towards an epoch that does not affect the predictions on PBH formation in the scalar sector. 
   This possibility can be achieved in scenarios 
   where the gauge field production is localized,
  leading to a 
pronounced   peak in the scalar power spectrum before
entering into a strong back-reaction regime (see Section \ref{s4p1p2}).
   Such a way out does not avoid the instability, but at least moves it towards epochs that do not affect our predictions for the production and properties of curvature perturbations.
  In Section
  \ref{s4p1p3} we instead discuss the possibility to avoid at all a
  strong vector back-reaction regime, in scenarios where the
  axion is a spectator field and does not drive inflation. Appropriate
  choices of the axion potential nevertheless lead to an enhancement of the scalar sector, at a level sufficient for producing PBH. All the opportunities we review next are possible in scenarios of inflation including axion-vector couplings, which have very rich ramifications,  and are well studied 
  given their particle physics motivations. 

\subsubsection{Bumpy axion inflation}\label{s4p1p2}

In this section, motivated by the instability arguments discussed above, 
we consider scenarios with localized production of gauge fields, able to  enhance the curvature spectrum and produce PBH before entering any  strong
back-reaction regime. As we will see, the mechanism is based on the ideas
of the previous Section \ref{s4p1p1}, but also uses tools for the activation
of the decaying mode that we introduced in Section \ref{s3p2}. 

For definiteness, we focus on a representative example, discussing
its properties in the main text and referring to Appendixes
for technical details. We assume that 
   the shift symmetry of the axion is broken both by non-perturbative instanton corrections, as well as by a non-periodic monomial term in 
 the potential~\footnote{In fact, the potential \eqref{VB} is not unique for the purpose of generating a localized particle production scenario. In this context, any potential that exhibits step-like feature(s) is sufficient, see \eg \cite{Cheng:2016qzb,Cheng:2018yyr}. The choice of the potential in \eqref{VB} does however provide some useful analytic control over the background evolution of the axion and hence for the resulting amplification of the power spectrum as discussed in \cite{Ozsoy:2020kat}.} 
 (see \eg \cite{McAllister:2008hb,McAllister:2014mpa,Flauger:2014ana,Kallosh:2014vja,Kobayashi:2015aaa,BLZ,Parameswaran:2016qqq,Bhattacharya:2022fze}), 
\begin{figure}
\begin{center}
\includegraphics[scale=0.7]{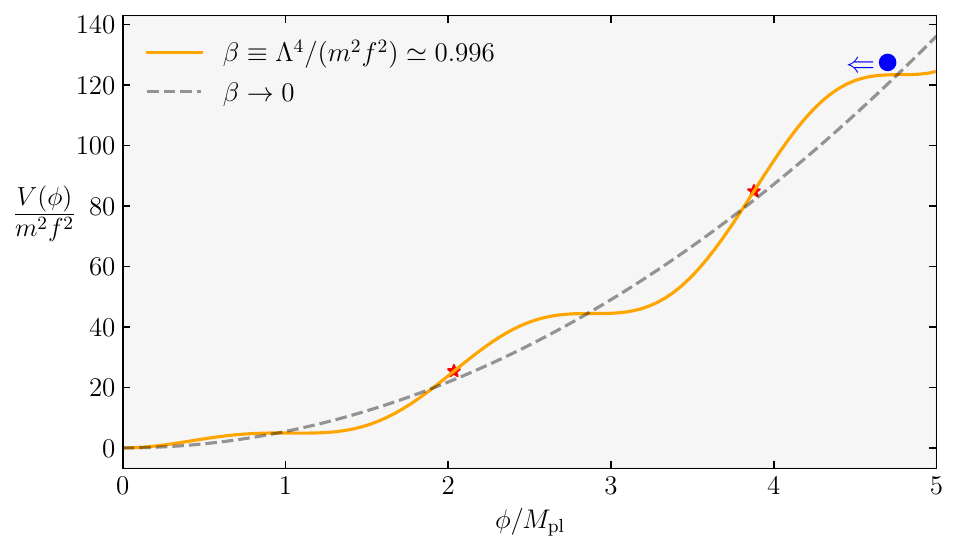}
\end{center}
\caption{The shape of the potential \eqref{VB} in the bumpy regime (orange curve) for $M_{\rm pl}/f = 3.3$. The red stars on the potential denotes the locations in field space where axion velocity and the slope of the potential is maximal. \label{fig:Vb}}
\end{figure}
\beq\label{VB}
V(\phi)=\frac{1}{2} m^2 \phi^2+\Lambda^4 \frac{\phi}{f} \sin \left(\frac{\phi}{f}\right),
\eeq
where $m,\Lambda, f$ are parameters of  unit mass dimension. In fact, potentials as the above, characterized by modulations  on top of a smooth profile, 
are well motivated and studied in theoretical
constructions of axion inflation models in a variety of situations.  
For these reasons, we find interesting to review their useful applications  for our aim of producing PBH.

We are interested in obtaining noticeable  modulations of the functional form of the potential, within the regime $\Lambda^4 \lesssim m^2 f^2$ (which we refer as ``bumpy" regime in what follows). 
 In particular,
we exploit the aforementioned non-perturbative corrections for generating a roller-coaster profile to the otherwise smooth potential,  where plateau-like regions are connected to each other by steep cliffs (See Fig. \ref{fig:Vb}). While plateau-like regions are suitable to sustain long enough inflation -- and to generate nearly scale invariant  fluctuations  at CMB scales -- when reaching the cliff-like regions the scalar velocity  speeds up intermittently.  

During such brief stages, a localized production of  gauge fields occurs. To illustrate   this phenomenon explicitly, we
rely on a numerical 
 procedure  along the same lines of what discussed in single-field models of inflation. We
numerically solve the background equations \eqref{mkgafe} in the bumpy regime of the potential \eqref{VB}, neglecting possible back-reaction effects induced by   gauge fields\footnote{As discussed in \cite{Ozsoy:2020kat}, this assumption 
 is actually self-consistent due to the localized nature of the particle production.}. 
 The resulting background evolution during inflation is represented 
 in Fig. \ref{fig:BB},
 in terms of the first two slow-roll parameters $\epsilon(N), \eta(N)$, for the 
 parameter choices used in Fig. \ref{fig:Vb}. 
\begin{figure}
\begin{center}
\includegraphics[scale=0.75]{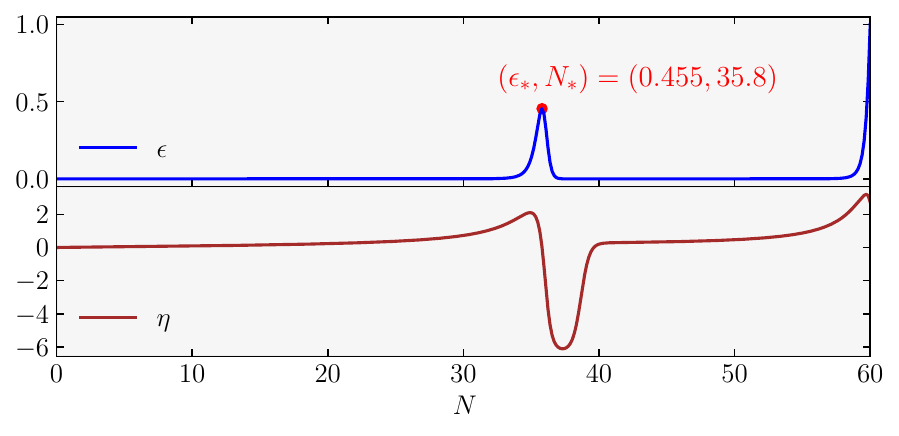}
\end{center}
\caption{Evolution of the Hubble slow-roll parameters $\epsilon$ and $\eta$ with respect to e-fold number, for the  bumpy axion inflation model 
described by potential \eqref{VB}. We
numerically evaluate \eqref{mkgafe} using the parameter choices in Fig. \ref{fig:Vb} with $\phi_{\rm in}\simeq 4.8$, ignoring back-reaction effects as explained in the text. \label{fig:BB}}
\end{figure}
Adopting the initial condition  $\phi_{\rm in} \simeq 4.8 \Mpl$, the scalar field completes a total of 60 e-folding of its evolution when traversing a single bump like region, around which the slow-roll parameter $\epsilon$ reaches its maximal value of $\epsilon_* = 0.455$ at $N_* = 35.8$. Following the stage of  maximal velocity for the scalar field, the background dynamics enters into a short non-attractor phase with $\eta \lesssim -6$, during which $\phi$ slows down before the dynamics re-enters into a final inflationary slow-roll stage $\eta \ll 1$. Since the inflationary background proceeds through an epoch of slow-roll violation,  in our calculations
we 
compute
the vacuum contribution to the scalar power spectrum using the numerical methods discussed in Section \ref{s3p2}, and in Appendix \ref{AppC}. In fact, as anticipated above  in developing this example we make use both of ideas based
on the conversion into curvature fluctuations of gauge-field modes (as in the previous section), and on the activation of the would-be decaying mode
of the inflaton scalar sector (as discussed in Section \ref{s3p2}).

Since while travelling the plateau-like regions the scalar  velocity is  small, the only contributions to the scalar power spectrum arise from the cliff-like region of the potential (see Fig \ref{fig:Vb}). Around the cliff, the scalar field velocity and the effective coupling $\xi$ acquire the following  profile \cite{Ozsoy:2020kat}: 
\beq\label{EC}
\xi(N) = \frac{g_{\rm cs}\,\delta}{1+\delta^2\left(N-N_*\right)^2}, \quad\quad\longrightarrow\quad\quad \xi_{*} \equiv \xi(N_*) = g_{\rm cs} \delta \,,
\eeq
where $\xi_*$ denotes the maximal value of the effective coupling.  The dimensionless parameter $\delta \simeq \sqrt{2/3} (\Mpl/f)(m/H)$ determines  the amount of e-folds during which the parameter $\xi$ maintains its maximal value: \ie $\Delta N \sim \delta^{-1}$. Being $\delta \propto m$, the duration of such epoch is inversely proportional to the restoring force provided by the potential:  hence it inversely depends  on the mass parameter $m$. A time-dependent profile for the effective coupling gives rise to a scale-dependent amplification of the gauge fields whose growing solutions correspond to the real part of the  mode functions, see Eq. \eqref{trAsol}, using Eq. \eqref{Nform}. (See  Appendix \ref{AppD1} for more details). 
 
 As expected, the resulting sourced contribution to the power spectrum induced by the $\delta A + \delta A \to \delta \phi$ inherits the scale dependence of the gauge fields, and  the curvature power spectrum   can be parametrized as \cite{Ozsoy:2020kat},
\beq\label{PSBAI}
\mathcal{P}_{\mathcal{R}}(k)=\mathcal{P}_{\mathcal{R}}^{(\rm v)}(k)\left[1+\frac{H^2}{64 \pi^2 M_{\mathrm{pl}}^2} f_{2, \mathcal{R}}\left(\xi_*, \frac{k}{k_*}, \delta\right)\right].
\eeq
 The vacuum contribution $\mathcal{P}^{(\rm v)}_{\mathcal{R}}$ can be numerically calculated  following  some of the methods explained in Section \ref{sec_single}. 
  Due to the presence of a brief non-attractor
  phase, it is amplified by the activation
  of the would-be scalar decaying mode (see Section \ref{s3p2}). 
 The dimensionless function $f_{2,\mathcal{R}}$ characterizes the contributions from the gauge fields, and has a log-normal shape \cite{Ozsoy:2020kat},
\beq\label{f2R}
f_{2, \mathcal{R}}\left(\xi_*,\frac{k}{k_*}, \delta\right) \simeq f_{2, \mathcal{R}}^c\left[\xi_*, \delta\right] \exp \left[-\frac{1}{2 \sigma_{2, \mathcal{R}}^2\left[\xi_*, \delta\right]} \ln ^2\left(\frac{k}{k_* x_{2, \mathcal{R}}^c\left[\xi_*, \delta\right]}\right)\right]\,.
\eeq
\begin{figure}[t!]
\begin{center}
\includegraphics[scale=0.7]{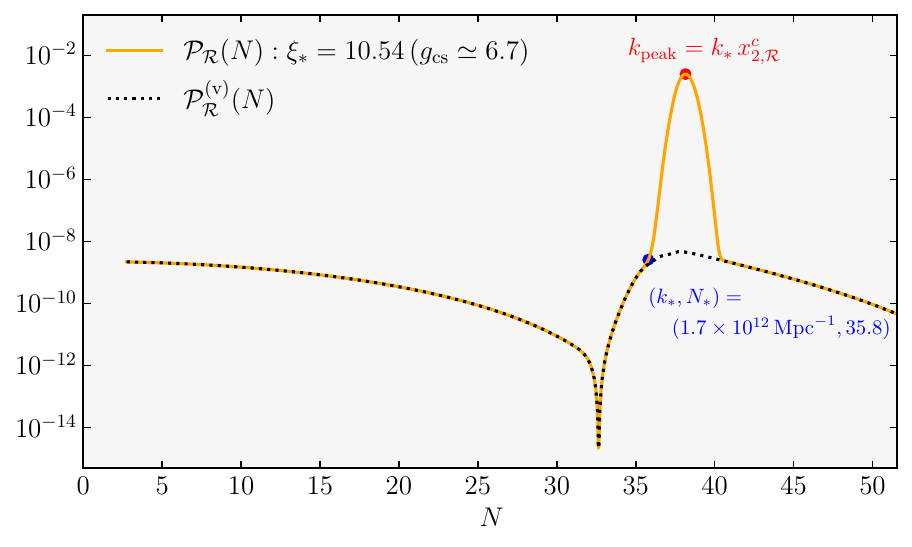}
\end{center}
\caption{The total curvature power spectrum of Eq. \eqref{PSBAI} as a function of $N$ (orange curve), for the bumpy axion inflation model whose  background evolution 
is 
shown in Fig. \ref{fig:BB}. We make  the following parameter choices: $\delta = 1.57$, $\xi_* = 10.54$ corresponding to $g_{\rm cs} \simeq 6.7$ (see \eg \eqref{EC}).\label{fig:PSTBAI}}
\end{figure}
As suggested by the expression \eqref{f2R}, the information about the location, width and 
height of the sourced signals in \eqref{PSBAI} depends on the motion of $\phi$ through the step-like features of its wiggly potential, particularly through $\xi_*$ and $\delta$ dependence of the functions $x^c_{2,\mathcal{R}}, \sigma_{2,\mathcal{R}}, f^c_{2,\mathcal{R}}$. We learn from \eqref{f2R} that the sourced signal is maximal at $k = k_* x^c_{2,\mathcal{R}}$, where it evaluates to $f^c_{2,\mathcal{R}}$ whereas $\sigma_{2,\mathcal{R}}$ controls
the width of the signal. For the background evolution presented in Fig. \ref{fig:BB}, the fitting functions describing the height, width and the location of the sourced signal is calculated in \cite{Ozsoy:2020ccy}, using the semi-analytical techniques developed in \cite{Namba:2015gja}. 

 Collecting all the results above, we present in Fig. \ref{fig:PSTBAI} the full power spectrum in Eq. \eqref{PSBAI} in terms of the e-folding number, by converting the scale dependence to e-fold dependence  using the horizon crossing condition  $k = a(N) H(N)$. Due to non-trivial background evolution, during which the slow-roll conditions are briefly violated, the vacuum contribution to the power spectrum (dotted lines) acquires  a non-trivial shape, which shares the characteristic features of the single-field scenarios we discussed in Section \ref{s3p2}. Notice in fact the
presence of the dip due to the interference between growing and decaying modes
of the curvature perturbation, that immediately precedes a phase of steady growth of the spectrum.

However, in the present example, the duration of the non-attractor era is not sufficient for  providing a prominent peak in the power spectrum solely by the presence of vacuum fluctuations of the curvature perturbation. At this point, the additional source provided by the gauge fields come to the rescue,  giving rise to an extra steep growth, leading to the required power to generate a sizeable peak at wave-numbers corresponding to $k_{\rm peak} = k_* x^c_{2,\mathcal{R}} \simeq 1.5 \times 10^{13}\, {\rm Mpc}^{-1}$ corresponding to peak PBH mass at the time of formation $M^{(\rm f)}_{\rm PBH} \simeq 2.2 \times 10^{-13}\, M_{\odot}$ (see Eq. \eqref{pbhmvskf}). 
Hence, in the presence of couplings to the Abelian gauge fields, the roller-coaster architecture of the potential  provides an assisted amplification mechanism of the power spectrum. The curvature perturbation sourced by the vector fields and its vacuum counterpart help each another, so to generate a sufficiently pronounced  peak to produce PBH. Interestingly, the profile of the curvature spectrum
as function of momenta, in particular 
 its growth rate and 
the details of the peak, are quite different from the corresponding predictions of single-field
inflation, as  investigated in Section \ref{sec_single}.
These differences make the two scenarios distinguishable. 

To conclude, the localized production scenario we presented in this section might be considered as a possible way  to address the instability issues associated with the background of the smooth axion inflation of Section \ref{s4p1p1}. In fact, 
  for the model we considered in this section, back-reaction effects  become prominent for $\xi_* > 15.6$ -- see \cite{Ozsoy:2020kat} for details --  a much larger value than the  phenomenological value $\xi_* \simeq 10.5$ needed for the amplification of the power spectrum. Hence, we work and make predictions in a safe region of cosmological evolution.
  
  Although this is promising towards building workable models of PBH production in axion inflation, the issue regarding the stability of the background will likely to re-appear towards the end of bumpy axion inflation -- 
   in particular towards its final stages when  $\epsilon \to 1$ where $\xi \equiv g_{\rm cs} \sqrt{\epsilon/2} (\Mpl /f) \simeq \mathcal{O}(10)$. In the next section, we discuss an extension  of the model  presented here, in which a spectator axion sector generates a localized scalar enhancement through its coupling to an Abelian gauge sector. Since  in this model the dynamical evolution  of the axion  field  stops long before  inflation ends, we can avoid the strong back-reaction regime which leads to  instabilities of the background configuration. 

\subsubsection{Spectator axion-gauge field dynamics}\label{s4p1p3}

We continue to discuss possible methods to locally enhance
the spectrum of scalar fluctuations in axion-gauge field models,
 with an eye in solving the dynamical instability problems
 mentioned above. We focus on
scenarios where the axion is a spectator field, i.e. does not directly drive inflation. We call $\sigma$ 
the spectator axion  field, and couple it to a Abelian gauge sector. The Lagrangian is
\beq\label{LMS}
 \mathcal{L}_{\rm m} = \mathcal{L}_{\rm inf}-\frac{1}{2}\partial_\mu \sigma\, \partial^{\mu} \sigma - U(\sigma) -\frac{1}{4}F_{\mu\nu}\,F^{\mu\nu} - \frac{g_{\rm cs}}{4 f}\,\sigma\, F_{\mu\nu}\,\tilde{F}^{\mu\nu},
\eeq
where $\mathcal{L}_{\rm inf} = -(\partial \phi)^2/2 - V(\phi)$ is the inflaton Lagrangian, with $V(\phi)$ its potential. We are interested on an inflationary set-up where the spectator sector  provides sub-leading contribution to the total energy density during inflation. This implies that the energy densities of the scalar fields in the model \eqref{LMS} satisfy $\bar{\rho}_\sigma \ll \bar{\rho}_\phi$. Assuming small backraction from the gauge field fluctuations ${\rho}_A \ll \bar{\rho}_{\sigma}$ (more on this later), the Friedmann equation simplifies to
\beq
3 H^2 M_{\mathrm{pl}}^2 \simeq \bar{\rho}_\phi+\bar{\rho}_{\sigma} \quad \longrightarrow \quad 3 H^2 M_{\mathrm{pl}}^2 \simeq V(\bar{\phi}),
\eeq
so that the inflationary background is completely dictated by the inflaton potential. We assume that $V(\phi)$ is flat enough to support sufficient 
inflation,   but otherwise we let it unspecified, as the fine details of the inflaton  dynamics are not essential for what 
 follows.  Since we work in a weak back-reaction system, we can avoid  instabilities
 in the inflaton sector, as the one reviewed in Section \ref{s4p1p1}.

\smallskip
 We  consider, from now on, representative
 examples of axion potentials and its corresponding dynamics.  The choice of our examples is motivated by their interesting ramifications for PBH production, and for their motivations
 from high-energy physics. 
If the spectator axion $\sigma$ is displaced from its global minimum, it can  be dynamically active during inflation. Thanks to the perturbative shift symmetry of the axion sector, we expect the axion potential to be nearly flat, hence the axion dynamics should occurr in a regime of slow-roll $\ddot{\bar{\sigma}}\ll 3H\dot{\bar{\sigma}}$. In such a case we can neglect back-reaction effects induced by gauge field fluctuations\footnote{For the phenomenological examples we will present in this section, this statement can be made precise and back-reaction can be shown to be negligible \cite{Peloso:2016gqs,Ozsoy:2020ccy}.} and the background evolution of the spectator axion can be studied through the following equation,
\begin{figure}
\begin{center}
\includegraphics[scale=0.54]{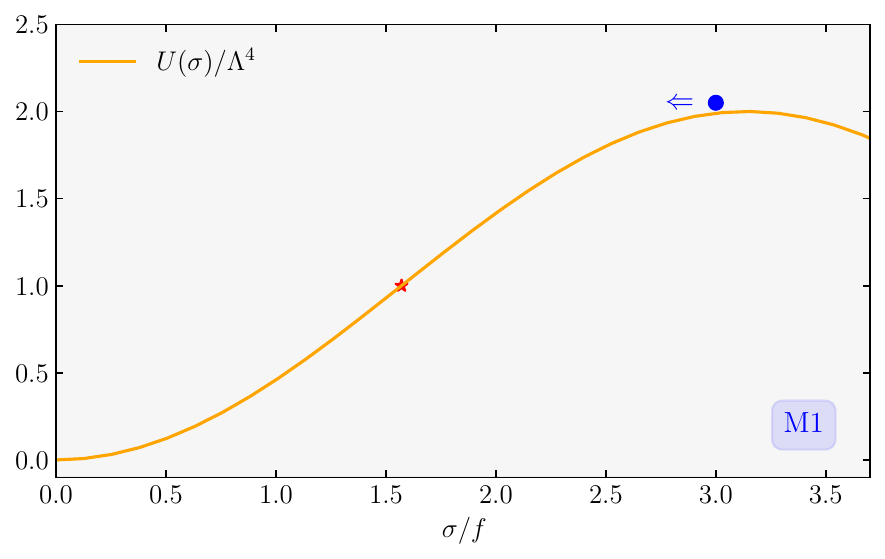}\includegraphics[scale=0.54]{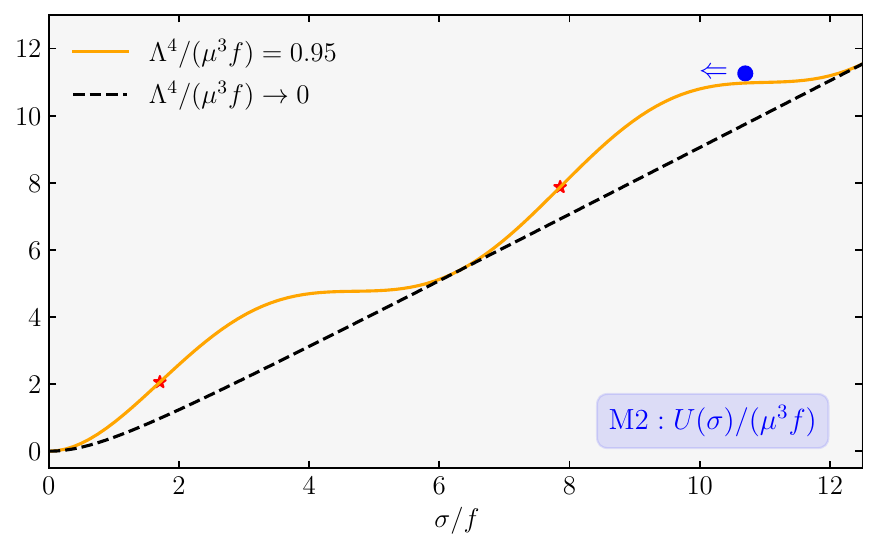}
\end{center}
\caption{The shape of spectator axion potentials for M1 (left) and M2 (right). For both panels, the red stars indicate the location of the inflections point(s) at which the slope of the potential $U'(\sigma)$ and hence the background velocity $\dot{\bar{\sigma}}$ of $\sigma$ becomes maximal (see \eg Eq. \eqref{KGSAM}).\label{fig:POTS}}
\end{figure}
\beq\label{KGSAM}
\ddot{\bar{\sigma}} + 3H\dot{\bar{\sigma}} + U'(\bar{\sigma}) = 0,\quad\quad \longrightarrow \quad\quad 3H\dot{\bar{\sigma}} + U'(\bar{\sigma}) \simeq 0.
\eeq
To realize localized gauge field production through the last term in \eqref{LMS}, we  consider two class of transiently rolling spectator axion models characterized by the following potentials \cite{Namba:2015gja,Ozsoy:2020ccy}
\beq\label{pots}
U(\sigma)=
 \begin{dcases} 
       \Lambda^4 \left[1-\cos\left(\frac{\sigma}{f}\right)\right],& \quad ({\rm M}1) \,,\\
        \mu^3\sigma + \Lambda^4 \left[1-\cos\left(\frac{\sigma}{f}\right)\right]\,\,\,\,{\rm and}\,\,\,\, \Lambda^4\lesssim \mu^3 f&  \quad({\rm M}2),
   \end{dcases}
\eeq
where $\mu$ and $\Lambda$ are parameters of mass-dimension one.

The first model, {\bf M1}, features a standard (discrete) shift symmetric potential akin to natural inflation \cite{Freese:1990rb} where the size of the axion modulations is set by $\Lambda$. In this model, the  axion dynamics spans within an interval bounded the maximum ($\sigma = \pi f$) and the minimum ($\sigma = 0$) of the potential. 
 Therefore, for large  and small field values (early and late times), the axion rolls with  small velocities.  However, $\dot{\bar{\sigma}}$ obtains relatively large values at an intermediate time when $\sigma$ passes through an inflection point $\sigma_* = \sigma(t_*)$ with $U''(\sigma_*) = 0$, where the slope of the potential $U'(\sigma)$ becomes maximal. 

In the second model, {\bf M2}, the axion field range is extended via a linear term \cite{McAllister:2008hb,McAllister:2014mpa} proportional to a soft symmetry breaking mass parameter $\mu$.
 We assume that the axion  $\sigma$ can  probe the corresponding bumpy potential\footnote{In this work, by an appropriate choice of initial conditions and model parameters, we assume that $\sigma$ traverses two step-like features on its potential before it settles to its global minimum.} in the $\Lambda^{4}\lesssim \mu^3 f$ regime, similar to the axion inflation model we discussed in  Section \ref{s4p1p3}. In the plateau-like region(s), and towards the global minimum ($\sigma = 0$) \footnote{The roll of $\sigma$ towards the global minimum can be captured by modifying the monomial term as $\mu^3 sigma \to \mu^3 f [\sqrt{1 + (\sigma/f)^2} -1 ]$, so that the axion potential \eqref{pots} interpolates between $\mu^3 \sigma$ and $(\mu^3/f) \sigma^2$ from large to small field ($\sigma/f \to 0$) values respectively.}, the spectator axion acquires very small velocities, 
  but obtains a relatively large transient peak when the slope of the potential $U'(\sigma)$ becomes maximal at the cliff-like region(s), in particular at the inflection point(s) denoted by $*$ in Fig. \ref{fig:POTS}. 

The background evolution $\sigma$ can be analytically derived  in the slow-roll regime \eqref{KGSAM}, starting from the expressions  for the scalar potentials of Eq. \eqref{pots}. For typical field ranges dictated by these potentials~\footnote{For {\bf M2}, this translates into a single step like region in the potential (see Fig. \ref{fig:POTS}).}, the spectator axion velocity $\dot{\bar{\sigma}}$ and the effective coupling strength $\xi = -g_{\rm cs}\dot{\bar{\sigma}}/(2Hf)$  obtain a peaked time-dependent profile \cite{Namba:2015gja,Ozsoy:2020ccy}:
\begin{equation}\label{Joep}
\xi(N)=
\begin{dcases} 
       \,\frac{2 \xi_{*}}{\mathrm{e}^{\delta(N -N_*)}+\mathrm{e}^{\delta(N_* - N)}}, \,\,  \,\,\, \delta \equiv \frac{\Lambda^4}{3H^2 f^2}\,\,{\rm and} \,\,\xi_*\equiv \frac{g_{\rm cs}\, \delta}{2} &\quad({\rm M}1) \\
     \, \frac{\xi_{*}}{1+\delta^2 (N-N_*)^2}, \,\, \,\,\,\,\,\quad \delta \equiv \frac{\mu^3}{3H^2 f}\,\,{\rm and} \,\, \xi_* \equiv g_{\rm cs}\, \delta\quad &\,\,\,\,\,({\rm M}2)
\end{dcases}
\end{equation}
where $N_*$ denotes the e-fold number when the axion passes through the inflection point (See Fig. \ref{fig:POTS}) and $\xi_*$ is the maximal value of the effective coupling parameter at $N_*$. As in the bumpy axion inflation model, the width of the time dependent peak in $\xi$ depends on the dimensionless ratio $\delta$ which  characterizes the mass of the spectator axion in its global minimum $\delta \approx m^2_{\sigma} /H^2$. For a heavier axion (corresponding to larger $\delta$), the restoring force towards the global minimum becomes very relevant:  the  axion $\sigma$ traverses the inflection point faster, the result being   a sharper peak in $\xi$. In other words, $\delta$ controls the acceleration ($\dot{\xi}/{(\xi H)} = \ddot{\bar{\sigma}}/(\dot{\bar{\sigma}}H) \sim \delta $) of $\sigma$ as it rolls down on its potential. Notice that given our  slow-roll approximation, we require $\delta \ll 3$.  

The peaked structure of $\xi$ profile controls  a critical scale $k_*  = a_* H_*$ characterizing the equation of motion \eqref{meqa}, corresponding to the scale of momenta leaving the horizon at an epoch when the axion velocity is maximal (i.e when $\xi = \xi_*$). Since the mass of the ${\rm U}(1)$ field in \eqref{meqa} is as  tachyonic as possible around this scale, it gives rise to a scale-dependent growth of the gauge field fluctuations where only modes whose size is comparable to the horizon size at $N =N_*$, i.e $k \sim \mathcal{O}(1) a_* H_*$, are efficiently amplified. In Appendix \ref{AppD1} we briefly review the parametric amplification of the gauge field modes corresponding to both models {\bf M1,2} we are presenting
in this section;  for more details we refer the reader to \cite{Namba:2015gja,Peloso:2016gqs,Ozsoy:2020ccy}.  

In the spectator axion models we are focusing, the impact of the vector field production on the visible scalar fluctuations is encoded only indirectly, by the presence of gravitational interactions \cite{Ferreira:2014zia}. In fact, although we consider a matter Lagrangian \eqref{LMS} where the spectator axion-gauge field sector is decoupled from the visible inflaton sector, when integrating out the non-dynamical lapse and shift metric fluctuations we find a mass mixing between inflaton $\delta \phi$ and spectator axion $\delta \sigma$ fluctuations. This phenomenon introduces the possibility that  the  gauge fields  influence the curvature perturbation
$\mathcal{R} \simeq {H\,\delta \phi}/{\dot{\bar{\phi}}}$
through a succession of inverse decays:
 $\delta A + \delta A \to \delta \sigma \to \delta \phi \propto \mathcal{R}$ (see Appendix \ref{AppE}). 
 
 Therefore, the dynamics of this sourced contribution can be understood by first studying the influence of particle production on the spectator axion fluctuations $\delta \sigma$ and then by computing their conversion to curvature perturbation $\delta \phi \propto \mathcal{R}$. We sketch some of the steps  in Appendix \ref{AppD3}; a more detailed computations can be found in \cite{Namba:2015gja,Ozsoy:2020ccy}. 

Taking into account the vacuum fluctuations within the inflaton sector, the total power spectrum of curvature perturbation in the spectator axion-gauge field model can be expressed as \cite{Namba:2015gja,Ozsoy:2020ccy},
\beq\label{PSSAM}
\mathcal{P}_{\mathcal{R}}(k)=\mathcal{P}_{\mathcal{R}}^{(\rm v)}(k)+\left[\epsilon_\phi \mathcal{P}_{\mathcal{R}}^{(\rm v)}(k)\right]^2 \sum_{i} f_{2, \mathcal{R}}^{(i)}\left(\xi_*, \frac{k}{k_*}, \delta\right),
\eeq
where $\mathcal{P}_{\mathcal{R}} = H^2 /(8\pi^2 \epsilon_\phi \Mpl^2)$ with $\epsilon_\phi \equiv \dot{\bar{\phi}}^2 / (2H^2\Mpl^2)$ and the sum over $i$ denotes contributions from each particle production location. {\it E.g.} for model {\bf M1} the sum is over only from one of such regions, while for {\bf M2} there are two of them (due to our assumptions about the initial conditions of $\sigma$ in the bumpy regime of the potential in \eqref{pots}), as the spectator rolls along the potential  towards its global minimum.  Eq. \eqref{PSSAM}
teaches us that the part of the spectrum sourced by the gauge fields has an extra slow-roll suppression $\epsilon_\phi$, in particular with respect to the direct coupling model we discussed in the previous section (see Eq. \eqref{PSBAI}). However, the part that parametrizes the scale dependence of the sourced contribution, \ie $f_{2,\mathcal{R}}$ in \eqref{PSSAM}, exhibits the same log-normal scale dependence of the bumpy axion model in \eqref{f2R} whose peak height, width and location ($f^c_{2,\mathcal{R}}, \sigma_{2,\mathcal{R}}, x^c_{2,\mathcal{R}}$) is calculated in \cite{Namba:2015gja,Ozsoy:2020ccy} in terms of $\xi_*$ for phenomenologically interesting values of $\delta$ that characterize the background evolution of the spectator axion. 

To understand the full scale dependence of the power spectrum (and in particular its vacuum contribution $\mathcal{P}^{(\rm v)}_{\mathcal{R}}$), we need to specify the scalar potential $V(\phi)$ in the inflationary sector. Instead of specifying $V(\phi)$,  we  take a phenomenological approach to determine the important set of parameters  summarizing the inflationary dynamics. For this purpose, first notice that assuming the effects introduced by the rolling of $\sigma$ is negligible at CMB scales (early times during inflation), we have $n_s -1 \simeq 2\eta_\phi -6\epsilon_\phi$ and $r \simeq 16\epsilon_\phi$. Considering the latest Planck results, we adopt $r \lesssim 10^{-2}$ at CMB scales \cite{Planck:2018jri}, to obtain $\epsilon_\phi \simeq 6.25 \times 10^{-4}$. However, the observed value of the spectral tilt gives $n_s -1 \simeq -0.035$ \cite{Planck:2018jri}. Furthermore, neglecting higher order slow-roll parameters, we  assume that $\epsilon_\phi$ remains constant throughout the inflation. These simplifying approximations enable us the describe total power spectrum in the multi-field scenarios we described above. Denoting the e-fold dependence of the Hubble parameter during slow-roll inflation as $H(N) = H_{\rm cmb}\, e^{-\epsilon_{\phi}N}$ (where the subscript ${\rm cmb}$ denotes the time at which the CMB pivot scale $k_{\rm cmb} = 0. 05 \, {\rm Mpc}^{-1}$ exits the horizon), we can turn the the scale dependence of the vacuum scalar power spectrum into e-fold dependence as
\begin{figure}
\begin{center}
\includegraphics[scale=0.52]{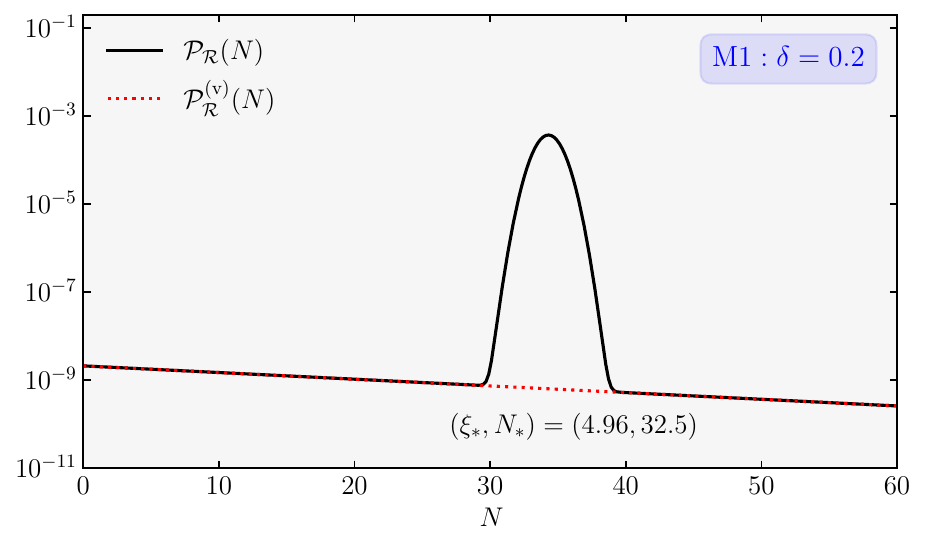}\includegraphics[scale=0.52]{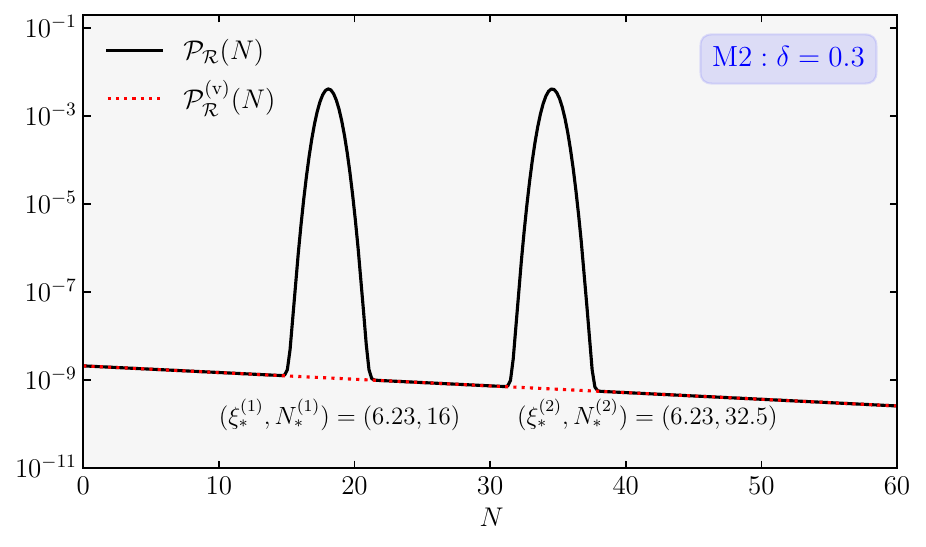}
\end{center}
\caption{Total scalar power spectrum (black curves) as a function of e-folds during inflation for the spectator axion-gauge field models {\bf M1} (Left) and {\bf M2} (Right). In the left panel, we assume $\dot{\bar{\sigma}}$ is maximal at $N_* = 32.5$ where the effective coupling reaches $\xi(N_*) = 4.96$ corresponding to $g_{\rm cs} = 49.6$ using Eq.\eqref{pots}. In the right panel, spectator traverses two bumps before it settles to its minimum and we choose the inflection points to occur at $N^{(1)}_* = 16$ and $N^{(1)}_* = 32.5$ where $\xi_* = 6.23$ ($g_{\rm cs} \simeq 20.7$). \label{fig:PSSAM}}
\end{figure}
\beq\label{psvacsm}
\mathcal{P}^{(\rm v)}_{\mathcal{R}}\left(k\right) \simeq \mathcal{A}_s\,  \mathrm{e}^{-\left(1-\epsilon_{\phi}\right)\left(1-n_{s}\right)N},
\eeq
where $\mathcal{A}_s \equiv \mathcal{P}^{(\rm v)}_{\mathcal{R}}\left(k_{\rm cmb}\right) = 2.1 \times 10^{-9}$ where $\epsilon_\phi \simeq 6.25 \times 10^{-4}$ and $1 -n_s \simeq 0.035$ as we discussed above. Collecting all the information we presented above, we plot in Fig. \ref{fig:PSSAM} the total scalar power spectrum (Eq. \eqref{PSSAM}) for a representative parameter choices, describing both of the spectator axion-gauge field models {\bf M1} and {\bf M2}. 

We learn from  Fig. \ref{fig:PSSAM} that since the spectator dynamics do not influence the inflationary background significantly, the vacuum contribution (dotted lines) has a smooth red tilted power law form, which should be contrasted with the bumpy axion inflaton model we discussed in the previous section. However, the transient roll of the spectator axion $\sigma$ before settling to its global minimum 
triggers gauge field production as it probes the inflection point(s) on its potential (see Fig. \ref{fig:POTS}). This phenomenon  generates an additional Gaussian bump in the scalar power spectrum. Notice that for larger $\delta$ the width of the corresponding bump decreases because $\dot{\bar{\sigma}} \propto \xi$ is maximal for a shorter time interval,  during which it can affect fewer gauge field modes. This explains why the sourced signals in the second model {\bf M2} has a narrower width. As in the direct coupling cases to the gauge fields we considered earlier, the peak in the power spectrum originates from a cubic term, \ie the last term in \eqref{LMS} and therefore the statistics of the curvature perturbation is expected to be non-Gaussian. This explains why the examplifying  scenarios we present in Fig. \ref{fig:PSSAM} with a peak power of $\mathcal{P}_{\mathcal{R}} \sim 10^{-3}$ is sufficient to generate large populations of PBHs at masses corresponding to the peak scales in Eq. \eqref{pbhmvsn}.

To close this section on
 axion-vector inflationary systems aimed at producing PBHs, we can conclude that this possibility requires a fair amount of complex model building. Nevertheless,  the physics of axion  reached so far a high degree of theoretical sophistication, thanks to the input and motivations
 from particle physics, quantum gravity, and cosmology.
 The distinctive predictions we explored for what respect to the properties of the curvature power spectrum, in particular
 its profile as function of momenta, and its shape around
 the peak, make these scenarios 
 distinguishable from single-field inflation, and with
 relevant ramifications for PBH phenomenology. 
 It is certainly worthwhile explore and apply existing efforts to explore their consequences for the physics of PBH formation.
 
 \subsection{Strong turns in the multi-scalar field space}\label{s4p2}

An alternative approach for enhancing scalar fluctuations at small scales is
to exploit the dynamics of multiple field inflation. Multi-field inflation a topic very well studied
in the inflationary literature with a variety of motivations, see e.g. \cite{Bassett:2005xm} for a review.
In reviewing the applications to PBH production, we focus on a general class of non-linear sigma models that are well motivated  when embedding    inflation in high energy physics. Although we go through representative concrete examples, the essence of the mechanism is again associated with a transient tachyonic instability in the scalar sector -- along a direction orthogonal to the inflationary one -- which is converted through suitable couplings into an amplification of curvature perturbations.
 
The generic two-derivative  Lagrangian describing the dynamics of such a system of scalar fields $\phi^{I}$ minimally coupled to gravity is given by ($ S \equiv \int \d^4 x \sqrt{-g} \,\mathcal{L}_m$),
\beq
\mathcal{L}_m = -\frac{1}{2}G_{IJ}(\phi) \partial_{\mu} \phi^{I} \partial^{\mu} \phi^{J} - V(\phi),
\eeq
where the fields may interact through the potential $V(\phi)$ and the field field space metric $G_{IJ}(\phi)$. For an FLRW background characterized by the scale factor $a(t)$ and Hubble parameter $H(t)$, the equation of motion of the homogeneous fields is described by
\beq\label{MFKG}
\mathcal{D}_t \dot{\bar{\phi}}^{I} + 3H\dot{\bar{\phi}}^{I} + G^{IJ} V_{,J} = 0,
\eeq
where the time field space covariant derivative of any field space vector $A^{I}$ is defined as $\mathcal{D}_t A^{I} = \dot{A}^{I} + \Gamma^{I}_{\,\,JK} \dot{\bar{\phi}}^{J} A^{K}$. Considering two field case for simplicity, the background trajectory can be split into an adiabatic $e_{\sigma}^{I} = \dot{\bar{\phi}}^{I} / \dot{\sigma}$ and entropic $e^{I}_s$ field bases that are orthogonal to each other. Here, $\sigma \equiv (G_{IJ} \dot{\bar{\phi}}^{I}\dot{\bar{\phi}}^{J})^{1/2}$ which is related to the Hubble slow-roll parameter as $\epsilon \equiv -\dot{H}/H^2 = \dot{\sigma}^2 / (2H^2\Mpl^2)$.

We find it worth mentioning that the vanilla multi-field slow-roll trajectory corresponds to fields following standard gradient flow of the potential $3 H \dot{\bar{\phi}}^{I} \simeq - V^{,I}$ corresponding to a field space trajectory that is approximately a geodesic with $\mathcal{D}_t \dot{\bar{\phi}}^{I} \simeq 0$ (see \eg \eqref{MFKG}). However, a sufficiently long phase of inflation only requires a small acceleration along the unit vector tangent to the inflationary trajectory, $e_{\sigma\, I} \mathcal{D}_t \dot{\bar{\phi}}^{I} \simeq 0 $ and the acceleration pointing in the perpendicular direction --which parametrizes the ``bending'' of the inflationary trajectory -- is generically not constrained and hence can be large. The level of ``bending'' can be described by the dimensionless parameter $\eta_{\perp}$ so that the orthogonal field bases evolve as
\beq
 \mathcal{D}_t\, e^{I}_{\sigma} = H \eta_{\perp} e^{I}_{s}, \quad\quad  \mathcal{D}_t\, e^{I}_{s} = - H \eta_{\perp} e^{I}_{\sigma},
\eeq
where $\eta_{\perp} = - e^{I}_s\, V_{,I} / (H \dot{\sigma})$ measures the acceleration of the trajectory perpendicular to its velocity \cite{GrootNibbelink:2000vx,GrootNibbelink:2001qt}. Combining it with Hubble rate, one can define a ``turning rate'' associated with the trajectory as $\Omega \equiv \eta_\perp H$. Therefore, an inflationary background with a strong turn $\eta_\perp \gg 1$ satisfies $\Omega \gg H$. As we will show below, this is the regime of interest for the amplification of the scalar fluctuations in this multi-field setup. 
\begin{figure}
\begin{center}
\includegraphics[scale=0.53]{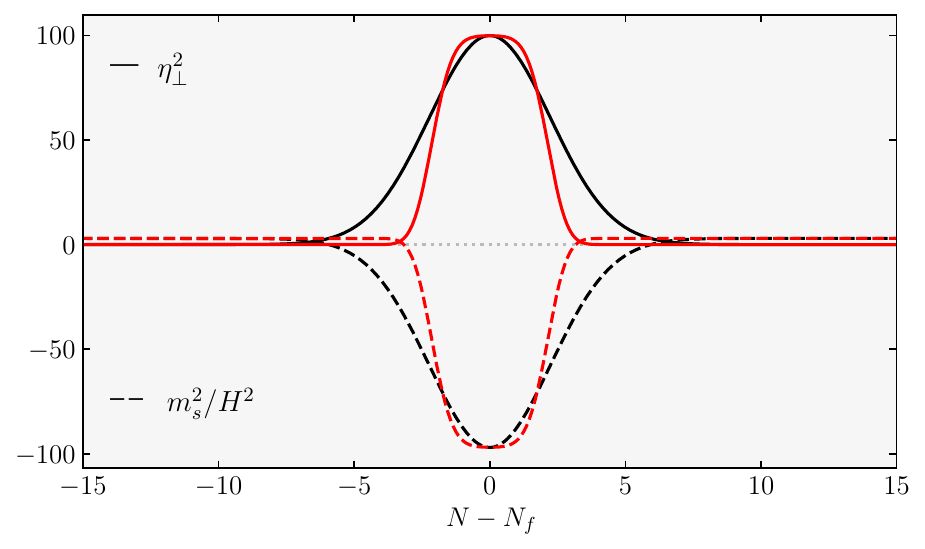}\includegraphics[scale=0.53]{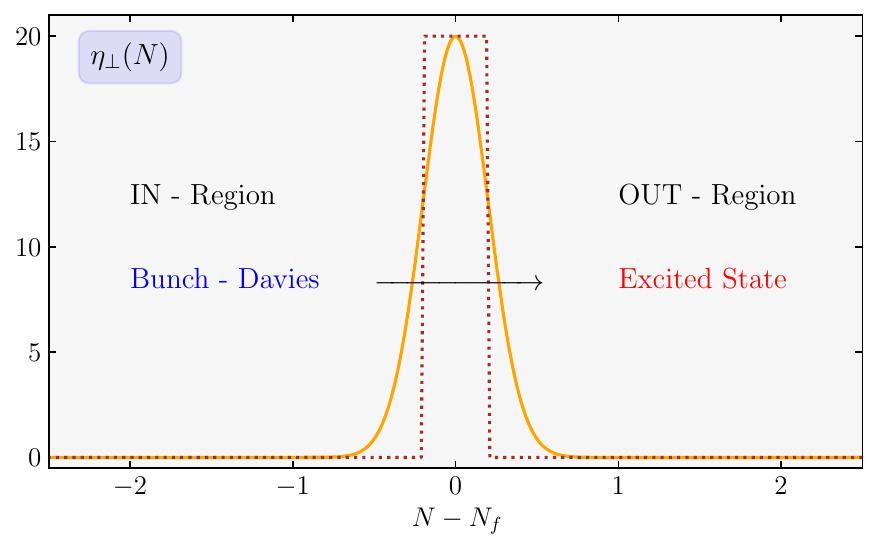}
\end{center}
\caption{Left: Schematic time evolution of the bending parameter $\eta_{\perp}^2$ and entropic mass $m_s^2 = b - \eta_{\perp}^2$ for a broad turn using a Gaussian $\eta_{\perp} = \eta^{\rm max}_{\perp} e^{-y^2/(2\Delta^2)}$, $(\eta^{\rm max}_{\perp},\Delta^2) = (10,10)$ and smoothed top-hat profile $\eta_{\perp} = \eta^{\rm max}_{\perp} ({\rm erf}(y-\delta/2) - {\rm erf}(y+\delta/2))/2$, $(\eta^{\rm max}_{\perp},\delta) = (10,5)$ with $y = N-N_f$. Right: Time dependence of bending parameter $\eta_{\perp}$ for representative sharp (rapid) turn examples with Gaussian and top-hat profile.\label{fig:turn}}
\end{figure}
For this purpose, we consider the linear fluctuations around the background we described above, which is given by the following action \cite{Sasaki:1995aw,GrootNibbelink:2001qt,Langlois:2008mn}
\beq\label{SMF}
S^{(2)} = \int \d^4 x\, a^3 \left[\Mpl^2 \epsilon \left(\dot{\mathcal{R}}^2 - \frac{(\partial \mathcal{R})^2}{a^2}\right)+ 2\dot{\sigma}\eta_{\perp}\dot{\mathcal{R}}Q_s + \frac{1}{2}\left(\dot{Q}_s^2 - \frac{(\partial Q_s)^2}{a^2} + m_s^2 Q_s^2 \right)\right],
\eeq
where $\mathcal{R}$ is the comoving curvature perturbation\footnote{In the comoving gauge we are operating, the field fluctuations are proportional to entropy fluctuation $\delta \phi^{I} = Q_s e^{I}_s$ while the spatial part of the metric takes the form $\hat{g}_{ij} = a^2 e^{-2\mathcal{R}} \delta_{ij}$.}, $Q_s$ is the instantaneous entropy perturbation whose mass is given by 
\beq
m_s^2 = V_{;\,ss} - H^2 \eta_{\perp}^2 + \epsilon H^2 \Mpl^2 R_{\rm fs}
\eeq
with $V_{;\,ss} = e^{I}_{s}e^{J}_{s} V_{; IJ}$ is the projection of the covariant Hessian of the potential along the entropic direction and $R_{\rm fs}$ is the field space curvature. As should be clear from the action \eqref{SMF}, the dynamics of the fluctuations in this two field model is completely dictated by three functions of time $N = \ln(a)$: namely Hubble rate $H(N)$, turning rate $\eta_{\perp}(N)$ and the mass $m_s^2(N)$ of the entropy mode. To demonstrate the amplification in the power spectrum via strong turns in curved multi-field space, we follow the effective approach presented in \cite{Fumagalli:2020adf} to consider a field space that undergoes a strong turn $\eta_{\perp} \gg 1$ around e-folding number $N_f$ during inflation, corresponding a time well after CMB scales exited the horizon and assume a featureless Hubble rate $H(N)$. We also assume that the time dependence of entropic mass is mainly controlled by the bending parameter, taking $m_s^2 = (b - \eta_{\perp}^2) H^2$ where $b$ is chosen to be a constant for simplicity. 

The behavior of the bending parameter (and that of the entropic mass $m_s^2$) for typical strong turn cases are shown in Fig. \ref{fig:turn}. An important quantity that determines the behavior of the fluctuations is the duration of the turn which leads to either broad (left panel) or rapid turn (right panel) cases as shown in the figure. Although the dynamics of the scalar perturbations are qualitatively different for broad and sudden (strong) turn cases, their behavior share some common characteristics that can be attributed to the transient nature of the strong turn in field space. First of all, the modes that are still deep in the horizon at the end of the turn do not feel the presence of the feature and hence their dynamics is standard single-field type where the power spectrum is given by
\beq\label{PS0}
\mathcal{P}^{(0)}_{\mathcal{R}}(k) = \frac{H^2}{8\pi^2\epsilon\Mpl^2}\,\, \bigg|_{k = aH},
\eeq
with a slight red tilted scale dependence. For this purpose, we will utilize the phenomenological model we discussed in the previous section to parametrize the scale dependence (time dependence) of the power spectrum as in Eq. \eqref{psvacsm} with a constant $\epsilon \simeq 6.25 \times 10^{-4}$ and $1-n_s \simeq 0.035$ along with $H_{\rm cmb} \simeq 10^{-5}$. On the other hand, large scales that leave the horizon well before the turn experience typical multi-field dynamics resulting with a transfer of power from entropic to curvature perturbations and  enhancement in the power spectrum \cite{Bassett:2005xm}. The corresponding amplification is sensitive to the value of the entropic mass $m_s$ from the time of horizon exit until the turn. In the following, we will assume a sizeable $m_s$ (\ie $b$) before the turn so that the entropic fluctuations have decayed sufficiently before the onset of the turn so that the power spectrum is given by the standard result \eqref{PS0} for these scales. For scales that exit the horizon around the time of the turn, the entropic fluctuations exhibit transient (exponential) growth due to their tachyonic mass $m_s^2 < 0$ around the turn, which is eventually transferred to the curvature perturbation through the kinetic coupling in \eqref{SMF} proportional to the bending $\eta_{\perp}$ parameter. The exponential amplification of $\mathcal{P}_{\mathcal{R}}$ associated with these scales is a typical observational feature of strong turns, which we discuss briefly focusing on broad and sharp turn cases separately. 

\medskip
\noindent{\bf Broad turns}
\medskip

\noindent
For a strong turn in field space that lasts sufficiently long time so that modes of interest (\ie modes that are enhanced) have already exited the Hubble radius at the end of the turn, the dynamics of the curvature perturbation can be
described by the single-field effective theory with an imaginary sound speed parametrizing the growth of the fluctuations \cite{Garcia-Saenz:2018ifx,Garcia-Saenz:2018vqf}. From the perspective of two-field description, imaginary sound speed is just a manifestation of the transient instability induced by the strong non-geodesic motion in field space and the associated transient negative entropic mass $m_s^2$ \cite{Bjorkmo:2019qno}. Therefore, the dynamics of mode functions are mainly characterized by two time scales: i) $\tilde{N}$ corresponding to entropic mass crossing and the onset of instability ii) $\bar{N}$ denoting the sound horizon crossing of fluctuations after which the curvature perturbation $\mathcal{R}$ becomes frozen:
\beq\label{emc}
\frac{k}{a(\tilde{N})}=\left|m_s(\tilde{N})\right|, \quad \& \quad \frac{k\left|c_s\right|}{a(\bar{N})}=H(\bar{N}),
\eeq
where $|c_s|$ denotes the absolute value of the imaginary speed of sound describing the instability/growth in the fluctuations ($\mathcal{R}$) in the single EFT language. Since the background dynamics change mildly between the time of entropic mass crossing $\tilde{N}$ and freeze-out $\bar{N}$ in the broad turn case, the exponential enhancement in the power spectrum can be parametrized as \cite{Bjorkmo:2019qno,Fumagalli:2020adf}
\beq\label{PSB}
\mathcal{P}_{\mathcal{R}}(k)= \mathcal{P}^{(0)}_\mathcal{R}(k)\, \exp\left[\pi \eta_{\perp} (2 - \sqrt{3 + b/\eta_{\perp}^2})\right]\,\,\bigg|_{\tilde{N}_k},
\eeq
for all $k$ modes that satisfies $ k = a(\tilde{N})|m_s|$ while $m_s^2 < 0$. Notice that exponential enhancement in the power spectrum with respect to base spectrum is proportional to bending parameter $\eta_{\perp} \gg 1$.
\begin{figure}
\begin{center}
\includegraphics[scale=0.6]{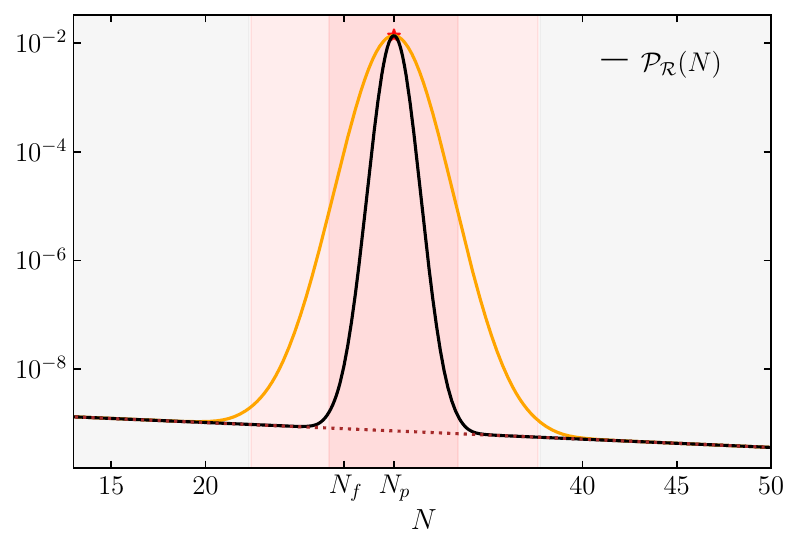}\includegraphics[scale=0.6]{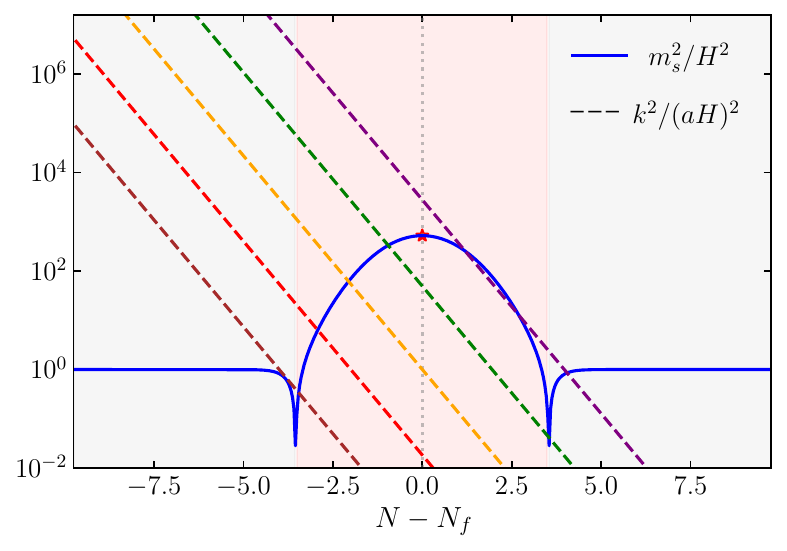}
\end{center}
\caption{Left: Curvature power spectrum (normalized to $\mathcal{P}_{\mathcal{R}}(N=0) = 2.1 \times 10^{-9}$) for broad strong turns and for two representative parameter choices using a Gaussian bending profile (see main text). Right: Migration of modes $k^2 / (aH)^2$ for equally spaced $\ln(k)$ values around the turn to confirm the expectation that power spectrum should fall more quickly after the peak (as compared to the left panel) because more modes experience entropic mass crossing for $N < N_f$ than for $N > N_f$ in the $m_s^2 < 0$ region highlighted by red.\label{fig:PSBaMig}}
\end{figure}
In Fig. \ref{fig:PSBaMig}, we show the power spectrum profile as a function of e-folds (left panel) using the analytic formula \eqref{PSB} for Gaussian bending profiles $\eta_{\perp} = \eta^{\rm max}_{\perp} e^{(N-N_f)^2/(2\Delta^2)}$ with $(\eta^{\rm max}_{\perp}, \Delta^2) = (20,10)$ (orange curve - Model 1) and $(\eta^{\rm max}_{\perp}, \Delta^2) = (20,2)$ (black curve - Model 2) and $N_f = 30$. For both models, the range of scales and times affected by the turn where $m_s^2 < 0$ are highlighted by red and correspond to $2.4 \times 10^{8}<k\, [{\rm Mpc}^{-1}]<1.2\times ^{15}$ (Model 1) $1.7 \times 10^{10}<k\, [{\rm Mpc}^{-1}]< 1.6 \times 10^{13}$ (Model 2). As should be clear from the expression \eqref{PSB}, the power spectrum reaches its maximal value for the scale whose entropic mass crossing overlaps with the peak of the bending parameter, \ie when $N_f = \tilde{N}_{k_p}$ corresponding to
\beq\label{kpb}
k_p \simeq k_f \, |(\eta^{\rm max}_{\perp})^2 - b|^{1/2},
\eeq
where $k_f = a(N_f) H(N_f) \simeq 5.24 \times 10^{11}\, {\rm Mpc}^{-1}$ is the scale that exits the horizon at $N_f$. To determine the slope of the power spectrum on the rise, one can compute the spectral index with respect to the base power spectrum to get 
\beq\label{BTnsm1}
\left(n_s-1\right)-\left(n_s-1\right)_0 \simeq \pi(2-\sqrt{3}) \frac{N_{\mathrm{f}}-\tilde{N}}{\Delta^2+N_{\mathrm{f}}-\tilde{N}} \eta_{\perp}
\eeq
which shows that stronger and/or less broad turns in field space lead to a steeper power spectrum. In particular, the expression \eqref{BTnsm1} tells us that for broad and strong turn in field space, the slope of the power spectrum towards its peak can be much larger than the single-field models where $n_s -1 \lesssim 4 $ \cite{Byrnes:2018txb,Carrilho:2019oqg,Ozsoy:2019lyy,Cole:2022xqc}.

We would like to note that for scales affected by the turn, the analytic profile presented in \eqref{PSB} is symmetric around the peak however numerical computations carried in \cite{Fumagalli:2020adf} shows that the power spectrum typically exhibit a much quicker fall of behaviour for scales following the peak compared to ones preceding it. This could be understood by first recalling that the modes that are enhanced are the ones that goes through entropic mass crossing \eqref{emc} while the entropic mass is negative $m_s^2 < 0$. Right panel of Fig. \ref{fig:PSBaMig} can then guide us for an intuitive understanding for the expected asymmetry in the power spectrum following its peak because a larger range of modes enjoys entropic mass crossing before the peak of $|m_s|$ \footnote{Recall from our discussion above that entropic mass crossing at $N_f$ corresponds to the peak of the power spectrum.} than afterwards. We therefore conclude that the expression \eqref{PSB} along with the power spectrum shown in the left panel of Fig. \ref{fig:PSBaMig} only characterize the behavior of the fluctuations qualitatively and for more accurate results numerical methods are required as shown in \cite{Fumagalli:2020adf}.

\medskip
\noindent{\bf Sudden turns}
\medskip

\noindent
When the duration of the turn is shorter (a statement that we will make more precise below), the scales that are maximally enhanced are still those that experience entropic mass crossing during the turn and therefore they are of order $k \sim k_f \eta^{\rm max}_{\perp}$ (see \eg Eq. \eqref{kpb}). However, differently from the broad turn case, these modes get caught in the (rapid turn) feature while they are still deep inside the horizon, \ie they satisfy $k > a H$ at the end of the turn. This situation effectively generates an initial excited state for these modes that can be studied analytically in the regime of sudden and strong (constant) turns $\eta_{\perp}$ \cite{Palma:2020ejf,Fumagalli:2020nvq}. Along with a localized exponential amplification, the resulting power spectrum of curvature perturbation exhibit order one oscillations in $k$ --which is a common characteristic of sharp features \cite{Chluba:2015bqa,Slosar:2019gvt} -- with a frequency is set by the time of the turn. We will review these features below by focusing on the analytic model presented in \cite{Fumagalli:2020nvq} \footnote{For a detailed analytic/numerical analysis on the behaviour of the power spectrum from strong sharp turns see also \cite{Fumagalli:2020adf,Palma:2020ejf}.}. For this purpose, we consider an inflationary trajectory with a top hat \footnote{More realistic models of the turn are expected to have a smooth time dependence of $\eta_{\perp}$ like a Gaussian profile we considered earlier. As shown by the numerical evaluations in \cite{Fumagalli:2020nvq}, the choice of top hat profile does not introduce any qualitative difference but appears to be a convenient choice for analytic manipulations. For a detailed study of sudden turn trajectories in conjunction with PBH formation, see also \cite{Anguelova:2020nzl}.}sudden turn profile (see \eg the right panel in Fig. \ref{fig:turn}) characterized by three parameters: $N_f$ the central time of its location (measured with respect to the horizon exit time of the CMB pivot scale), its duration $\delta$ and typical (constant) value $\eta^{\rm max}_{\perp}$ around $N_f$: 
\beq\label{ST}
\eta_{\perp}(N)=\eta_{\perp}^{\max }\left[\theta\left(N-\left(N_{f}-\frac{\delta}{2}\right)\right)-\theta\left(N-\left(N_f+\frac{\delta}{2}\right)\right)\right],
\eeq
along with an entropic mass profile during the turn that is given by 
\beq
\frac{m_s^2}{H^2}=(\xi-1) \eta_{\perp}^2(N)
\eeq
where $\xi < 1$ is a constant parameter. Note from the parametrization of the bending parameter \eqref{ST} that the sudden turn case we consider corresponds to $\delta > \ln(\eta^{\rm max}_{\perp})$.
\begin{figure}[t!]
\begin{center}
\includegraphics[scale=0.62]{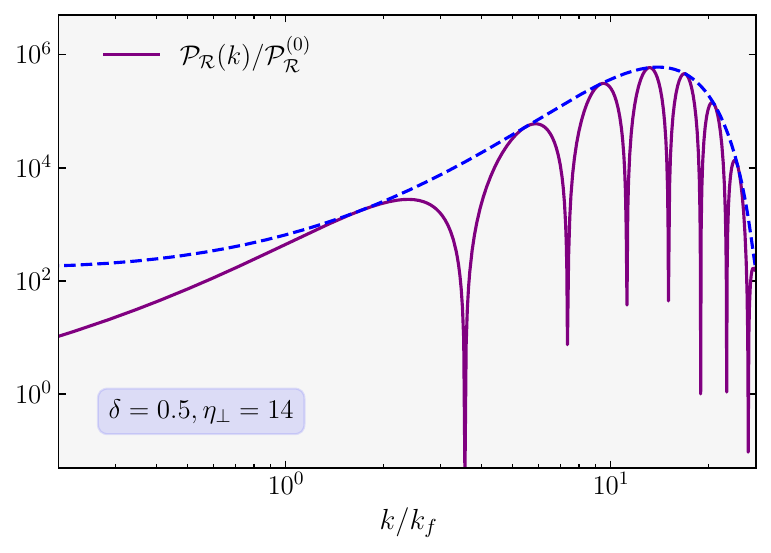}\includegraphics[scale=0.62]{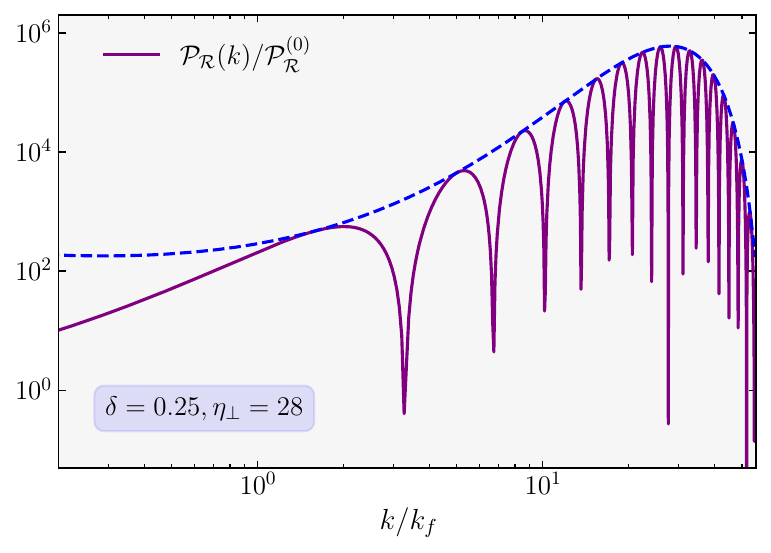}
\end{center}
\caption{Shape of the curvature power spectrum using the analytic formula in Eq. \eqref{psst}, for two representative sharp turn cases where the turn in parametrized with a top-hat bending parameter of Eq. \eqref{ST}. Dashed blue line present the envelope characterized by terms highlighted in Eq. \eqref{psst}.\label{fig:PSst}}
\end{figure}
In this setup, one can determine the power spectrum generated by sharp turns, by studying scattering problem of the two-field system using WKB methods \cite{Palma:2020ejf}. In particular, this procedure includes matching operators (and their derivatives) describing the perturbations, from the
IN region, where the Bunch-Davies vacuum is assumed, passing through the turn, and to the
OUT region, resulting in an excited “initial” state as sketched in the right panel of Fig. \ref{fig:turn}. The behaviour of the entropic and curvature perturbations in both regions (IN and OUT) are
standard where the fields are dynamically decoupled from each other, while the turn sandwiched between
the IN and OUT region results with their exponential growth. Following this procedure, an analytic expression for the power spectrum has been derived in \cite{Fumagalli:2020nvq} for scales satisfying $k/k_f < \sqrt{1-\xi} \eta_{\perp}$ as:

\beq\label{psst}
\frac{\mathcal{P}_\mathcal{R}(k)}{\mathcal{P}^{(0)}_{\mathcal{R}}}=\underbrace{\frac{e^{2 \eta_{\perp} \delta\,\, S}}{2 S^2(1+X+\sqrt{X(1+X)})}}_{\textrm{envelope}} \times \,{\sin ^2\left(e^{-\delta / 2} \kappa \eta_{\perp}+\arctan (\kappa / S)\right)}
\eeq
where
\beq\label{SkXk}
S(k)=\sqrt{\sqrt{4 \kappa^2+\frac{(3+\xi)^2}{4}}-\left(\kappa^2+\frac{(3+\xi)}{2}\right)}, \quad\quad {\rm and} \quad X(k)=\frac{(3+\xi)^2}{16 \kappa^2}
\eeq
with $\kappa \equiv k / (k_f \eta_{\perp})$ and $\eta_{\perp}$ denotes the typical constant value of the bending parameter corresponding to $\eta^{\rm max}_{\perp}$ in the top hat profile \eqref{ST}.

For the choice of $\xi = -3$ ---which corresponds to an effectively massless entropy mode on super-horizon scales (see Section 2 of \cite{Fumagalli:2020nvq} for a detailed discussion on this) --- the profile of the curvature power spectrum \footnote{We also note that the presence of very light entropy mode (for the $\xi = -3$ case) leads to a non negligible contribution to the power spectrum for scales that crosses the horizon before the turn, \ie for $k/k_f \to 0$. In particular, the analytic expression \eqref{psst} do not perform well on these scales. For a more complete analytic formula including numerical analysis, see \cite{Fumagalli:2020nvq}.} for two representative parameter choices characterizing the turn, is shown in Fig. \ref{fig:PSst}. As described by the analytic expression \eqref{psst}, the power spectrum features characteristic oscillations ($\sin^2$ term) modulated by an envelope that is described as highlighted in the equation. The peak value of the envelope (see \eg blue dashed lines in Fig. \ref{fig:PSst}) controls the maximal enhancement of the power spectrum with respect to baseline power spectrum \eqref{PS0} (we assume it to be scale invariant for simplicity) which will occur roughly at the maximum of the function $S(k)$ defined  in \eqref{SkXk}. Arguably, the oscillatory pattern of the power spectrum around the peak is much more interesting. In particular, as $\eta_{\perp} \delta \ll e^{-\delta/2} \eta_{\perp}$ $\sin^2$ term changes much faster than the envelope. On the other hand, in sharp turn regime $e^{-\delta/2} \eta_{\perp} \gg 1$ so that the first term in the argument of the $\sin^2$ term also changes much faster than the second. These two observations imply that oscillations in the power spectrum is periodic to a good degree. The period of the maxima in the oscillations occurs roughly in $\Delta \kappa \approx \pi e^{\delta/2} / \eta_{\perp}$ corresponding a linear frequency of
\beq\label{freqst}
\Delta k \approx \pi  e^{\delta/2} k_f \quad \longrightarrow \quad \omega \equiv \frac{2\pi}{\Delta k} \approx \frac{2 e^{-\delta/2}}{k_f}.
\eeq
Therefore, effectively the scale $k_f $ that exits the horizon at the center of the feature sets roughly the frequency of oscillations in the power spectrum. 

To summarize, strong-sudden turns in field space leads to a power spectrum that exhibits order one oscillations whose amplitude is modulated by an exponentially enhanced envelope \eqref{psst}. The oscillations are fast, making their peaks almost periodic with a frequency given by \eqref{freqst}. We note that these oscillations do not have an impact on the PBH mass spectrum $\beta(M)$ (see \eg \eqref{betaG}) because the calculation of the variance $\sigma^2(M)$ (of the density contrast) involves a smoothing procedure over scales comparable to the width of peaks in the power spectrum $\mathcal{P}_{\mathcal{R}}$ \cite{Fumagalli:2020adf}. Another issue of interest from the perspective of model building is the influence of non-Gaussianity. The periods of strong turns during inflation are known to generate non-Gaussianity of flattened type \cite{Garcia-Saenz:2018vqf,Fumagalli:2019noh,Ferreira:2020qkf}. The implications of this on the PBH distribution and model building (\ie the required peak amplitude of $\mathcal{P}_{\mathcal{R}}$) is not known and subject to future research. We therefore would like to emphasize that in the strong turn examples we discussed in this section (see Fig. \ref{fig:PSBaMig} and \ref{fig:PSst}), the peak amplitudes we provided are not chosen guided by a particular bias on the non-Gaussianity present in these models. 

\section{Outlook}\label{sec_out}

Primordial black holes (PBHs), if they exist, can shed light
on long-standing questions on the nature of dark
matter, and on the mechanisms driving cosmic inflation. 
They can provide distinctive sources of gravitational
waves, potentially detectable with current or forthcoming
gravitational wave experiments. For these reasons, the
physics of PBH  offer promising opportunities of collaboration between  cosmologists, astronomers, and gravitational wave scientists. 

\smallskip
In this review, we focused on theoretical  aspects
of PBH inflationary model building. 
We learned that generating PBH  from inflation is hard, but possible. We reviewed  conceptual ideas
for amplifying the primordial curvature spectrum
at a level sufficient to trigger black hole formation. These mechanisms find realizations within single-field inflation, through a conversion of pronounced  gradients  of homogeneous quantities  into curvature perturbations or within multiple fields inflation, where curvature fluctuations are instead amplified through appropriate couplings with additional sectors, which are 
characterized by 
tachyonic instabilities. The
required tuning on model parameters, or the
degree of model sophistication for  realizing
these ideas can be  demanding. But it is certainly 
a worthwhile effort, given that experimental
probes of PBH are   sensitive 
to the details  of the curvature power spectrum, for example, through the properties of the resulting PBH population, or through    an induced stochastic gravitational wave background sourced by curvature fluctuations.  In fact,
different categories of models lead to distinct, potentially
distinguishable predictions for the statistics of primordial
fluctuations.
Hence, a  detailed  analysis relating theoretical
scenarios with cosmological and gravitational wave probes offers
new  precise tests of inflationary mechanisms, complementary to   traditional ones associated with the  physics of the cosmic microwave background
and of the large-scale structures of our universe. 

\smallskip
Although much theoretical work has been done so far by the community, much more   is 
needed for further investigating and clarifying different aspects
of the physics of inflation leading to PBH. At the level of model building, it will be important
 to clarify and address  challenges associated with a severe tuning of model parameters, or with the dynamical stability of
 PBH models requiring  non-attractor, non--slow-roll phases of inflationary evolution. It will  also be crucial 
 to continue to characterize  the rich and subtle properties of primordial
 fluctuations in PBH models of inflation  with   large enhancements of the primordial power spectrum at small scales. The to-do list includes 
 a deeper analytic understanding of the properties of the curvature spectrum profile, as well as 
 non-linearities and  non-Gaussianities and their consequences for observable quantities.
 Such theoretical analysis will have  relevant ramifications
 for designing appropriate cosmological probes of the physics of PBH. Hopefully, a concerted effort of theory and experiments, motivated by a deeper understanding of PBH physics, will   allow us to set new bounds, or possibly make new discoveries, on the mechanisms driving inflation and on the nature of dark matter.

\acknowledgments
We would like to thank Cristiano Germani and Shi Pi for discussions and comments. O\"O is  supported by the “Juan de la Cierva” fellowship IJC2020-045803-I,  by the European Structural and Investment Funds and the Czech Ministry
of Education, Youth and Sports (Project CoGraDS-CZ.02.1.01/0.0/0.0/15003/0000437),  and by the Spanish Research Agency (Agencia Estatal de Investigación) 
 through the Grant IFT Centro de Excelencia Severo Ochoa No CEX2020-001007-S, funded by MCIN/AEI/10.13039/501100011033.  GT is partially funded by the STFC grant ST/T000813/1.  For the purpose of open access, the authors have applied a Creative Commons Attribution licence to any Author Accepted Manuscript version arising.

\begin{appendix}
\section{Background Cosmology: Mini-Review}\label{AppA}

\noindent{\bf Space-time metric.} The central premise in modern cosmology is that as we look at the Universe on large enough scales, it appears to be simpler and more uniform compared to the
small scales. In other words, if we focus on sufficiently large scales, clumpy regions like the distribution of galaxies appear to be isotropic and homogeneous. The large scale spatial homogeneity and isotropy of the universe has been tested by a variety of observations such as the Large Scale Structure (LSS) surveys \cite{Wu:1998ad,Yadav:2005vv} but perhaps the most important evidence supporting this claim comes from the almost uniform temperature of the CMB originating from different parts of the sky. To first approximation, we can therefore assume the Universe to be isotropic and homogeneous. The high degree of spatial symmetry uniquely determines the structure of space-time geometry where physical distances are measured by the so called Friedmann \cite{Friedmann:1924bb}, Lemaître \cite{Lemaitre:1933gd}, Robertson \cite{Robertson:1935zz} and Walker \cite{1937PLMS...42...90W} (FLRW) line element \footnote{Note that for flat spatial hyper-surfaces $K = 0$, we can define new coordinates by $x = r \cos \phi \sin \theta$, $y = r \sin \phi \sin \theta$ and $z = r \cos \theta$ to turn the metric into a commonly used form $\d s^2 = -\d t^2 + a^2(t) \delta_{ij} \d x^{i} \d x^{j}$ in the literature.},
\beq\label{flrw}
\mathrm{d} s^{2}=-\mathrm{d} t^{2}+a^{2}(t)\left(\frac{\mathrm{d} r^{2}}{1-K r^{2}}+r^{2} \mathrm{~d} \Omega_{2}\right),
\eeq
where $\mathrm{d} \Omega_{2}=\left(d \theta^{2}+\sin^{2}\theta\, \mathrm{d} \phi^{2}\right)$ is the line element on the two dimensional sphere $S_{2}$ and $K=$ $\{-1,0,1\}$ represents negative, zero and positive curvature of constant-time hyper-surfaces respectively. Note that the symmetries of the Universe allow us to describe the metric by just a single function of time $a(t)$ and a constant parameter $K$. The function $a(t)$ is called the scale factor which parametrizes the size of the spatial slices at a given moment in time and the Hubble rate $H(t)\equiv\dot{a}(t) / a(t)$ describes the speed of expansion at a given moment of time. Here, an expanding universe $H(t)>0$ corresponds to monotonically increasing scale factor $a(t)$.

\smallskip
\noindent{\bf Dynamics of the universe.} Dynamics of space-time is governed by the Einstein field equation,
\beq\label{Ee}
R_{\mu \nu}-\frac{1}{2} g_{\mu \nu} R=\frac{1}{M_{\mathrm{pl}}^{2}} T_{\mu \nu}, 
\eeq
where $R_{\mu \nu} \equiv \partial_{\lambda} \Gamma_{\mu \nu}^{\lambda}-\partial_{\nu} \Gamma_{\mu \lambda}^{\lambda}+\Gamma_{\lambda \rho}^{\lambda} \Gamma_{\mu \nu}^{\rho}-\Gamma_{\mu \lambda}^{\rho} \Gamma_{\nu \rho}^{\lambda}$ is the Ricci tensor build out of Christoffel symbols, $R \equiv R^{\mu}_{\mu} = g^{\mu\nu} R_{\mu\nu}$ is the Ricci scalar and in terms of the metric, Christoffel symbols is given by 
\beq
\Gamma_{\mu \nu}^{\sigma} \equiv \frac{1}{2} g^{\sigma \rho}\left(\partial_{\mu} g_{\rho \nu}+\partial_{\nu} g_{\mu \rho}-\partial_{\rho} g_{\mu \nu}\right).
\eeq
The Einstein equation in \eqref{Ee} relates the geometry of space-time on the \emph{l.h.s} to the matter content in the universe through the appearance of energy momentum tensor $T_{\mu\nu}$ on the \emph{r.h.s}. As in the case of the space-time metric, homogeneity and isotropy restrict the possible choices matter content and enforce the energy-momentum tensor to take the perfect fluid form,
\beq
\bar{T}_{\mu \nu}=(\bar{\rho}+\bar{P})\, U_{\mu}\, U_{\nu}+\bar{P}\, g_{\mu \nu},
\eeq
where $\bar{\rho}$ and $\bar{P}$ are the background energy density and the pressure in the rest frame of the fluid and $U_{\mu} = (-1,0,0,0)$ is its time like 4-velocity of the fluid relative to the observer. The evolution equation for the energy density deriving the expansion of the universe can be derived from the $\nu=0$ component of covariant conservation law of energy momentum tensor: 
\beq\label{ce}
\nabla_{\mu} \bar{T}^{\mu}_{\,\,\,\,0}=0\quad \Rightarrow \quad \dot{\bar{\rho}}+3\, \fr{\dot{a}}{a}\,(\bar{\rho}+\bar{P})=0. 
\eeq
On the other hand, using the FLRW metric \eqref{flrw}, Einstein field equations gives us information about how space-time reacts to the matter content in the universe through the Friedmann,
\beq\label{fe}
H^{2}=\frac{\bar{\rho}}{3 \Mpl^{2}}-\frac{K}{a^{2}},
\eeq
and acceleration equation,
\beq\label{ae}
\frac{\ddot{a}}{a}=-\frac{1}{6 \Mpl^{2}}(\bar{\rho}+3 \bar{P}).
\eeq
The equations \eqref{ce}-\eqref{ae} are the key equations in determining the evolution of the universe and its constituents at large scales, namely $a(t)$ and $\bar{\rho}(t),\bar{P}(t)$ (assuming knowledge on spatial curvature $K$). An important aspect of these equations is that they are not independent, for example it is possible to combine the last two to obtain the first one. In practice, this implies that we need another ingredient to solve for three variables $\bar{\rho} (t)$, $\bar{P} (t)$ and $a(t)$ (or equivalently $H(t)$). A quantity that comes to the rescue in this context is the equation of state (EoS) parameter which provides a linear relation between pressure and energy density of the fluid(s) constituting the universe:
\beq\label{P}
\bar{P}(t) = w\, \bar{\rho}(t), 
\eeq
where $w = {\rm constant}$ for each fluid contributing to the total pressure. Although this relation is not the most general form of $P(\rho)$ that is available to us, this parametrization is perfectly adequate in providing an accurate course grained description of our universe through most of its history. In particular, it provides a simple analytic control in determining the dynamics of the different components of the universe and the evolution of the Hubble parameter (scale factor). In order to describe the history of the universe in a continuous manner, we typically consider multiple fluids that contributes to the energy density of the universe while satisfying the relation \eqref{P}. For example labeling each EoS by $w_i$ and assuming a single component dominates the energy density for a given moment of time in the universe, one can easily integrate \eqref{ce} to obtain
\beq\label{rhoev}
\bar{\rho}_i(t) \propto a(t)^{-3(1+w_i)}.
\eeq 
\noindent{\bf Cosmological inventory.} We can classify different sources of cosmological evolution by their contribution to the pressure: (i) For relativistic gas of particles (radiation) pressure is about one third of energy density, with $w_r = 1/3$. Popular constituents of such a fluid are photons, neutrinos, gravitons.  (ii) Non-relativistic pressureless dust (matter) $w_m = 0$ such as dark matter (\eg PBHs) and baryons, (iii) Cosmological constant $w_\Lambda = -1$ such as vacuum energy. From Eq. \eqref{rhoev}, energy density of each of these fluids therefore evolve as $\bar{\rho}_{r} \propto a^{-4}, \,\, \bar{\rho}_{m} \propto a^{-3}, \,\, \bar{\rho}_{\Lambda} \propto a^{0}$. Therefore in the future, cosmological constant will dominate, while in the far past $a \to 0$ the universe was radiation dominated and in between these two stages there is a period of matter dominated stage. Due to inadequacy of explaining initial conditions of the observed CMB anisotropies, cosmologist tend to complete the picture we described above with a very early phase of accelerated expansion $\ddot{a} > 0$ called inflation (see \eg \cite{Baumann:2009ds} for a comprehensive review). During this phase, EoS parameter also closely mimics the behavior of a cosmological constant $w_{\rm inf} \simeq -1$. More precisely, the departure of the EoS from the value $-1$ during inflation is characterized by a time dependent slow-roll parameter $\epsilon(t) = - \dot{H}/H^2 \ll 1$ where $w_{\rm inf} = -1 + 2\epsilon(t)/3$. In this framework, during inflation $\epsilon \ll 1$ while inflation terminates when $\epsilon = 1$. After this stage, it is generically assumed that the constituent(s) (\ie a scalar field or fields) that drives inflation is considered to decay to relativistic particles in a process called (p)reheating \cite{Kofman:1997yn,Amin:2014eta}. In this review, we will assume this process proceeds very efficiently so that soon after inflation ends the universe is filled with relativistic particles and consequently, evolves through the radiation dominated (RD), matter dominated (MD) and finally dark energy dominated (DED) phases. On the other hand, we refer the full picture that arise by the inclusion of an early accelerated expansion as the \emph{inflationary universe} which is composed of following consequent phases: Inflation $\to$ RD $\to$ MD $\to$ DED.

\smallskip
\noindent{\bf Evolution of the comoving Hubble horizon.} As we described in the main text, the evolution of the comoving horizon is crucial to understand causal evolution of fluctuation modes in the inflationary universe. To understand its time evolution in the picture we described above,  we focus on its time derivative
\beq
\fr{\d}{\d t}  (aH)^{-1} = -\fr{1}{aH^2} (H^2 + \dot{H}) = - \fr{\ddot{a}}{a^2 H^2} = \fr{1}{a}\fr{(1+3w)}{2}
\eeq
where we used the acceleration equation \eqref{ae} together with Friedmann equation \eqref{fe} focusing on flat FLRW slicing $K = 0$.\footnote{Indeed, CMB data informs us that spatial geometry of our universe is flat on large cosmological scales (see \eqref{PP}). On the other hand, since it dilutes slowly $\propto a^{-2}$ compared to the radiation and matter, we would expect it dominate the energy density before DED but this did not happen. This implies that initial value of the curvature must either be tuned to be extremely small or it should relax to small values through a dynamical mechanism. Inflation could be also a solution to this puzzle, because during such an exponential expansion any initially large curvature would be inflated away.} The expression above can be shaped into a form suitable for integration as $\d \ln ((a H)^{-1}) = [(1+3w)/2 ]\,\d(\ln a)$ which immediately gives 
\beq
(aH)^{-1} = H_0^{-1}\, a^{(1+3w)/2}, 
\eeq
where we normalized the scale factor today as $a_0 = 1$. Notice that during inflation $w_{\rm inf} \simeq -1$, comoving horizon decreases while during the subsequent RD ($w_r = 1/3$), MD ($w_m = 0$) phases it grows with scale factor. 

\smallskip
\noindent{\bf Density parameters $\Omega$.} To describe different energy components in the universe, cosmologist often parametrize radiation, matter and dark energy density relative to the critical energy density of spatially flat hyper-surfaces using the following definitions (and dropping the over-bar notation to describe background quantities)
\beq
\rho_{_{c,0}} \equiv 3 H_0^2 \Mpl^2 \quad \rightarrow \quad \Omega_{i} \equiv \fr{{\rho}_{_{i,0}}}{{\rho}_{_{c,0}}}
\eeq
with subscript ``$0$" denoting quantities evaluated today and ``$i$" labels different kinds of fluids. Focusing on the post-inflationary era, we can then re-write the Friedmann equation in terms of dimensionless density parameters as 
\beq
3H(a)^2 \Mpl^2 = \rho_{r,0}\, a^{-4} + \rho_{m,0}\, a^{-3} + \rho_{\Lambda, 0}\,\,  \rightarrow \,\,\frac{H^{2}(a)}{H_{0}^{2}}=\Omega_{r}\, a^{-4}+\Omega_{m}\, a^{-3}\, +\, \Omega_{\Lambda}.
\eeq
From the latest observations of the CMB anisotropies by the Planck collaboration, matter and dark energy density is determined to be \cite{Planck:2018vyg},
\beq\label{omMandL}
 \Omega_{\Lambda} = 0.6847 \pm 0.0073, \quad \Omega_m = 0.3153 \pm 0.0073. 
\eeq
On the other hand, we can infer radiation density today by utilizing the transition time of the universe from RD to MD. This moment in our universe is commonly referred to as matter-radiation equality which is defined by 
\beq
\rho_{r} (a_{\rm eq}) = \rho_{m} (a_{\rm eq})\quad \Rightarrow \quad a_{\rm eq} = \fr{\Omega_{r}}{\Omega_{m}}. 
\eeq
Noting the relation between the red-shift parameter \footnote{The red-shift parameter can be defined as the fractional shift in the physical wavelength $\lambda$ of a photon emitted at a distant point and time $t$ in the universe until today, \ie $z(t) \equiv (\lambda_0 - \lambda(t))/ \lambda(t) = a(t_0)/a(t) - 1 = 1/a(t) -1$.} $z(t)$ and scale factor $a(t) = (1+z(t))^{-1}$ and Planck's prediction  $z_{\rm eq} = 3402 \pm 26$ \cite{Planck:2018vyg} together with Eq. \eqref{omMandL} therefore gives
\beq\label{omR}
\Omega_r \simeq 9.26535 \times 10^{-5},
\eeq
where we used the central values of the quantities $\Omega_m, z_{\rm eq}$ determined by Planck. 

Making a two component fluid approximation around the time of matter radiation equality, $a \approx a_{\rm eq}$, the total energy density of the universe at $a = a_{\rm eq}$ can be related to matter density today via 
\beq
\rho(a_{\rm eq}) \simeq \fr{\rho_{r,0}}{a_{\rm eq}^{4}} + \fr{\rho_{m,0}}{a_{\rm eq}^{3}} \simeq 2\,\, \fr{\rho_{m,0}}{a_{\rm eq}^{3}}.
\eeq

\smallskip
\noindent{\bf Cosmological parameters.} Other cosmological parameters determined by the latest Planck 2018 data (TT,TE,EE + low E + lensing, \% 68 CL) are as follows \cite{Planck:2018vyg}: 
\begin{align}\label{PP}
\nn \Omega_{K} &= 0.001 \pm 0.002, \quad\quad H_0 = 67.36 \pm 0.54\, [\rm km\, s^{-1} Mpc^{-1}],\\
& \Omega_m h^2 = 0.1430 \pm 0.0011, \quad\quad k_{\rm eq} = 0.010384 \pm 0.000081\,\, [\rm Mpc^{-1}].
\end{align}
So far we did not mention thermodynamical properties such as temperature and entropy in the universe following inflation. In what follows, we will note some of the key formulas we will use in Section \ref{S2} without getting into the details on the derivation of these formulas. For an in depth discussion on the contents we will introduce below, we refer the reader to chapter 3 of the seminal books by Kolb and Turner \cite{Kolb:1990vq} and Mukhanov \cite{Mukhanov:2005sc} or to Baumann's recent book \cite{Baumann:2022mni}. 

The total energy density of relativistic degrees of freedom that are in thermal equilibrium with the plasma, can be related to the temperature of the plasma $T$ (\ie temperature of the photon gas) as
\beq
\rho_r  = \fr{\pi^2}{30} g_* (T)\, T^4,
\eeq
where $g_* (T)$ is the effective number of relativistic degrees of freedom at the temperature $T$ which is defined as 
\beq
g_{*}(T)=\sum_{\rm bos.} g_{b}+\frac{7}{8} \sum_{\rm fermi.} g_{f},
\eeq
with $g_b$ and $g_f$ denoting the intrinsic degree of freedom (\ie spin) for bosonic and fermionic species, respectively. The total entropy density of relativistic species is defined as
\beq
s(T) \equiv \fr{2\pi^2}{45} g_s (T)\, T^3,
\eeq
where $g_s$ is the effective number of degrees of freedom in entropy. For species in thermal equilibrium (\ie species that have the same temperature $T_{\rm b} = T_{\rm f} =T$), $g_s (T) = g_* (T)$ because for each bosonic/fermionic degree of freedom, entropy density is defined to be $s \equiv (\rho + P) / T$ with a common denominator. 

A very useful formula can be derived by using the conservation of total entropy in the universe, $S = s V \propto s a^3 \simeq constant.$ which can be utilized to relate the scale factor and the temperature of the plasma as
\beq
g_s (T)\, T^3\, a^3 \simeq constant. \quad \Longrightarrow \quad \fr{a(t_1)}{a(t_2)} = \fr{T_2}{T_1} \left(\fr{g_s (T_2)}{g_s(T_1)}\right)^{1/3}.
\eeq
\section{A simple estimate for the threshold of collapse}\label{AppB}
In this short appendix, we present a simple analytic estimate on the characteristic value of collapse threshold for PBH formation during RDU based on the Jeans length argument in Newtonian gravity as first discussed in \cite{Carr:1975qj}, closely following \cite{Sasaki:2018dmp,Franciolini:2021nvv}. For this purpose, we take the background space-time after inflation to have the spatially flat ($K = 0$) FLRW form \eqref{flrw} and so the evolution of the scale factor described by the Friedmann equation: $3H^2 \Mpl^2 = \bar{\rho} (t)$. Now consider a locally perturbed, spherically symmetric region in the universe that is initially outside the horizon and could eventually collapse to form a PBH upon horizon re-entry. The metric describing such a region can be written as
\beq\label{flrwf}
\mathrm{d} s^{2} =-\mathrm{d} t^{2}+a^{2}(t)\, e^{-2 \mathcal{R}(\hat{r})}\left[\mathrm{d} \hat{r}^{2}+\hat{r}^{2} \mathrm{~d} \Omega_2\right] \\
\eeq
where $a(t)$ is the scale factor and $\mathcal{R} < 0$ is the non-linear generalization of the conserved comoving curvature perturbation defined on a super-Hubble scales \cite{Lyth:2004gb}. At large distances $\hat{r} \to \infty$, curvature perturbation assumed to vanish ($\mathcal{R}\to 0$) so that the universe is described by the spatially flat FLRW metric. By making a coordinate redefinition, $r = \hat{r} e^{-\mathcal{R}(\hat{r})}$, the metric describing the spherical perturbed region \eqref{flrwf} can be transformed into the one describing a closed universe with positive spatial curvature (as in \eqref{flrw}),
\beq
\d s^2=-\mathrm{d} t^{2}+a^{2}(t)\left(\frac{\mathrm{d} r^{2}}{1-K(r)\, r^{2}}+r^{2} \mathrm{~d} \Omega_{2}\right),
\eeq
where the relation between perturbations of the two metric is given by $ K(r)\, r^2 = \hat{r} \mathcal{R}'(\hat{r})(2 - \hat{r} \mathcal{R}'(\hat{r}))$ showing that the local spatial curvature is given in terms of the first order spatial (leading order) derivatives of the curvature perturbation $\mathcal{R}(\hat{r})$. Ignoring higher order spatial derivatives of $K$ ($\ie K' \sim \mathcal{R}''$) on sufficiently large scales, the evolution of the spherical region is given by the 00 component of the Einstein equations \eqref{Ee} 
\beq\label{Hpbh}
H^{2} = \frac{\rho_{\rm tot}}{3\Mpl^2}-\frac{K(r)}{a^{2}} 
\eeq
which is equivalent to the evolution of a closed universe (see \eg \eqref{fe}) with a perturbation $\delta \rho$ induced by spatial curvature $K(r)$: 
\beq\label{ddef}
 \frac{\delta \rho}{{\rho}} \equiv \frac{\rho_{\rm tot} - {\rho}}{{\rho}}  = \frac{3\Mpl^2 K}{\bar{\rho} a^2} = \frac{K}{H^{2} a^{2}}, 
\eeq
where we make use of  $\bar{\rho}(t) = 3H^2 \Mpl^2$. From \eqref{ddef}, since $\bar{\rho} \propto a^{-4}$ during RDU, perturbations are initially $a \to 0$ small, $\delta \rho/\rho \ll 1$. As the universe evolves, a local spherical region with $K > 0$ will eventually stop expanding and collapse to form a PBH. This happens precisely when the right hand side of \eqref{Hpbh} becomes negative, \ie at $t = t_{\rm c}$ where $3\Mpl^2 K = a^2 \rho_{\rm tot}$ corresponding to $\delta \rho /\rho = 1$ in \eqref{ddef} assuming a homogeneous and isotropic universe. In RDU, since perturbation modes that have a length scale smaller than the Jeans Length ($k_J^{-1} \equiv  c_s / a H $) cannot collapse, the smallest comoving scale that can undergo collapse at $t = t_{\rm c}$ is given by $k^2 = (a H)^2 /c_s^2$ where $c_s$ is the speed  propagation of perturbation. Therefore, we have 
\beq
\fr{\delta \rho}{\rho}(t_c) = \fr{K}{k^2} \fr{k^2}{a^2 H^2} = \fr{K}{c_s^2 k^2} = 1, 
\eeq
which suggests us to identify $K = c_s^2 k^2$. Therefore, at the time of horizon re-entry ($k = a_{\rm f} H_{\rm f}$), the perturbations relevant for PBH formation should have a density contrast larger than
\beq
\left(\fr{\delta \rho}{\rho}(t_{\rm f})\right)_{\rm c}=\frac{K}{\left(a_{\rm f} H_{\rm f}\right)^2}= c_{s}^{2} \left(\frac{k}{a_{\rm f} H_{\rm f}} \right)^2=c_{s}^{2} = w.
\eeq
\section{Solving the Mukhanov-Sasaki equation: Numerical procedure}\label{AppC}
In this Appendix, we discuss  important steps  for the numerical evaluation of the Mukhanov-Sasaki (MS) equation. The aim is to obtain the power spectrum of curvature perturbation in the inflationary scenarios we discussed in Sections \ref{s3p2} and \ref{s3p3}. Note that scenarios discussed in those sections can be captured by the generalized MS equation \eqref{MS} utilizing \eqref{zppoz} and \eqref{bds} as we describe below.  

We start by scaling away the highly oscillatory contributions from the Bunch Davies (BD) initial conditions \eqref{bds}. For this purpose, we define a dimensionless variable $\bar{v}_k$ through 
\beq
v_k(\bar{\tau}) = \fr{\bar{v}_k(\bar{\tau})}{\sqrt{2k}}\,\, e^{-i k\bar{\tau}},
\eeq
so 
to rewrite the MS equation \eqref{MS} as 
\beq\label{MSm}
\bar{v}_k''(\bar{\tau}) - 2 i k\, \bar{v}_k'(\bar{\tau}) - \fr{z''}{z}\bar{v}_k(\bar{\tau}) = 0.
\eeq
For mode by mode evaluation, the e-folds $N$ is a more suitable  variable for numerical integration so we turn the derivatives w.r.t $\bar{\tau}$ in \eqref{MSm} to e-folds using $\d N = ({aH}/{c_s}) \d \bar{\tau}$, and  obtain 
\beq\label{MSn}
\fr{\d^2\, \bar{v}_k}{\d N^2} + \left(\Big[1 - \epsilon-s\Big]- 2i\,\fr{c_s k }{aH}\right)\fr{\d\, \bar{v}_k}{\d N} - \fr{c_s^2}{(aH)^2}\fr{z''}{z}\bar{v}_k = 0\,.
\eeq
We parametrize the scale factor and Hubble rate as
\beq 
a(N) = a_{\rm end}\, e^{N-60},\quad\quad H(N) = H_{\rm end}\, e^{-\int_{60}^{N} \epsilon(N') \d N'}\,,
\eeq
and note that
\beq\label{zppozm}
\fr{z''}{z} = \left(\fr{a H}{c_s}\right)^2 \left[(1 - \epsilon - s)\left(1 + \fr{\eta - s +\mu}{2}\right) + \left(1 + \fr{\eta - s +\mu}{2}\right)^2 + \fr{1}{2}\left(\fr{\d \eta}{\d N}-\fr{\d s}{\d N} + \fr{\d \mu}{\d N}\right) \right]. 
\eeq
In terms of the new variable, the Bunch-Davies initial conditions simplifies considerably and can described as 
\beq\label{BDm}
\bar{v}_k(N) \big|_{\rm in} = 1, \quad\quad\quad \bar{v}_k'(N) \big|_{\rm in} = 0.
\eeq
Provided that the background evolution (\ie $\epsilon, \eta, \mu$ etc) is known in terms of e-fold number, the MS equation can be solved numerically for each $k$ mode, in terms of the rescaled variable \eqref{MSn},  using the initial conditions \eqref{BDm}. In this respect, some care must be taken for the initialization time of individual modes because deep inside the horizon $k^2 \gg z''/z$, the solution to \eqref{MSn} is highly oscillatory which would be costly for the numerical computation. Typically, it is enough to initialize the modes at some time such that they are sufficiently inside the horizon. For this purpose, we choose to evolve each mode by setting the initial time as 
\beq\label{Ninit}
N^{(k)}_{\rm in} = N^{(k)}_{0} - 4,
\eeq
where the $``k"$ super-script indicates the intrinsic mode dependence for the choice of $N_{\rm in}$ and $N^{(k)}_0$ denotes the horizon crossing time \footnote{Unless modes undergo resonance and get excited deep inside the horizon, the choice of initial conditions in Eqs. \eqref{Ninit} and Eq. \eqref{BDm} provide an accurate prescription for the initialization of the numerical evaluation. For the models we consider in Sections \ref{s3p2} and \ref{s3p3} this is indeed the case. For a model that leads to excited states inside the horizon see \cite{Flauger:2009ab,Flauger:2010ja}.}. As argued in the main text (see the discussion around \eqref{mt}), the latter can be obtained via
\beq
k^2 = \fr{z''}{2 z} \bigg|_{N^{(k)}_0} 
\eeq
using \eqref{zppozm} (or equivalently \eqref{zppoz}) provided the background solution is known. Having obtained the individual mode evaluation from $N^{(k)}_{\rm in}$ to $N_{\rm end} =60$, the power spectrum \eqref{psf} can be described as
\beq
\mathcal{P}_{\mathcal{R}}(k,N_{\rm end}) = \fr{H_{\rm end}^2}{8\pi^2 \epsilon_{\rm end} c_{s,\rm end}\Mpl^2} \left(\fr{{c_{s,\rm end}}}{\tilde{M}_{\rm end}}\right)^2 \left(\fr{k}{k_{\rm end}}\right)^2 \big|\bar{v}_k(N_{\rm end})\big|^2,
\eeq
where we defined $\tilde{M} \equiv M(N)/\Mpl$, while $k_{\rm end} = a_{\rm end} H_{\rm end}$ is the mode that exits the horizon at the end of inflation. 

In order to set the overall normalization of the power spectrum we need to determine $H_{\rm end}$ w.r.t $\Mpl$. We do so by requiring the normalization of the power spectrum indicated by Planck at the pivot scale $k_{\rm cmb} = 0.05\, {\rm Mpc}^{-1}$ using $\mathcal{P}_{\mathcal{R}} (k_{\rm cmb}, N_{\rm end}) \simeq 2.1 \times 10^{-9}$. Notice that, from the general formulas we presented above, canonical single-field scenarios (see Section \ref{s3p2}) can be recovered by making the following replacements $c_s \to 1, s \to 0$, $M \to \Mpl, \mu \to 0$ and $\d \bar{\tau} \to \d \tau$ and $\d N = (a H)\,\d \tau$. A pedagogical Python notebook file that calculates the power spectrum using the prescription described above can be found through the \href{https://github.com/oozsoy/SingleFieldINF_Powerspec_PBH}{github} link. 

\smallskip
\noindent{\bf The role of decaying modes in the canonical single-field scenarios.}
We can now verify if the general expression \eqref{cpgs} provides an accurate description for the enhancement of the power spectrum in the canonical single-field model  discussed in Section \ref{s3p2}. For this purpose, we assume that  the leading solution to the curvature perturbation at super-horizon scales is given by the constant solution at horizon crossing $\mathcal{R}_k(\tau) = \mathcal{R}^{(0)}_k$ and plugging it into the last term in \eqref{cpgs}, we can then generate an iterative solution for the curvature perturbation as 
\beq\label{cpgec}
\mathcal{R}_k(N_{\rm end})=\mathcal{R}^{(0)}_{k}\left[1+v_{\mathcal{R}}(k) \bigintsss_{N_0}^{N_{\rm end}} \fr{\d N'}{\tilde{z}^{2}\,(\tilde{a}\tilde{H})}- k^{2} \bigintsss_{N_{0}}^{N_{\rm end}} \fr{\d N'}{\tilde{z}^{2}\, (\tilde{a}\tilde{H})} \bigintsss_{N_{0}}^{N'} \mathrm{d} N'' \,\fr{\tilde{z}^{2}}{(\tilde{a}\tilde{H})} \right],
\eeq
where we switched to e-folds as time variable, $v_{\mathcal{R}} \equiv \mathcal{R}_k'(N_0)/\mathcal{R}_k(N_0)$ is the fractional velocity of curvature perturbation at horizon crossing epoch $N_0$, and tilde over quantities denotes a normalization with respect to their values at $N_0$, \ie $\tilde{X} \equiv X(N)/\tilde{X}(N_0)$. Using the standard solution to the curvature perturbation during the initial slow-roll era, an analytic formula for the fractional velocity of curvature perturbation is obtained \cite{Ozsoy:2019lyy}, valid  for modes that leave the horizon during the initial slow-roll era:
\beq\label{vr}
v_{\mathcal{R}}(k) = - \fr{x_0^2}{1 + x_0^2} - i \fr{x_0^3}{1+x_0^2}\,\,, \quad\quad\quad \textrm{with}\,\,\quad\quad x_0 \equiv \fr{k}{a_0 H_0}.
\eeq
Similarly, for modes that crosses the horizon during the initial slow-roll era, the amplitude of the curvature perturbation at horizon crossing results 
\beq\label{ampsq}
\big|\mathcal{R}^{(0)}_k\big|^2 = \fr{H_0^2}{8\pi^2 \epsilon_0 \Mpl^2}\left(1 + \fr{k^2}{(a_0H_0)^2}\right). 
\eeq
Combining these results together with a given background evolution (\ie $\tilde{z},\tilde{a}\tilde{H}$), the power spectrum of curvature perturbation for modes that leaves the horizon can be described via the definition 
\beq\label{c13}
\mathcal{P}_{\mathcal{R}}(k,N_{\rm end}) = \fr{k^3}{2\pi^2} \big|\mathcal{R}_k(N_{\rm end})\big|^2\,,
\eeq
using \eqref{cpgec}, \eqref{vr} and \eqref{ampsq}. For a set of modes that exit the horizon during Phase 1 (the initial slow-roll era), the amplitude of power spectrum obtained using the procedure we outlined  is shown by blue dots in Fig. \ref{fig:psc}. 
\section{Details of the axion-gauge field dynamics}\label{AppD}
In this appendix, we focus on the dynamics of the gauge fields by the presence rolling scalar $\chi(t)$ and the influence of this dynamics on the scalar perturbations for the models we discuss in Sections \ref{s4p1p1}, \ref{s4p1p2} and \ref{s4p1p3}. The part of the action \eqref{LAG} relevant for this purpose is given by
\beq\label{SGF}
S_{\rm GF}=\int \mathrm{d}^4 x \sqrt{-g}\left[-\frac{1}{4} F_{\mu \nu} F^{\mu \nu}-\frac{g_{\rm cs}\, \chi}{8 f}  \frac{\eta^{\mu \nu \rho \sigma}}{\sqrt{-g}} F_{\mu \nu} F_{\rho \sigma}\right],
\eeq
where using the definition of the field strength tensor $F_{\mu\nu} = \partial_\mu A_\nu - \partial_\nu A_\mu$ and the totally antisymmetric nature of the symbol $\eta^{\mu\nu\rho\sigma}$, we note the following identities 
\begin{align}
    F_{\mu\nu}F^{\mu\nu} &= 2\,(g^{\mu\rho}g^{\nu\sigma} - g^{\nu\rho}g^{\mu\sigma})\,\partial_\mu A_\nu\, \partial_\rho A_\sigma,\\
    \eta^{\mu\nu\rho\sigma} F_{\mu\nu} F_{\rho\sigma} &= 4\, \eta^{\mu\nu\rho\sigma}\, \partial_\mu A_\nu\, \partial_\rho A_\sigma.
\end{align}
Notice that apart from the gauge fields, these expressions involve the metric. Therefore, it is convenient to characterize the metric fluctuations in its most general form using ADM decomposition as
\beq
\mathrm{d} s^2=-N^2 \mathrm{~d} t^2+\hat{g}_{i j}\left(\mathrm{~d} x^i+N^i \mathrm{~d} t\right)\left(\mathrm{d} x^j+N^j \mathrm{~d} t\right),
\eeq
where $\hat{g}_{ij}$ is the spatial 3-metric on constant time hyper-surfaces, $N$ is the lapse function and $N^{i}$ is the shift vector. In terms of these variables, the component of the metric with the upper indices can be expressed as 
\beq
g^{00} = -\frac{1}{N^2},\quad g^{0i} = g^{i0} = \frac{N^{i}}{N^2}, \quad g^{ij} = \hat{g}^{ij} - \frac{N^{i}N^{j}}{N^2},
\eeq
where $\hat{g}^{ij}$ is the inverse of the spatial metric. 
Including the Einstein-Hilbert term $\mathcal{L}_{\rm EH} = \Mpl^2 R / 2$ to the action \eqref{SGF} and the matter action associated with the scalar field(s) $\chi$, one can show that the second order terms that include $N,N^{i}$ do not contain more than one time derivative, implying that they can be identified as Lagrange multipliers to be solved in terms of the physical fluctuations of $\chi$ and $A_\mu$ \cite{Maldacena:2002vr,Ozsoy:2017blg}.

To characterize the dynamics of gauge fields we assume vector fields start linear order in perturbations (so that they do not exhibit a time dependent background vev $\langle \bar{A}_\mu(t) \rangle = 0$) and adopt the Coulomb gauge \footnote{We note that at linear order in fluctuations, the Coulomb gauge condition $\partial_i A_i = 0$ is equivalent to setting the temporal component of the gauge field to zero, $A_0 = 0$. This can be seen by explicitly expanding the gauge field action \eqref{SGF} to second order in $A_0$ and $A_i$ and then solving for the non-dynamical $A_0$ mode (see also the discussion in Section 3 of \cite{Adshead:2015pva}).} $\partial_i A_i = 0$ in the gauge sector along with the flat gauge choice in the scalar - gravitational sector
\begin{align}
\nn \chi(t, \vec{x}) &=\bar{\chi}(t)+\delta \chi(t, \vec{x}), \\
\nn \hat{g}_{i j} &=a^2(t)\left[\,\delta_{i j}+h_{i j}\,\right], \\
\partial_i A_i &=0,
\end{align}
where $h_{ij}$ is the transverse, traceless tensor fluctuation of the metric, $\partial_i h_{ij} = h_{ii} = 0$. In this appendix, we will present the dynamics of the gauge fields by expanding the action \eqref{SGF} up to third order in fluctuations. Noting that the shift vector is order one $N^{i}$ in perturbations and expanding the lapse $N = 1 + \delta N$, we compile all the information above to obtain the second and third order action involving gauge field fluctuations as
\begin{align}
\label{S2GF}&S_{\rm GF}^{(2)}=\int \mathrm{d}^4 x \,a^3\,\left\{\frac{1}{2 a^2} \dot{A}_i \dot{A}_i-\frac{1}{2 a^4} \partial_j A_i\, \partial_j A_i+\frac{g_{\rm cs}\,\dot{\bar{\chi}}}{a^3 f}\, \epsilon_{i j k}\, A_i\, \partial_j A_k \right\},\\
\label{S31}&S^{(3,1)}_{\rm GF}=\int \mathrm{d}^4 x \,a^3\,\left\{g_{\rm cs}\frac{\delta \chi}{f} \,\left[\vec{E}.\vec{B} -{\langle \vec{E}.\vec{B}\rangle}\right]\right\} \\
\label{S32}&S^{(3,2)}_{\rm GF}=\int \mathrm{d}^4 x \,a^3\,\left\{-\frac{\delta N}{2} \left[\vec{E}^2+\vec{B}^2 - {\langle\vec{E}^2+\vec{B}^2\rangle}\right]-\frac{N^i}{a^2} \dot{A}_j F_{i j}\right\} \\
\label{S33}&S^{(3,3)}_{\rm GF}=\int \mathrm{d}^4 x \,a^3\,\left\{-\frac{h_{i j}}{2}\bigg[E_i E_j + B_i B_j\bigg]\right\},
\end{align}
where we introduced electric and magnetic field notation using $E_i = - \dot{A}_i/a$, $B_i = \epsilon_{ijk}\partial_j A_k/a^2$. Note that we artificially introduced tadpole terms proportional to the expectation values involving $\vec{E}$ and $\vec{B}$ fields in the expressions \eqref{S31} and \eqref{S32}. Later we will subtract these terms in the action describing the inflation (gravity plus inflaton sector) in \eqref{Sinf1} to consistently take into account the modifications that might arise to the background equations of the scalar field(s) $\chi = \{\phi,\sigma\}$ and the Friedmann equation (see \eg Eq. \eqref{mkgafe}).

The action labeled by $S^{(3,2)}_{\rm GF}$ parametrizes the cubic interactions (see \eg the first and the third term in \eqref{S32}) induced by the gravitational fluctuations. The influence of these terms on the observable scalar sector has been studied in \cite{Barnaby:2011vw, Barnaby:2012xt} with the conclusion that they can be neglected compared to the direct interaction term in \eqref{S31}. This is because the vertices associated with the gravitationally induced cubic terms in \eqref{S32} is suppressed  $\mathcal{L}^{(3,2)}_{\rm GF} \propto (\sqrt{\epsilon}/\Mpl)\, \delta \chi \, \mathcal{O}(A^2)$ with respect to the one in \eqref{S31} unless $f \simeq \Mpl$ and $g_{\rm cs} \ll 1$ \footnote{This regime is not interesting from a phenomenological point of view as the gauge field production will be very weak in this case.} and therefore can be safely ignored compared to the latter. Finally, the cubic action in \eqref{S33} parametrizes the influence of the gauge fields on the tensor part of the metric. In this review, we will not study these effects. For the impact of gauge field production on the tensor fluctuations during inflation and its interesting parity violating effects, see \cite{Sorbo:2011rz,Cook:2011hg,Namba:2015gja,Ozsoy:2020ccy, Ozsoy:2020kat,Ozsoy:2021onx,Campeti:2022acx} and the references therein. 

To summarize, in the presence of rolling axion-like fields during inflation and the interaction $\mathcal{L}_{\rm int} \propto \chi F \tilde{F}$, the dynamics of the gauge fields can be studied by focusing on the second order action $S^{(2)}_{\rm GF}$ \eqref{S2GF}. On the other hand, the influence of the gauge fields on the scalar sector depends on whether we identify $\chi$ as the inflaton (Sections \ref{s4p1p1} and \ref{s4p1p2}) or as a spectator scalar rolling during inflation (Section \ref{s4p1p3}). In the following, we will first introduce the basics of the gauge field production in the presence of a rolling scalar discussing the different cases we cover in this review. The nature of the subsequent sourcing of (visible sector) scalar fluctuations by the vector fields depends whether the scalar $\chi$ directly interacts with ${\rm U}(1)$ fields or not. We will cover each case following our discussion on the vector field production. 

\subsection{Gauge field production by rolling scalars}\label{AppD1}
To understand the gauge field production by a rolling scalar (either an inflaton or a spectator scalar), we focus on the second order action \eqref{S2GF} and decompose the gauge field into its helicity modes $\lambda = \pm$ in Fourier space using conformal time $\d \tau = \d t/a$ as
\beq\label{GFME}
 \hat{A}_i(\tau, \vec{x}) = \int \frac{\mathrm{d}^3 k}{(2 \pi)^{3 / 2}}\, \mathrm{e}^{i \vec{k} \cdot \vec{x}}\, \sum_{\lambda=\pm} \epsilon_i^{(\lambda)}(\vec{k})\, \hat{A}_\lambda (\tau, \vec{k}),
 \eeq
where the polarization vectors obey
\begin{align}\label{hv}
\nn &k_i\, \epsilon_i^{\pm}(\vec{k})=0,\quad\quad\quad\quad\quad \epsilon_{i j k}\, k_j\, \epsilon_k^{\pm}(\vec{k})=\mp i \,|\vec{k}|\, \epsilon_i^{\pm}(\vec{k}),\\& \epsilon_i^{\lambda}(\vec{k})\, \epsilon_i^{\lambda'}(\vec{k})^{*}=\delta^{\lambda\lambda'}, \quad\quad \epsilon_i^\lambda(\vec{k})^{*}=\epsilon_i^\lambda(-\vec{k})=\epsilon_i^{-\lambda}(\vec{k})
\end{align}
and we defined
\beq\label{Akhat}
\hat{A}_\lambda (\tau, \vec{k}) = \left[A_{\lambda}(\tau, k) \,a_{\lambda}(\vec{k}) + A^{*}_{\lambda}(\tau, k)\, a^{\dagger}_{\lambda}(-\vec{k})\right],
\eeq
which satisfies $\hat{A}_\lambda (\tau, \vec{k})^{\dagger} = \hat{A}_\lambda (\tau, -\vec{k})$ so that $\hat{A}_i(\tau,\vec{x})$ is a hermitian operator. Finally, annihilation and creation operators satisfy the standard commutation relations
\beq
\left[\hat{a}_\lambda(\vec{k}), \hat{a}_{\lambda'}^{\dagger}(\vec{k}')\right]=\delta_{\lambda \lambda^{\prime}} \delta(\vec{k}-\vec{k}^{\prime}).
\eeq
Plugging the decomposition in \eqref{S2GF} and varying the action, the mode functions of the gauge field can be shown to satisfy 
\beq\label{meqa}
A''_{\pm} + k^2\left( 1 \pm \frac{aH}{k} 2\xi\right) A_{\pm} = 0, \quad\quad \xi \equiv -\frac{g_{\rm cs}\,\dot{\bar{\chi}}}{2Hf}
\eeq
It is clear from Eq. \eqref{meqa} that the dispersion relation of the gauge fields are modified in the presence of the last term in \eqref{SGF}. More importantly, the negative helicity modes $A_{-}$ can experience an tachyonic instability for modes satisfying $k/(aH) < - \dot{\bar{\chi}}/(Hf)$ while positive helicity modes stay in their vacuum. These facts reflect the parity violating nature of the interaction $\mathcal{L}_{\rm int} \propto \chi F \tilde{F}$. The behavior of the solutions to the Eq. \eqref{meqa} is sensitive to the velocity profile $\dot{\bar{\chi}}$ of the rolling scalar. In what follows we will discuss the different cases and the corresponding solutions within the context of Sections \ref{s4p1p1}, \ref{s4p1p2} and \ref{s4p1p3}.
\smallskip

\noindent{\bf Production by a slowly-rolling scalar.} For a slowly rolling scalar field, we can treat $g_{\rm cs}\dot{\chi}/(Hf)$ as constant per Hubble time. In terms of the effective coupling $\xi$, this adiabaticity condition can be parametrized as
\beq
\frac{\dot{\xi}}{\xi H} = \frac{\ddot{\bar{\chi}}}{\dot{\bar{\chi}}H} - \frac{\dot{H}}{H^2} \ll 1.
\eeq
In this case, the solution to the Eq. \eqref{meqa} that reduces to the Bunch Davies vacuum solutions $A_\pm = \mathrm{e}^{-i k \tau}/\sqrt{2k}$ deep inside the horizon $k \gg aH$ can be written in terms of the Coulomb functions
\beq
A_{-}(\tau,k) \simeq \frac{1}{\sqrt{2k}} \left[G_0 (\xi, -k\tau) + iF_0(\xi, -k\tau)\right]
\eeq
where $\xi = -g_{\rm cs} \dot{\bar{\chi}}/(2Hf)$ and $-\tau = (aH)^{-1}$. Another simplification can be made focusing on the $\xi \gg -k\tau$ regime (as we will see, particle production is only efficient for $\xi \sim \mathcal{O}(1)$ and takes place as $-k\tau \to 0$)
\beq\label{solBessel}
A_{-}(\tau,k) \simeq \sqrt{\frac{-k\tau}{2k}} \left[2 \mathrm{e}^{\pi \xi} \pi^{-1/2} K_1(\sqrt{-8\xi k\tau}) + i \mathrm{e}^{-\pi \xi} \pi^{1/2} I_1(\sqrt{-8\xi k\tau}) \right],
\eeq
where $I_1$ and $K_1$ are modified Bessel functions of the first and second kind. From the solution above, one can realize that for the interesting case of $\xi \sim \mathcal{O}(1)$, field amplification occurs shortly after horizon crossing $-k \tau \sim \mathcal{O}(1)$. Therefore for a final simplification we can take the large argument limit of the Bessel functions in \eqref{solBessel} to get
\beq\label{srAsol}
A_{-}(\tau,k) \simeq \frac{1}{\sqrt{2k}} \left(\frac{-k\tau}{2\xi}\right)^{1/4} \mathrm{e}^{\pi\xi -2 \sqrt{-2\xi k\tau}}\,\left[1 + \frac{i}{2}\, \mathrm{e}^{-2\pi\xi + 4\sqrt{-2\xi k\tau}} \right], \quad\quad (8\xi)^{-1} \ll -k\tau < 2\xi.
\eeq
The real part of the solution \eqref{srAsol} is the growing solution as $-k\tau \to 0$ and encodes the physical amplification of the negative helicity mode by the presence of a slowly rolling axion-like field. Within the approximations we undertake, the imaginary part of the solution \eqref{srAsol} represents the UV divergent part as $-k\tau$ grows which should precede the vacuum solution $A_{-} = \mathrm{e}^{-i k \tau}/\sqrt{2k}$ deep inside the horizon. Therefore, ignoring the imaginary part typically amounts to throwing away the standard divergent (also present in flat space) peace of quantities like $\langle \vec{E}.\vec{B}\rangle$ and $\langle \vec{E}^2 + \vec{B}^2\rangle$ (see below). For a detailed discussion on these issues we refer the reader to \cite{Peloso:2016gqs} and the Appendices of \cite{Ozsoy:2020ccy}. 
\smallskip

\noindent{\bf Production by a transiently rolling scalar.} For the models we discuss in Sections \ref{s4p1p2} and \ref{s4p1p2}, the potential of the scalar $\chi$ has a feature around which the background velocity $\dot{\bar{\chi}}$ and the effective coupling $\xi$ in the equation of motion of the gauge field modes \eqref{meqa} have a transient peak with a maximal value $\xi_*$ at $\tau = \tau_*$. A peaked time dependent profile of $\xi = \xi(\tau)$ in turn translates into a scale dependent growth of the mode functions $A_{-}$ where only modes that are in the vicinity of the scale $k_* = a_* H_* = (-\tau_*)^{-1}$ that exits the horizon at $\tau = \tau_*$ are maximally amplified. For the time dependent profiles we study in Section \ref{s4p1p2} and \ref{s4p1p3}, it is hard to obtain a fully analytic solution describing the amplification of the gauge modes. However, an accurate description of the mode function at late times can be obtained employing WKB approximation methods supplemented with numerical analysis \cite{Namba:2015gja} as we mention below. In particular, at late times $\tau/\tau_* < 1$, the amplification of the mode functions can be parametrized in terms of a (real and positive) normalization factor as 
\begin{align}\label{trAsol}
A_{-}(\tau, k) &\simeq \fr{1}{\sqrt{2k}} \left(\fr{-k\tau}{2\xi(\tau)}\right)^{1/4} \, N_A(\xi_*,x_*,\delta)\, \mathrm{e}^{- 2E(\tau){\sqrt{-2\xi_* k \tau}}}\Bigg\{ 1 +  \fr{i\,\mathrm{e}^{4E(\tau){\sqrt{-2\xi_* k \tau}}}}{2N_A(\xi_*,x_*,\delta)^2}  \Bigg\}
\end{align}
where we defined $x_* = -k \tau_* = k/k_*$ and $E(\tau)$ is a time dependent function that asymptotes to zero at late times $\tau/\tau_* \to 0$ whose functional form depends on the model under consideration. For example for the bumpy axion inflation of Section \ref{s4p1p2} and its cousin spectator model (see Section \ref{s4p1p3}), it is given by $E(\tau) = 1/(\delta |\ln(\tau/\tau_*)|)$. On the other hand, for the transiently rolling axion model with the standard cosine potential one gets $E(\tau) = \sqrt{2}(\tau/\tau_*)^{\delta/2}/(1+\delta)$. Notice that the solution \eqref{trAsol} reduces to \eqref{srAsol} of constant $\xi$ if we make the following  replacements $E(\tau) \to 1$ and $N_A \to \mathrm{e}^{\pi\xi}$. An important point that should be observed from the form of the solution \eqref{trAsol} is the dependence of the normalization (amplification) factor on the dimensionless parameter $\delta$ which can be derived in terms of the physical model parameters (see Sections \ref{s4p1p2} and \ref{s4p1p3}). As we explain in the main text this parameter is a measure of the scalar's mass around global minimum $\delta \approx m_\chi^2 / H^2$ and hence determines the rate at which the field rolls towards its minimum. In this sense it determines the width of gauge field modes that takes part in the particle production: the faster the scalar traverses the region in its potential where the velocity is maximal (\ie with larger acceleration $\dot{\xi}/(\xi H) \propto \ddot{\bar{\chi}}/(\dot{\bar{\chi}} H) \sim \delta$), each gauge field mode spends less time in the tachyonic region and fewer of them will be excited by the rolling scalar, leading to a sharper distribution of excited modes. These arguments clarifies the dependence of the Normalization factor $N_A$ in \eqref{trAsol} on the parameter $\delta$ along with the $\xi_*$ and $x_* = k/k_*$ that characterize the amplification and scale dependence of the particle production process.

At fixed $\xi_*$ and $\delta$, the scale dependence of the normalization factor can then be obtained by solving numerically \eqref{meqa} for a grid of $x_* = k/k_*$ values and matching these solutions to the WKB solution given at late times. In this way, one can confirm that $N_A$ is given by a log-normal distribution for the models we consider in Sections \ref{s4p1p2} and \ref{s4p1p3}:
\beq\label{Nform}
N_A\left(\xi_{*}, x_*, \delta\right) \simeq N_A^{c}\left[\xi_{*}, \delta\right] \exp \left(-\frac{1}{2 \sigma_A^{2}\left[\xi_{*}, \delta\right]} \ln ^{2}\left(\frac{x_*}{q_A^{c}\left[\xi_{*}, \delta\right]}\right)\right),
\eeq
where the functions $N_A^{c}, q_A^c$ and $\sgm_A$ parametrizes the background dependence of gauge field production through their dependence on $\xi_*$ and $\delta$. In particular, accurate fitting formulas for these quantities can be obtained at fixed $\delta$ in terms of $\xi_*$. For the parameter choices we adopt in this review, these formulas can be found in \cite{Ozsoy:2020ccy,Ozsoy:2020kat,Peloso:2016gqs}.
\smallskip

\noindent{\bf The ‘‘Electric'' and ‘‘Magnetic'' fields as sources.} For future reference we also note the electric and magnetic fields which are related to the auxiliary potential $A_i$ as
\beq\label{EandB}
E_i (\tau, \vec{x}) = -\frac{1}{a^2} \partial_\tau A_i(\tau, \vec{x}), \quad\quad B_i(\tau, \vec{x}) = \frac{1}{a^2} (\,\vec{\nabla} \times \vec{A}(\tau,\vec{x})\,)_i = \frac{1}{a^2} \epsilon_{ijk}\, \partial_j A_k(\tau,\vec{x}).
\eeq
Utilizing the decomposition \eqref{GFME} together with \eqref{hv} and \eqref{Akhat}, we then take into account only the growing part (\ie real part) of the solutions (corresponding to the physical amplification of vector fields caused by a rolling scalar) we derived in \eqref{srAsol} and \eqref{trAsol} to express the Fourier decomposition of the $\vec{E}$ and $\vec{B}$ fields as follows
\beq
\hat{E}_i(\tau,\vec{x}) = \int \frac{\d^3 k}{(2\pi)^{3/2}}\, \mathrm{e}^{i\vec{k}\cdot\vec{x}} \,\hat{E}_{i}(\tau,\vec{k}), \quad\quad\quad \hat{B}_i(\tau,\vec{x}) = \int \frac{\d^3 k}{(2\pi)^{3/2}}\, \mathrm{e}^{i\vec{k}\cdot\vec{x}}\, \hat{B}_{i}(\tau,\vec{k}),
\eeq
where
\begin{align}\label{EandBs}
    \nn \hat{E}_i(\tau,\vec{k}) &= - \sqrt{\fr{k}{2}}\,\fr{\epsilon_{i}^{-}(\vec{k})}{a(\tau)^{2}} \left(\fr{2\xi(\tau)}{-k\tau}\right)^{1/4}  N_A(\xi_*,-k\tau_*,\delta) \exp \left[-E(\tau)\sqrt{-2\xi_*k\tau}  \right]\hat{\mathcal{O}}_{-}(\vec{k}),\\
\hat{B}_i(\tau,\vec{k}) &= -\sqrt{\fr{k}{2}}\,\fr{\epsilon_{i}^{-}(\vec{k})}{a(\tau)^{2}}\left(\fr{-k \tau}{2\xi(\tau)}\right)^{1/4} N_A(\xi_*,-k\tau_*,\delta) \exp \left[-E(\tau){\sqrt{-2\xi_*k\tau}}\right]\hat{\mathcal{O}}_{-}(\vec{k}),
\end{align}
where we defined the short-hand notation $\hat{\mathcal{O}}_{\lambda}(\vec{k}) = [a_{\lambda}(\vec{k}) + a^{\dagger}_{\lambda}(-\vec{k})]$. Note that the case of particle production through a slowly-rolling scalar with $\xi \simeq {\rm cons.}$ can be recovered from the formulas by making the following replacements $\xi_* \to \xi$, $E(\tau) \to 1$ and $N_A \to \mathrm{e}^{\pi\xi}$. Having studied the particle production in the gauge field sector, we now study the expectation values including electromagnetic fields before we discuss how these sources influence the fluctuations of the scalar sector.
\smallskip

\noindent{\bf Expectation values involving gauge fields.} Noting again the decomposition of the vector fields \eqref{GFME} (along with \eqref{hv} and \eqref{Akhat}) and the definition of their electromagnetic counterparts \eqref{EandB}, the energy density of the gauge fields $\rho_A$ and $\langle \vec{E}.\vec{B}\rangle$ can be expressed as
\begin{align}\label{Int}
\nn\rho_A \equiv \fr{1}{2}\langle\vec{E}^2+\vec{B^2}\rangle &= \int \d \ln k \, \frac{\d \rho_A}{\d \ln k}, \\
\langle \vec{E}.\vec{B}\rangle &= \int \d \ln k \,\frac{\d \langle \vec{E}.\vec{B}\rangle}{\d \ln k}.
\end{align}
Taking into account only the amplified mode function $A_{-}$ of the gauge field, the energy density $\rho_A$ and $\langle \vec{E}.\vec{B}\rangle$ per logarithmic wave-number is given by
\begin{align}\label{Intg}
\nn \frac{\d \rho_A}{\d \ln k} &\simeq \frac{H^4}{8\pi^2}\,x^4\, \left(\frac{2\xi}{x} +  1\right)\,\left|\tilde{A}_{-}(x)\right|^2, \\
\frac{\d \langle \vec{E}.\vec{B}\rangle}{\d \ln k} &\simeq - \frac{H^4}{8\pi^2}\, x^4 \, \frac{\d}{\d x} \left|\tilde{A}_{-}(x)\right|^2,
\end{align}
where we utilized \cite{Peloso:2016gqs}, 
\beq
A'_{-} = \sqrt{\frac{2k \xi}{-\tau}} A_{-}^{*} \quad \to \quad \frac{\d \tilde{A}_{-}}{\d x} = - \sqrt{\frac{2\xi}{x}} \tilde{A}_{-}^{*}
\eeq
defining the the following dimensionless variables: $x = - k\tau$ and $\sqrt{2k}\, A_{-}(\tau, k) \equiv \tilde{A}_{-}(x)$. Employing the solutions for $A_{-}$ we derived in \eqref{srAsol} and \eqref{trAsol}, one can use the formulas \eqref{Intg} and \eqref{Int} to obtain $\rho_A$ and $\langle\vec{E}.\vec{B}\rangle$. In particular for the localized gauge field production models presented in Section \ref{s4p1p2} (see \cite{Ozsoy:2020kat}) and \eqref{s4p1p3} (see \cite{Ozsoy:2020ccy}), these formulas can be used to justify the negligible back-reaction of the gauge fields on the background evolution. We will not repeat these calculations here for the localized production case, however below we will derive the relevant formulas for the slowly-rolling smooth axion inflation of Section \ref{s4p1p1}. For this purpose, we focus on the real part of the solution \eqref{srAsol} that corresponds to the physical amplification of the gauge field fluctuations. Plugging the solution in \eqref{Intg} and \eqref{Int}, we obtain 
\begin{align}\label{rhoAandEBsr}
\nn \rho_A &= \fr{H^4~ e^{2\pi\xi}}{8\pi^2 (2\xi)^{1/2}}\int_0^{2\xi} \d x ~x^{7/2} \left(\fr{2\xi}{x}+1\right) e^{-4\sqrt{2\xi x}},\\
\langle \vec{E}.\vec{B}\rangle &= \fr{H^4~ e^{2\pi\xi}}{4\pi^2} \int_0^{2\xi} \d x ~x^{3} \left(1 - \frac{1}{4\sqrt{2\xi x}}\right) e^{-4\sqrt{2\xi x}},
\end{align}
where we send the lower limits of the integrals to zero as the integrands converge in the $x \to 0$ limit. Similarly, since we are only focusing on the physical amplification of the gauge fields by throwing away the imaginary part of the solution \eqref{srAsol}, the integrands vanish quickly deep in the UV $x \to \infty$ and so we can also send the upper limit of the integrals in \eqref{rhoAandEBsr} to infinity, $2\xi \to \infty$. Finally, by making a change of variable to $ 4 \sqrt{2\xi x} = y$, one can realize that the resulting integrals can be carried analytically and in fact they are proportional to Gamma functions with integer arguments. In particular, we get 
\begin{align}\label{evfsr}
\nn \rho_A &= \fr{H^4}{\xi^3} ~ e^{2\pi\xi}~\fr{\Gamma(7)}{2^{19}\pi^2} \left[1  + \fr{1}{2^6 \xi^2}\fr{\Gamma(9)}{\Gamma(7)} \right], \\   
\langle \vec{E}.\vec{B}\rangle &= \fr{H^4}{\xi^4}\, e^{2\pi\xi} ~ \frac{\left[\, \Gamma(8)  - \Gamma(7)\,\right]}{2^{21}\pi^2}.
\end{align}
Inserting the numerical values of the Gamma functions, these expressions give rise to the Eq. \eqref{ebarhoA} we provide in the main text. 
\subsection{Scalars sourced by vector fields, the direct coupling case: $\chi = \phi$.}\label{AppD2}
To understand the sourcing of scalar fluctuations by the gauge fields,  we should consider the gravitational and inflaton (that we refer to $S_{\rm inf}$) action in addition to $S_{\rm GF}$ : $S_{\rm tot} = S_{\rm inf} + S_{\rm GF}$ where 
\beq
S_{\rm inf} = \int \d^4 x \sqrt{-g} \left\{\frac{\Mpl^2}{2}R - \frac{1}{2}\partial_\mu \phi \partial^{\mu} \phi - V(\phi)\right\}.  
\eeq
Expanding the action in terms of the scalar fluctuations $\delta N$, $N^{i}$ and $\delta \phi$ around a background solution, one gets the following linear and second order actions,
\begin{align}
\label{Sinf1}    S^{(1)}_{\rm inf} &= \int \d^4 x\,a^3\, \bigg\{\left[3H^2\Mpl^2 - \frac{1}{2}\dot{\bar{\phi}}^2-V(\bar{\phi}) -{\frac{1}{2}\langle\vec{E}^2+\vec{B}^2\rangle}\right]\delta N\\\nn&\quad\quad\quad\quad\quad\quad\quad\quad\quad\quad\quad\quad\quad\quad\quad\quad\,\,\,-\left[\ddot{\bar{\phi}} +  3H\dot{\bar{\phi}}+V'(\bar{\phi}) -{\fr{g_{\rm cs}}{f} \langle\vec{E}.\vec{B}\rangle} \right]\delta \phi\bigg\},\\
\label{Sinf2}      S^{(2)}_{\rm inf} &=\frac{1}{2} \int \d^4 x\,a^3\, \bigg\{\delta\dot{\phi}^2 - \frac{(\partial_i \delta \phi)^2}{a^2} - V''(\bar{\phi})\,\delta \phi^2 -3H^2\Mpl^2 \delta N^2-2H\Mpl^2 \partial_i N^{i} \delta N\\\nn& 
    \quad\quad\quad\quad\quad\quad\quad\quad\quad\quad\quad\quad\quad - 2V'(\bar{\phi})\,\delta\phi\,\delta N - 2\dot{\bar{\phi}}\, \delta\dot{\phi}\,\delta N -  2\dot{\bar{\phi}}\, N^{i} \partial_i \delta\phi + \dot{\bar{\phi}}^2 \delta N^2  \bigg\}.
\end{align}
Notice that in $S^{(1)}_{\rm inf}$ \eqref{Sinf1}, we subtracted the tadpole terms $\propto \langle \vec{E}.\vec{B}\rangle, \langle \vec{E}^2 + \vec{B}^2 \rangle$ that we introduced earlier in Eqs. \eqref{S31} and \eqref{S32}. By virtue of the background equations presented in Eq. \eqref{mkgafe}, these terms precisely cancel. The terms that contains gravitational fluctuations $\delta N$ and $N^{i}$ in the second order action $S^{(1)}_{\rm inf}$ \eqref{Sinf2} induces additional mass contribution to the inflaton fluctuations which can be seen by varying this action with respect to $\delta N$ and $N^{i}$ and solving them in terms of $\delta \phi$ as 
\beq\label{solgrav}
\delta N=-\sqrt{\frac{\epsilon}{2}}\frac{\delta\phi}{\Mpl} , \quad \partial_i N^i=\sqrt{\frac{\epsilon}{2}}\frac{1}{\Mpl}\left(\delta\dot{\phi}-\frac{\eta H}{2} \delta\phi\right),
\eeq
where we defined the Hubble slow-roll parameters as 
\beq
\epsilon=-\frac{\dot{H}}{H^2}=\frac{\dot{\bar{\phi}}^2}{2 H^2 M_{\mathrm{pl}}^2}, \quad \text { and } \quad \eta=\frac{\dot{\epsilon}}{\epsilon H}. 
\eeq
Similar to our discussion regarding the cubic terms induced by gravity within the gauge field sector (see \eg \eqref{S32}), the influence of the lapse and shift on the second order action of the inflaton fluctuations can be typically ignored in the slow-roll regime (as well as for the non-attractor era associated with the model in Section \ref{s4p1p2} where $\epsilon \to 0$, $|\eta| \simeq \mathcal{O}(1)$) which is a situation that is generically referred as the decoupling limit of gravity within the literature. For completeness however we will keep them here. Plugging \eqref{solgrav} in the action \eqref{Sinf2}, we perform several integration by parts and taking into account the source terms induced by the gauge fields ($\delta \chi \to \delta \phi$ in Eq. \eqref{S31}), the equation of motion obeyed by the inflaton fluctuations can be written as
\beq\label{eomdphi}
\delta\ddot{\phi} + 3H\delta\dot{\phi} -\left(\vec{\nabla}^2 - m_{\rm eff}^2(t)\right)\delta \phi = \frac{g_{\rm cs}}{f} \left[\vec{E}.\vec{B} - \langle \vec{E}.\vec{B}\rangle\right] + {\frac{g_{\rm cs}}{f}\frac{\partial \langle\vec{E}.\vec{B}\rangle}{\partial \dot{\bar{\phi}}} \delta \dot{\phi}}
\eeq
where $\vec{\nabla}^2 = \partial_i \partial_i$ is the Laplacian in flat Euclidean space and the effective time-dependent mass is defined as $m_{\rm eff}^2 = V''(\bar{\phi}) - (2\epsilon \eta + 6\epsilon - 2\epsilon^2) H^2$. Noting the second derivative of the potential in terms of the slow-roll parameters:
\beq
V''(\bar{\phi}) = H^2 \left[ - \frac{3\eta}{2} + \frac{5\epsilon\eta}{2}-\frac{1}{4}\eta^2 -\frac{\dot{\eta}}{2H}- 2\epsilon^2+6\epsilon\right],
\eeq
the effective time dependent mass can be described fully in terms of the slow-roll parameters as 
\beq
m_{\rm eff}^2 =  H^2 \left[\frac{9}{4} - \frac{1}{4}(\eta + 3)^2 +  \frac{\epsilon \eta}{2} - \frac{\dot{\eta}}{2H}\right].
\eeq
In \eqref{eomdphi}, the last term is introduced to parametrize the additional sources that might arise in the strong back-reaction regime. In particular, as $\langle \vec{E}.\vec{B} \rangle$ grows during inflation (\ie as the effective coupling $\xi$ increases during inflation, see Eq. \eqref{ebarhoA}), this will first have the effect of sourcing the inflaton perturbations through the first term in the right hand side of \eqref{eomdphi}. As a result the inflaton perturbations starts to grow, and eventually the solution of the gauge field modes obtained by assuming a homogeneous inflaton will no longer be valid, and expected to go from the solution \eqref{srAsol} to a more general solution  $A_{-}[\bar{\phi} + \delta \phi]$. The additional term in the right hand side of \eqref{eomdphi} precisely introduced to capture the influence of this modified solution of the gauge fields on the inflaton perturbations. Since gauge field production (and $\langle \vec{E}.\vec{B} \rangle$) is sensitive to the velocity of the homogeneous inflaton mode, it is reasonable to expect the influence of this additional source to be proportional to $(\partial \langle \vec{E}.\vec{B} \rangle/\partial \dot{\bar{\phi}}) \delta \dot{\phi}$ as shown in \eqref{eomdphi} (see also \cite{Anber:2009ua,Garcia-Bellido:2016dkw,Linde:2012bt} for a detailed discussion on this point). Recalling $\xi = -g_{\rm cs} \dot{\bar{\phi}}/(2Hf)$ and \eqref{ebarhoA}, we note 
\beq
\frac{g_{\rm cs}}{f}\frac{\partial \langle\vec{E}.\vec{B}\rangle}{\partial \dot{\bar{\phi}}} \simeq  \frac{g_{\rm cs}}{f}\, \frac{\langle\vec{E}.\vec{B}\rangle}{\dot{\bar{\phi}}}\, {2\pi \xi},
\eeq
and so the influence of the source term can be parametrized as an additional damping term in the equation of motion of the inflaton fluctuations as
\beq
\delta\ddot{\phi} + 3H{{\beta}}\,\delta\dot{\phi} -\left(\vec{\nabla}^2 - m_{\rm eff}^2(t)\right)\delta \phi = \frac{g_{\rm cs}}{f} \left[\vec{E}.\vec{B} - \langle \vec{E}.\vec{B}\rangle\right],
\eeq
where 
\beq\label{betaff}
\beta = 1 - \frac{g_{\rm cs}}{f}\, \frac{\langle\vec{E}.\vec{B}\rangle}{3 H \dot{\bar{\phi}}}\, {2\pi \xi}.
\eeq
Assuming $\beta = 1$ amounts to neglecting the influence back-reaction effects of the inflaton fluctuations on the gauge field solutions. This is for example the approach taken in \cite{Ozsoy:2020kat}, for the bumpy axion inflation we discuss in Section \ref{s4p1p2} with the reasoning that back-reaction effects are mild due to the localized nature of gauge field production (and so does the resulting increase in inflaton fluctuations) in the presence of a transiently increasing effective coupling $\xi$ between scalar and gauge field sector. In the smooth axion inflation (Section \ref{s4p1p1}) however, gauge field sources and the resulting scalar fluctuations continuously grow which can eventually influence the dynamics of the fluctuations through an additional (positive) friction term in $\beta$ \eqref{betaff} as soon as the homogeneous dynamics of the inflaton enters into the back-reaction regime with $g_{\rm cs} \langle \vec{E}.\vec{B} \rangle / (3H|\dot{\bar{\phi}}|f) \sim \mathcal{O}(1)$.  

Armed with the equations of motion in real space, one can study vacuum and sourced solutions of the inflaton perturbations in Fourier space by using the splitting $\delta \phi = \delta\phi_{\rm v} + \delta\phi_{\rm s}$ and utilizing the standard Green function methods. We will not repeat these computations here, for the calculation of the scalar power spectrum relevant for PBH formation, interested readers can follow the works of \cite{Anber:2009ua,Barnaby:2011vw,Linde:2012bt, Garcia-Bellido:2016dkw} in the context of smooth axion inflation and \cite{Ozsoy:2020kat} in the context of bumpy axion inflation we discuss in Sections \ref{s4p1p1} and \ref{s4p1p2}.

\subsection{Scalars sourced by vector fields, the indirect coupling case: $\chi = \sigma$.}\label{AppD3}
To capture the dynamics of scalar fluctuations in the models studied in Section \ref{s4p1p3}, we extend the inflationary action with a spectator sector $\sigma$ that interacts with the gauge fields as in \eqref{SGF}. The action that describes inflation is therefore given by
\beq\label{SSinf}
S_{\rm inf} = \int \d^4 x \sqrt{-g} \left\{\frac{\Mpl^2}{2}R - \frac{1}{2}\partial_\mu \phi \partial^{\mu} \phi - V(\phi)- \frac{1}{2}\partial_\mu \sigma \partial^{\mu} \sigma - U(\sigma)\right\},  
\eeq
where we assume that the spectator axion-like field do not contribute significantly to the background evolution during inflation which is mainly driven by a flat enough inflaton potential $V$ that we will leave unspecified. 

Expanding the action in terms of the scalar fluctuations $\delta N$, $N^{i}$ and $\delta \phi,\delta \sigma$ around a background solution, we obtain the following linear and second order actions,
\begin{align}
\label{SSinf1}    S^{(1)}_{\rm inf} &= \int \d^4 x\,a^3\, \bigg\{\left[3H^2\Mpl^2 - \sum_{a}\left(\,\frac{1}{2}\dot{\bar{\phi}}_a^2-V_a(\bar{\phi})\right) -{\frac{1}{2}\langle\vec{E}^2+\vec{B}^2\rangle}\right]\delta N\\\nn
&\quad\quad\quad\quad\quad\quad -\left[\ddot{\bar{\phi}} +  3H\dot{\bar{\phi}}+V'(\bar{\phi}) \right]\delta \phi -\left[\ddot{\bar{\sigma}} +  3H\dot{\bar{\sigma}}+U'(\bar{\sigma}) -{\fr{g_{\rm cs}}{f} \langle\vec{E}.\vec{B}\rangle} \right]\delta \sigma\bigg\},\\
\label{SSinf2}      S^{(2)}_{\rm inf} &=\frac{1}{2}\sum_a \int \d^4 x\,a^3\, \bigg\{\delta\dot{\phi}_a^2 - \frac{(\partial_i \delta \phi_a)^2}{a^2} - V''_a(\bar{\phi}_a)\,\delta \phi_a^2 -3H^2\Mpl^2 \delta N^2-2H\Mpl^2 \partial_i N^{i} \delta N\\\nn& 
    \quad\quad\quad\quad\quad\quad\quad\quad\quad\quad\quad\quad\quad - 2V'_a(\bar{\phi}_a)\,\delta\phi_a\,\delta N - 2\dot{\bar{\phi}}_a\, \delta\dot{\phi}_a\,\delta N -  2\dot{\bar{\phi}}_a\, N^{i} \partial_i \delta\phi_a + \dot{\bar{\phi}}_a^2 \delta N^2  \bigg\},
\end{align}
where the summation over $a$ runs over fluctuations and background values of the two fields: $\phi_a = \{\phi,\sigma\}$ and $V_a = \{V(\bar{\phi}),U(\bar{\sigma})\}$. Varying \ref{SSinf2} with respect to Lagrange multipliers $\delta N$ and $N^{i}$, we obtain 
\begin{align}
\label{LS1}2H\Mpl^2 \, \delta N &= \sum_a \dot{\bar{\phi}}_a \delta \phi_a ,\\
\label{LS2} -2 H \Mpl^2 \,\partial_i N^{i} &= \sum_a \left(\dot{\bar{\phi}}_a\delta \dot{\phi}_a + V_a'(\bar{\phi}_a)\delta \phi_a\right) +\left(6H^2\Mpl^2 -\sum_a \dot{\bar{\phi}}_a^2\right)\delta N.
\end{align}
Plugging these solutions back in the actions, we obtain the following second order action for scalar fluctuations \cite{Ferreira:2014zia,Ozsoy:2017blg},
\begin{align}
\label{SSinf2f}S_{\rm inf}^{(2)} &=\frac{1}{2}\int  \d^4 x \, a^3 \, \Bigg\{ \left[\delta \dot{\phi}_a^2-\frac{\left(\partial_i\delta \phi_a\right)^2}{a^2}\right]-m_{ab}^2 \delta \phi_a\delta \phi_b\Bigg\},\\
\label{mm}m_{ab}^2 &= V^{(\rm tot)}_{,ab} - \frac{1}{a^3}\frac{\d}{\d t}\left(\frac{a^3}{H}\frac{\dot{\bar{\phi}}_a \dot{\bar{\phi}}_b}{\Mpl^2}\right).
\end{align}
where summation over repeated indices $a,b$ is implied and $V^{(\rm tot)}_{,ab}\equiv \partial^2 V^{(\rm tot)}/(\partial\bar{\phi}_a\partial\bar{\phi}_b)$ with $V^{(\rm tot)} = \sum_a V_a(\bar{\phi}_a) = V(\bar{\phi}) + U(\bar{\sigma})$. We note that in deriving the expressions above we used background equations of motion ignoring the artificially introduced mean field expectation values involving gauge fields (noticing that these terms that appear in \eqref{S31}, \eqref{S32} and \eqref{SSinf1} sums up to zero): 
\begin{align}\label{BGX}
\ddot{\bar{\phi}}_a + 3H\dot{\bar{\phi}}_a &+ V_a'(\bar{\phi}_a)=0, \\
-2 \dot{H}\Mpl^2 &= \sum_a \dot{\bar{\phi}}_a^2 .
\end{align} 
Indeed for the spectator models discussed in Section \ref{s4p1p3}, due to localized nature of the gauge field production back-reaction effects is small and can be ignored \cite{Peloso:2016gqs,Garcia-Bellido:2016dkw,Ozsoy:2020ccy}, and therefore the background evolution can be studied in a consistent way by focusing on the equations in \eqref{BGX} along with the Friedmann equation $3H^2\Mpl^2 = \sum_a \dot{\bar{\phi}}_a^2 + V_a(\bar{\phi}_a)$. Taking into account the effects of gauge field sources ($\delta \chi \to \delta \sigma$ in Eq. \eqref{S31}) on the spectator scalar fluctuations, the equations of motion for the scalar perturbations read as 
\beq\label{SS}
\delta\ddot{\phi}_a + 3H\delta\dot{\phi}_a - \left(\vec{\nabla}^2 - V''_a\right)\delta\phi_a - \sum_b\frac{1}{a^3}\frac{\d}{\d t}\left(\frac{a^3}{H}\frac{\dot{\bar{\phi}}_a \dot{\bar{\phi}}_b}{\Mpl^2}\right) \delta\phi_b = J_a(\vec{E},\vec{B}),
\eeq
where $\delta \phi_a = (\delta \phi, \delta \sigma)^T$ and $J_a = (0, \frac{g_{\rm cs}}{f}\vec{E}.\vec{B})^T$. 
Notice that since we consider sum separable potentials $V^{(\rm tot)} = V + U$, the first term in the mass matrix \eqref{mm} makes a diagonal contribution to the mass of the each field in the equations of motion \eqref{SS}. However, the presence of the second term in the mass matrix \eqref{mm}, which is induced by the presence of gravitational fluctuations $\delta N$ and $N^{i}$, introduces a mass mixing between scalar fluctuations $\delta\phi_a - \delta\phi_b$ ($a \neq b$)(\ie through the last term in \eqref{SS}). In other words, although we consider a Lagrangian \eqref{SSinf} where the two scalar fields appear to be decoupled from each other, gravitational interaction will inevitably introduce a minimal communication channel between the physical fluctuations of the two scalar sectors. At leading order in the slow-roll expansion, the mass mixing ($a \neq b$) originates from the following terms in the Lagrangian 
\beq
\mathcal{L}_{\rm mix} = \frac{1}{2a^3}\frac{\d}{\d t}\left(\frac{a^3}{H}\frac{\dot{\bar{\phi}}_a \dot{\bar{\phi}}_b}{\Mpl^2}\right) \delta\phi_a \delta\phi_b \quad \longrightarrow \quad \mathcal{L}_{\rm mix} \simeq 6 \sqrt{\epsilon_\phi \epsilon_\sigma}\, H^2 \delta \phi \delta \sigma,
\eeq
where $\epsilon_a = \dot{\bar{\phi}}_a^2 / (2H^2\Mpl^2)$ is the slow-roll parameter of the each field. Therefore as long as both fields roll with a non-vanishing velocity $\dot{\bar{\sigma}}, \dot{\bar{\phi}} \neq 0$ during inflation, their fluctuations can be converted to one another. In the presence of particle production in the gauge field sector, the mixing between the two scalar sectors is crucial in understanding the influence of the gauge field sources on the visible scalar sector fluctuations $\delta \phi$. In order to see this explicitly, we focus on the leading order mixing in the slow-roll expansion to rewrite the system of equations in \eqref{SS} as 
\begin{align}\label{SSEQ}
\nn \delta\ddot{\phi} + 3H\delta\dot{\phi} -\left(\vec{\nabla}^2 - m_\phi^2\right)\delta \phi \simeq 6 \sqrt{\epsilon_\phi \epsilon_\sigma}\, H^2 \delta \sigma,\\
\delta\ddot{\sigma} + 3H\delta\dot{\sigma} -\left(\vec{\nabla}^2 - m_\sigma^2\right)\delta \sigma\simeq \frac{g_{\rm cs}}{f} \vec{E}.\vec{B},
\end{align}
where $m_a^2 \simeq V''_a - 6 \epsilon_a H^2$ at leading order in the slow-roll expansion. We note that in the second line of \eqref{SSEQ}, we ignored mixing terms that can source spectator fluctuations $\delta \sigma$ through $\delta \phi$ which is a sub-leading effect compared to sourcing of $\delta \sigma$ by the enhanced gauge fields. This amounts to considering the main production channel of the observable scalar fluctuations schematically as: $\delta A + \delta A \to \delta \sigma \to \delta \phi$. Using the equations of motion in real space \eqref{SSEQ} and considering its sources in \eqref{EandBs}, the observable power spectrum of scalar fluctuations can be carried by using Green's function methods in Fourier space following detailed calculations presented in \cite{Namba:2015gja, Ozsoy:2020ccy} along with the prescription we present in Appendix \ref{AppE} regarding the curvature perturbation. 

\section{Curvature perturbation}\label{AppE}
In this short appendix, we define the curvature perturbation $\mathcal{R}$ on comoving slices and study its parametric dependence on the matter fluctuations for some of the scenarios we consider in the main text. We begin by noting the definition of comoving curvature perturbation in flat gauge which reads as \cite{Malik:2008im,Baumann:2009ds}:
\beq\label{R}
\mathcal{R} = - \frac{H}{(\bar{\rho}+\bar{P})} \,\delta q_{\rm flat}
\eeq
where $\bar{\rho}$ and $\bar{P}$ is the total background energy density and pressure (see Appendix \ref{AppA}) and $\delta q_{\rm flat}$ is the scalar momentum density in flat gauge. In terms of the perturbed energy momentum tensor, $\delta q_{\rm flat}$ is given by
\beq\label{defT}
\delta T^{0}_i = \partial_i \delta q_{\rm flat} \quad \longleftarrow \quad T_{\mu \nu} \equiv - \frac{2}{\sqrt{-g}} \frac{\delta S_m}{\delta g^{\mu\nu}},
\eeq
where the definition of the unperturbed energy momentum tensor provided on the right hand side of \eqref{defT} can be used for a given matter action describing the system. We note that in this definition $\delta S_m /\delta g^{\mu\nu}$ represents the variation of the matter action with respect to the metric field. Anticipating that we will focus on different limits of a more complicated/general model, we start by parametrizing perturbed $\delta T^{0}_i$ for the spectator axion model we discuss in Section \ref{s4p1p3} with the matter action given by the sum of \eqref{SSinf} and \eqref{SGF}. Noting that the last term in \eqref{SGF} is topological and does not gravitate, we have
\beq\label{dT0i}
\delta T^{0}_i = g^{0\mu} \left(\partial_\mu \phi \partial_i \delta \phi + \partial_\mu \sigma \partial_i \delta \sigma + g^{\rho\sigma}F_{\mu\rho}F_{i\sigma}\right),
\eeq
where the last term characterize the contribution from gauge fields which is second order in fluctuations. Following \eqref{dT0i}, \eqref{defT} and \eqref{R}, it is therefore clear that for a multi-sector inflationary model, curvature perturbation can in principle obtain contributions from all fields present in the matter Lagrangian. In what follows, we take \eqref{dT0i} as the main reference point to study different limits of it to derive an expression for the curvature perturbation $\mathcal{R}$ relevant for some of the models we study in the main text. 
\smallskip

\noindent{\bf Canonical single-field inflation.} This case corresponds to neglecting terms proportional to spectator and gauge field fluctuations in \eqref{dT0i}. Noting that $\bar{\rho} + \bar{P} \to \dot{\bar{\phi}}^2$ for single-field canonical inflation, to linear order in inflaton fluctuations, the curvature perturbation \eqref{R} is given by 
\beq\label{RSF}
\mathcal{R} = \frac{H}{a\dot{\bar{\phi}}} (a \delta \phi) \quad \leftrightarrow \quad \mathcal{R} = \frac{v}{z},
\eeq
where we made the connection between the MS variable (see Section \ref{S3p1}) and inflaton fluctuations clear $v = a \delta \phi$ using the definition of pump field in this case $z = -a\dot{\bar{\phi}}/H = a \sqrt{2\epsilon} \Mpl$. Upon canonical quantization of $\delta\phi$ (and hence $\mathcal{R}$), their corresponding Fourier space variables satisfy the same relation in \eqref{RSF} and the dimensionless power spectrum at the end of inflation can be computed via \eqref{c13}.
\smallskip

\noindent{\bf Smooth Axion Inflation and Bumpy Axion Inflation.} For the models we study in Sections \ref{s4p1p1} and \eqref{s4p1p2}, we instead focus on the curvature perturbation $\zeta$ on uniform density gauge which can be related to the density perturbation in flat gauge $\delta \rho_{\rm flat}$ and comoving curvature perturbation $\mathcal{R}$ on super-horizon scales as \cite{Malik:2008im}

\beq
\zeta \simeq -\mathcal{R} =  -\frac{H}{\dot{\bar{\rho}}} \delta \rho_{\rm flat},
\eeq
where $\bar{\rho}$ is the total energy density of the axion gauge field system. Using background equations of the system, time derivative of the total energy density is given by $\dot{\bar{\rho}} = -3 H \dot{\bar{\phi}}^2 - 4 H \rho_A$ \cite{Notari:2016npn} where $\rho_A$ is the energy density of the gauge field as defined in \eqref{ebarhoA}. Noting this relation, to linear order \footnote{We note that $\mathcal{R}$ also takes up a contribution bilinear in gauge fields, proportional to the absolute value of the Poynting vector $\mathcal{R}_{(AA)} \propto a |\vec{E}\times \vec{B}|$. We expect this contribution to be negligible at the end of inflation (\ie at the time we are interested in the correlators of $\mathcal{R}$) because the gauge field production saturates on super-horizon scales and the  corresponding electromagnetic fields decay as  $\vec{E}, \vec{B} \propto a^{-2}$. See \eg the discussion presented in \cite{Caprini:2017vnn,Ozsoy:2020ccy} within similar contexts.} in the cosmological fluctuations, the comoving curvature perturbation $\mathcal{R}$ can be obtained as \cite{Garcia-Bellido:2016dkw}
\beq\label{RAGF}
\mathcal{R} = \frac{H}{\dot{\bar{\phi}}} \delta \phi \,\, \mathcal{F}, \quad\quad \mathcal{F} \equiv \frac{-\dot{\bar{\phi}}\, V'(\bar{\phi})}{H (3 \dot{\bar{\phi}}^2 + 4 \rho_A)},
\eeq
where we used $\delta \rho_{\rm flat} \simeq V'(\bar{\phi}) \delta \phi$ neglecting contributions to the total energy proportional to the kinetic energy of the inflaton as they should be small both in the slow-roll and the inflationary regime dominated by the friction provided by the gauge fields. In \eqref{RAGF}, the factor $\mathcal{F}$ parametrizes the correction to the definition of the curvature perturbation in the strong back-reaction regime which must be taken account in the smooth axion inflation model we present in the main text. On the other hand, the standard relation in \eqref{RSF} applies in the regime of negligible back-reaction of the gauge field on the evolution of the inflaton and expansion of the universe, \ie when $\rho_A$ is negligible and the relation $-3H\dot{\bar{\phi}} \simeq V'(\bar{\phi})$ is satisfied. This situation applies both at early times during the smooth axion model we consider as well as within the bumpy axion inflation model of Section \ref{s4p1p2} where back-reaction effects have shown to be small around the peak of the scalar power spectrum. To summarize, in the strong back-reaction regime, the power spectrum of curvature perturbation can bu computed using the standard relation \eqref{RSF} times a factor of $\mathcal{F}$ correction.
\smallskip

\noindent{\bf Spectator axion-gauge field model.} In the model presented in \ref{s4p1p3}, in principle we need to consider all the contributions to the curvature perturbation using the formulas \eqref{dT0i}, \eqref{defT} and \eqref{R}. However, as explicitly checked in \cite{Ozsoy:2020ccy}, the contribution to the curvature perturbation that is bilinear in the gauge fields can be safely ignored at late times during which we are interested in the correlators of $\mathcal{R}$. Further simplifications on the functional form of $\mathcal{R}$ arise due to the spectator nature of the axion sector $\sigma$ and due to the assumption that it settles back to its global minimum long before the end of inflation where $\dot{\bar{\sigma}} \to 0$. Noting $\bar{\rho}+\bar{P} = \dot{\bar{\phi}}^2+\dot{\bar{\sigma}}^2 \simeq \dot{\bar{\phi}}^2$ and the fact that $\dot{\bar{\sigma}} \to 0$ long before inflation ends, the late time curvature perturbation obtains the following form 
\beq\label{Rtot}
\mathcal{R}=\fr{H}{a(\dot{\bar{\phi}}^{2}+\dot{\bar{\sigma}}^{2})}\left(\dot{\bar{\phi}}\, (a \delta \phi)+\dot{\bar{\sigma}}\,(a \delta \sigma)\right) \simeq \frac{H}{\dot{\bar{\phi}}}\delta \phi.
\eeq
Therefore the standard expression $\mathcal{R} = (H/\dot{\bar{\phi}}) \delta \phi$, valid in single-field inflation still provides a very good approximation for the computation of late time correlators of the curvature perturbation in this model.
\end{appendix}
{\small
\addcontentsline{toc}{section}{References}
\bibliographystyle{utphys}
\bibliography{paper2}
}
\end{document}